\documentclass[11pt,letterpaper]{article}
\usepackage{amsmath,amssymb,amsthm,amscd,a4wide}
\usepackage{epsfig,subfigure,graphics,graphicx}
\usepackage{color,xcolor}
\usepackage{kotex}
\usepackage{bm}
\usepackage{epstopdf}
\usepackage{booktabs}
\usepackage{ifpdf}
\usepackage{tikz}
\usepackage{comment}
\usepackage{natbib}
\usepackage{varioref,xr-hyper,hyperref}
\usetikzlibrary{calc}

\providecommand{\lt}{<}
\providecommand{\gt}{>}

\hypersetup{colorlinks,linkcolor=blue,citecolor=blue,filecolor=blue,urlcolor=blue}

\newtheorem{pro}{Proposition}[section]
\newtheorem{dfn}{Definition}[section]
\newtheorem{thm}{Theorem}[section]
\newtheorem{cor}{Corollary}[section]
\newtheorem{lem}{Lemma}[section]

\newtheorem{rem}{Remark}[section]

\definecolor{retirementblue}{RGB}{225,238,248}
\definecolor{continuationgray}{RGB}{174,174,174}
\definecolor{guidegray}{RGB}{165,165,165}

\numberwithin{equation}{section}

\def\bc{\bm{c}}
\def\bpi{\bm{\pi}}
\newcommand{\Vt}{\widetilde V}
\newcommand{\Vh}{\widehat V}
\newcommand{\Lt}{\widetilde{\mathcal L}}
\newcommand{\Ut}{\widetilde U_1}
\newcommand{\E}{\mathbb E}

\begin{document}

\title{\Large\textbf{Retirement Planning with\\
Borrowing Constraints and Income Support}\footnote{{\bf Funding}: 
Tae Ung Gang was supported by the National Research Foundation of Korea (NRF) Grant [Grant No.s RS-2026-25468875, RS-2026-25487734]. Yong Hyun Shin was supported by the NRF Grant [Grant No.s RS-2026-25477813, RS-2026-25580186]. }}

\author{Tae Ung Gang\footnote{Department of Mathematical Sciences, Ulsan National Institute of Science and Technology (UNIST), Ulsan 44919, Republic of Korea. \;E-mail: {\tt gahnghansahnae@unist.ac.kr}}\and
\and
Seyoung Park\footnote{Department of Finance, Accounting and Banking, Nottingham University Business School, Nottingham, UK. \;E-mail: {\tt seyoung.park@nottingham.ac.uk}
}\and
\and
Yong Hyun Shin\footnote{Department of Mathematics \& Research Institute of Natural Sciences, Sookmyung Women's University, Seoul 04310, Republic of Korea. Corresponding author.\;E-mail: {\tt yhshin@sookmyung.ac.kr}
}}

\date{\today}

\maketitle \pagestyle{plain} \pagenumbering{arabic}

\thispagestyle{empty}

\clearpage
\newpage

\titlepage

\begin{center}
\Large\textbf{Retirement Planning with\\
Borrowing Constraints and Income Support}
\end{center}

\thispagestyle{empty}

\begin{abstract}
We study how pension income, its pledgeability, and government support that supplements effective consumption shape an irreversible retirement decision. In a continuous-time life-cycle model, a worker chooses consumption, risky investment, and the timing of retirement, subject to a tighter borrowing limit after retirement. When retirement reduces the utility value of a given level of effective consumption, immediate retirement can be optimal on a bounded wealth interval: a worker below the interval retires when wealth rises to its lower boundary, while a worker above it retires when wealth falls to its upper boundary. Two scalar boundary equations characterize these distinct entry points. Risky investment drops discretely upon retirement at either interior boundary, even when it rises with wealth along the low-wealth working branch. In our two-boundary baseline, greater income support lowers the lower retirement threshold and raises the upper one, whereas greater pension pledgeability shifts both thresholds downward. Finally, because the residuals of the boundary equations represent weighted tail integrals of the instantaneous gain from postponing retirement, we completely characterize the optimal retirement region across all levels of post-retirement quality of life.
\end{abstract}

\bigskip




\clearpage
\newpage
\setcounter{page}{1}


\section{Introduction}

Retirement constitutes a critical, irreversible life-cycle regime shift that fundamentally alters an individual's income dynamics, financial flexibility, and credit accessibility. In the pre-retirement phase, labor income serves not only as a continuous income stream but also anchors the individual's borrowing capacity via the capitalization of future earnings. Post-retirement, this labor income is replaced by non-discretionary, less pledgeable retirement cash flows (e.g., pensions, annuities), thereby introducing more stringent state-dependent borrowing constraints. Consequently, the determination of optimal retirement timing hinges crucially on the structural interplay among income reduction, borrowing boundaries, and the provision of post-retirement income support.

Our framework internalizes the mechanism that retirement choices depend not only on the absolute level of retirement wealth, but also on the degree to which post-retirement resources mitigate the decline in an individual's standard of living. Whereas pension income provides liquid cash, a consumption-equivalent benefit -- such as an in-kind service -- supports living standards without increasing funds available for financial investment or debt repayment. Rather than explicitly modeling the institutional details of such programs, we introduce a parsimonious, reduced-form specification featuring an exogenous support parameter that augments effective consumption post-retirement. This approach captures the welfare dynamics of social safety nets while preserving the analytical tractability of the underlying stochastic control problem.

This paper analyzes a continuous-time optimal consumption, investment, and retirement planning problem where post-retirement income support is asymmetrically embedded into the preference structure \citep{Gang,Kim et al.}. Prior to labor market exit, the agent is endowed with a constant stream of labor income. Post-retirement, the agent receives consumption-equivalent income support that directly augments effective consumption. Crucially, unlike traditional subsistence-consumption models \citep{Achury,Chen} that impose a strict minimum consumption requirement, our formulation eschews an exogenous consumption floor. Consequently, the agent may optimally select a corner solution of zero consumption after retirement, because positive effective consumption funded by income support yields finite utility even when out-of-pocket consumption is zero. This approach isolates the welfare-enhancing effects of retirement support without enforcing an artificially restrictive consumption floor.

We further extend our analysis to encompass a less favorable post-retirement environment, a scenario highly relevant for low-income or financially vulnerable households whose labor market exit is frequently compounded by degraded health, restricted leisure opportunities, and institutional frictions. Within this adverse preference regime, we rigorously investigate the non-monotonic structural interaction between binding borrowing constraints and post-retirement quality of life. Specifically, we ask at which wealth levels the agent chooses to retire and how those thresholds respond to changes in pension income, income support, and post-retirement borrowing capacity.

First, we provide an analytical solution to the post-retirement portfolio and consumption problem under joint borrowing constraints and income support mechanisms. When retirement income support is sufficiently generous relative to the agent's effective borrowing capacity, a zero-consumption region emerges near the borrowing boundary, wherein the agent optimally sets consumption to zero while continuously deriving utility from the income support component. We focus on this regime throughout the paper and establish closed-form solutions together with a verification theorem; the complementary regime of smaller support is left for future research.

Second, we characterize the optimal retirement decision within an optimal stopping framework, where the post-retirement value function serves as the termination payoff. In inverse marginal-value coordinates, separate scalar boundary equations identify retirement entry from low and high wealth levels. Because their residuals represent weighted tail integrals of the instantaneous gain from postponing retirement, we obtain a complete characterization of the retirement region for every level of post-retirement quality of life without requiring additional verification conditions. When post-retirement life improves, the retirement region is always an upper half-line of wealth, and the threshold lies above the zero-consumption region if and only if labor income exceeds the critical level. When post-retirement life deteriorates, sufficiently wealthy agents prefer to remain in the labor force. Voluntary retirement can nevertheless be optimal over an intermediate wealth interval, where income support contributes more substantially to effective consumption relative to out-of-pocket consumption.

Third, distinguishing between the two retirement entry boundaries carries important policy implications. In our two-boundary baseline, greater income support lowers the wealth threshold required to retire from below and raises the wealth level at which a worker retires from above. Both shifts expand the bounded interval of wealth levels over which immediate retirement is optimal. Furthermore, risky investment increases with wealth along the low-wealth working branch and drops discretely upon retirement, regardless of whether entry occurs from below or from above.

This paper contributes to the growing literature on endogenous retirement and optimal stopping in continuous-time financial environments. Following the seminal formulation of discretionary stopping paired with optimal portfolio selection by \citet{Karatzas}, subsequent frameworks pioneered by \citet{Choi}, \citet{Farhi}, \citet{Choi2008}, \citet{Dybvig}, and \citet{Dybvig2011} systematically integrated labor income dynamics into retirement models. This stream of literature was further generalized by \citet{Jang2013}, \citet{Bensoussan}, \citet{Jang2020}, and \citet{Bensoussan2025} to capture the structural impacts of labor income risk. More recent theoretical dimensions include finite-horizon constraints yielding time-dependent stopping rules by \citet{Yang}, partial-information frameworks by \citet{Chen2022}, heterogeneous consumption by \citet{Jang2024}, and the incorporation of long-run income risks by \citet{Huang}. We adopt the consumption-equivalent support specification of \citet{Gang}. The central question here is how this support interacts with the loss of borrowing capacity at retirement: for the same preferences and income streams, can a rise in wealth induce one worker to retire while a fall in wealth induces another to do so?

In addition to the literature on endogenous retirement, we contribute significantly to the portfolio and wealth management paradigm by exploring the joint structural effects of credit constraints and institutional safety nets. Following the foundational framework of borrowing constraints by \citet{HP}, credit frictions have been extensively evaluated in asset allocation models by \citet{Dybvig}, \citet{Dybvig2011}, \citet{PJ}, \citet{Ahn}, \citet{Kim}, \citet{Choi2025}, and \citet{Lee}. Our paper departs structurally from these canonical portfolio choice models by introducing a state-dependent tightening of borrowing capacity upon retirement, reflecting the empirical reality that labor earnings are replaced by lower, less pledgeable retirement streams. Crucially, we intertwine this borrowing friction with the role of government income support embedded directly into the agent's preferences. This formulation establishes a distinction from traditional subsistence-consumption models \citep{Achury,Chen}, which characteristically enforce a binding consumption floor.

The remainder of this paper is organized as follows. Section 2 formulates the life-cycle optimal retirement planning model under joint credit constraints and preference-based income support. Section 3 solves the post-retirement regime, deriving analytical closed-form solutions for the value function and associated policies. Section 4 addresses the pre-retirement phase, characterizing the optimal retirement wealth thresholds and establishing existence and uniqueness via free-boundary analysis, and presents numerical results. Finally, Section 5 provides concluding remarks. All proofs are collected in the Appendix.


\section{The Model}\label{sec:model}

We consider an infinite-horizon optimal consumption, investment, and retirement planning problem in continuous time. Let $(\Omega,\mathcal{F},\mathbb{F},\mathbb{P})$ be a complete filtered probability space, where $\mathbb{F}:=(\mathcal{F}_{t})_{t\geq 0}$ is the augmented filtration generated by a standard one-dimensional Brownian motion $\beta=(\beta_t)_{t\geq 0}$.

The financial market consists of one risk-free bond and one risky stock. The asset price processes are given by
$$
	\mathrm{d}P_t=rP_t\mathrm{d}t,
$$
and
$$
	\mathrm{d}S_t=\mu S_t\mathrm{d}t+\sigma S_t\mathrm{d}\beta_t,
$$
where $r>0$ is the risk-free interest rate, $\mu>r$ is the expected stock return, and $\sigma>0$ is the stock volatility. We define the Sharpe ratio of the risky stock as follows:
$$
	\theta:=\frac{\mu-r}{\sigma}.
$$

We consider a representative agent who optimally chooses a consumption process $\bc=(c_t)_{t\geq 0}$, a portfolio process $\bpi=(\pi_t)_{t\geq 0}$, and a (voluntary) retirement time $\tau$. The process $c_t$ represents the consumption rate at time $t$, and $\pi_t$ denotes the dollar amount invested in the risky stock. The retirement time $\tau$ is an $\mathbb{F}$-stopping time. The retirement considered in this paper is irreversible, so the agent cannot return to the workforce after retirement.

Before retirement, the agent receives a constant labor income $Y_1>0$. Labor income provides both current resources for consumption and investment, and guarantees borrowing capacity through its future stream. After retirement, the agent receives a constant retirement income $0<Y_2<Y_1$, which is lower than labor income and a less pledgeable source of income. Hence, the agent's borrowing is assumed to be restricted after retirement. Defining the agent's income process $Y_{t}$ as
$$
	Y_t:=Y_1\mathbf{1}_{\{0\leq t<\tau\}}+Y_2\mathbf{1}_{\{t\geq \tau\}},
$$
the agent accumulates her wealth $W_{t}$ according to the following wealth dynamics:
\begin{equation} \label{eq:wealth1}
	\mathrm{d}W_t=\left[rW_t+Y_t+\pi_t(\mu-r)-c_t\right]\mathrm{d}t+\sigma\pi_t \mathrm{d}\beta_t,~~W_{0}=w.
\end{equation}
Reflecting the restrictive borrowing capacity after retirement, we consider different borrowing constraints before and after retirement as follows:
\begin{equation} \label{eq:BC1}
	W_t\geq
	\begin{cases}
		-\dfrac{Y_1}{r}\equiv\underline{w}_{1},    & 0\leq t<\tau, \\
		-\nu\dfrac{Y_2}{r}\equiv\underline{w}_{2}, & t\geq \tau,
	\end{cases}
\end{equation}
where $Y_{t}/r$ represents the present value of income discounted at the risk-free interest rate, and $\nu\in[0,1)$ determines the extent to which the agent is borrowing constrained.

Prior to retirement, the agent is assumed to exhibit standard constant relative risk aversion (CRRA) utility preferences:
$$
	U_1(c_{t})=\frac{c^{1-\gamma}_{t}}{1-\gamma},
$$
where $\gamma>0$ ($\gamma\neq 1$) is the agent's constant coefficient of relative risk aversion. Following \citet{Gang} and \citet{Kim et al.}, the presence of post-retirement income support for low-income or financially vulnerable people allows the agent to exhibit different utility preferences as follows:
$$
U_2(c_{t})=\frac{\left[\kappa(c_{t}+L)\right]^{1-\gamma}}{1-\gamma},
$$
where $\kappa>0$ determines the quality of post-retirement life in terms of leisure and financial situations, and $L\geq 0$ represents government support measured in units of consumption: $L$ enters preferences, while pension income $Y_2$ enters the wealth equation. Different from subsistence-consumption models \citep{Achury,Chen}, the agent may optimally choose zero consumption after retirement while continuing to obtain utility from the income support element via effective consumption.

For an admissible strategy $(\bc,\bpi,\tau)$, the agent's payoff is given by
$$
	J(w;\bc,\bpi,\tau)=\mathbb{E}^{w}\left[\int_0^\tau e^{-\rho t}U_1(c_t)\mathrm{d}t+\int_\tau^\infty e^{-\rho t}U_2(c_t)\mathrm{d}t\right],
$$
where $\rho>0$ is the subjective discount rate and $\mathbb{E}^{w}$ denotes the conditional expectation under $W_0=w$. We assume throughout that the Merton constant $K:=r+(\rho-r)/\gamma+(\gamma-1)\theta^2/(2\gamma^2)$ is positive. The value function is then defined as
\begin{equation}\label{value_function}
	V(w)=\sup_{(\bc,\bpi,\tau)\in\mathcal{A}(w)}J(w;\bc,\bpi,\tau),
\end{equation}
where $\mathcal{A}(w)$ is the set of admissible consumption, portfolio and retirement policies that satisfy the wealth dynamics in \eqref{eq:wealth1} and the borrowing constraints in (\ref{eq:BC1}), with the following standard integrability conditions:
\begin{equation*}
	\int_0^t c_s\mathrm{d}s<\infty, \qquad \int_0^t \pi_s^2\mathrm{d}s<\infty, \qquad \text{a.s. for all } t>0.
\end{equation*}
The expected discounted utility is also assumed to be well-defined:
\begin{equation*}
	\mathbb{E}^{w} \left[ \int_0^\tau e^{-\rho t}\left|U_1(c_t)\right|\mathrm{d}t + \int_\tau^\infty e^{-\rho t}\left|U_2(c_t)\right|\mathrm{d}t \right] <\infty.
\end{equation*}

We first consider the post-retirement problem in the value function, i.e., when $t\geq\tau$. The post-retirement problem can be formulated as
$$
	V_2(w):=\sup_{(\bc,\bpi)\in\mathcal{A}_2(w)} \mathbb{E}^{w} \left[ \int_0^\infty e^{-\rho t}\frac{[\kappa(c_t+L)]^{1-\gamma}}{1-\gamma}\mathrm{d}t \right],~~w\geq\underline{w}_2,
$$
subject to the following wealth dynamics:
$$
	\mathrm{d}W_t=\left[rW_t+Y_2+\pi_t(\mu-r)-c_t\right]\mathrm{d}t+\sigma\pi_t \mathrm{d}\beta_t, \qquad W_t\geq\underline{w}_2.
$$
Here, $\mathcal{A}_2(w)$ consists of progressively measurable consumption processes $c_t\ge0$ and portfolio processes $\pi_t$ with $\int_0^t c_s\mathrm{d}s<\infty$ and $\int_0^t\pi_s^2\mathrm{d}s<\infty$ a.s.\ for all $t>0$ such that the above wealth process satisfies $W_t\ge\underline{w}_2$ for all $t\ge0$. Since $U_2$ has the same sign as $1-\gamma$, the expectation is well defined (possibly $-\infty$ if $\gamma>1$).

We next consider the pre-retirement problem, i.e., when $t<\tau$. The pre-retirement problem can be formulated as
\begin{equation}\label{pre-retirement value function}
V_1(w):=\sup_{(\bc,\bpi,\tau)\in\mathcal{A}_1(w)}\mathbb{E}^{w}\left[\int_0^\tau e^{-\rho t}\frac{c_t^{1-\gamma}}{1-\gamma}\mathrm{d}t+e^{-\rho\tau}V_2(W_\tau)\right],\qquad w>\underline{w}_1.
\end{equation}
Here, the stopping time $\tau$ should satisfy the feasibility condition $W_\tau\ge \underline{w}_2$ whenever retirement occurs. The terminal payoff is taken to be zero on $\{\tau=\infty\}$. Precisely, $\mathcal{A}_1(w)$ consists of progressively measurable consumption processes $c_t\ge0$ and portfolio processes $\pi_t$ with $\int_0^t c_s\mathrm{d}s<\infty$ and $\int_0^t\pi_s^2\mathrm{d}s<\infty$ a.s.\ for all $t>0$, and of stopping times $\tau$ such that the wealth process with income $Y_1$ satisfies $W_t\ge\underline{w}_1$ for $t\le\tau$ and $W_\tau\ge \underline{w}_2$ on $\{\tau<\infty\}$. Since $U_1$ and $V_2$ have the same sign as $1-\gamma$ (Lemma~\ref{lem:V2bounds}), the expectation in \eqref{pre-retirement value function} is always well defined (possibly $-\infty$ if $\gamma>1$), so the integrability requirement imposed on $\mathcal{A}(w)$ is not needed in \eqref{pre-retirement value function}. Proposition~\ref{prop:reduction} and Corollary~\ref{cor:original} show that $V=V_1$.



\section{Post-Retirement Solutions} \label{sec:post}

We begin by introducing the two real roots $m_+ > 0$ and $m_- < -1/\gamma$ of the characteristic equation:
\begin{equation} \label{eq:characteristic_eq}
CE(m)\equiv\dfrac{1}{2}\theta^{2}m^{2}+\Big(\rho-r+\dfrac{1}{2}\theta^{2}\Big)m-r=0.
\end{equation}
To characterize whether the agent optimally chooses zero consumption near the post-retirement borrowing limit $\underline{w}_2$, we define the critical support threshold
\begin{equation*}
	\Delta_L:=-(\gamma m_-+1)(1-\nu)Y_2 > 0,
\end{equation*}
which represents the minimum level of income support for which the construction below yields a zero-consumption region at the credit boundary.\footnote{Since $m_-<-1/\gamma$, we have $\gamma m_-+1<0$, ensuring $\Delta_L>0$.}

For $w>\underline{w}_2$, the post-retirement value function $V_2(w)$ satisfies the Hamilton--Jacobi--Bellman (HJB) equation:
\begin{equation} \label{eq:Bellman1}
	\rho V_2(w)=\sup_{c\geq 0,\, \pi\in\mathbb{R}}\left[\left(rw+Y_2+\pi(\mu-r)-c\right)V_2'(w)+\frac{1}{2}\sigma^2\pi^2V_2''(w)+\frac{[\kappa(c+L)]^{1-\gamma}}{1-\gamma}\right].
\end{equation}
Assuming $V_2(w)$ is strictly increasing and strictly concave, the first-order conditions with respect to $\pi$ and $c$ yield the feedback controls
\begin{equation}\label{FOC_pi}
	\pi^*(w)=-\frac{\theta}{\sigma} \frac{V_2'(w)}{V_2''(w)}
\end{equation}
and
\begin{equation}\label{FOC post-retirement consumption}
	c^*(w)=\max \left\{ A \big(V_2'(w)\big)^{-1/\gamma} -L,\, 0 \right\},
\end{equation}
where
\begin{equation}\label{A_kappa}
	A:=\kappa^{\frac{1-\gamma}{\gamma}}.
\end{equation}
Let $\psi(w):=V_2'(w)$ denote the marginal value of wealth, and define the critical marginal-value threshold at which private consumption hits zero:
\begin{equation*}
    \psi_L := \kappa^{1-\gamma}L^{-\gamma}.
\end{equation*}
Substituting \eqref{FOC_pi} into \eqref{eq:Bellman1} reduces the HJB equation to
\begin{equation} \label{eq:post_HJB_after_pi}
	\rho V_2(w)=(rw+Y_2)\psi(w) - \frac{1}{2}\theta^2 \frac{\{\psi(w)\}^2}{\psi'(w)} + \mathcal{M}(\psi(w)),
\end{equation}
where the maximized consumption dual $\mathcal{M}(\psi):=\sup_{c\geq 0} \left[ \frac{[\kappa(c+L)]^{1-\gamma}}{1-\gamma}-c\psi \right]$ is explicitly given by
\begin{equation} \label{eq:M_solution}
	\mathcal{M}(\psi)=
	\begin{cases}
		\dfrac{(\kappa L)^{1-\gamma}}{1-\gamma}, & \psi\geq \psi_L, \\[1.0em]
		\dfrac{\gamma}{1-\gamma}A\psi^{-\frac{1-\gamma}{\gamma}} + L\psi, & 0<\psi<\psi_L.
	\end{cases}
\end{equation}

The following theorem characterizes the existence and closed-form expression of the wealth threshold separating the zero-consumption and positive-consumption regimes.

\begin{thm}[Post-retirement consumption regimes]\label{theorem wealth threshold}
Suppose that post-retirement income support is sufficiently generous relative to the unpledgeable portion of retirement income, specifically
\begin{equation} \label{eq:post_case1_condition_intro}
	L\geq \Delta_L.
\end{equation}
Then there exists a unique wealth threshold $w_{0,2} \ge \underline{w}_2$, corresponding to $\psi(w_{0,2})=\psi_L$, such that optimal consumption is zero on $[\underline{w}_2, w_{0,2}]$ and strictly positive on $(w_{0,2},\infty)$. Moreover, this threshold is explicitly given by
\begin{equation*}
	w_{0,2}=C_1\psi_L^{m_+} + C_2\psi_L^{m_-} - \frac{Y_2}{r},
\end{equation*}
where $\psi_\nu := \psi(\underline{w}_2)$ and the positive constants $C_1, C_2$ are defined as
\begin{equation*}
    C_1=-\frac{m_-(1-\nu)Y_2}{(m_+-m_-)r}\psi_\nu^{-m_+}>0, \qquad C_2=\frac{m_+(1-\nu)Y_2}{(m_+-m_-)r}\psi_\nu^{-m_-}>0.
\end{equation*}
The marginal value of wealth at the borrowing limit is given explicitly by
\begin{equation}\label{eq:psi_nu_main}
	\psi_\nu=\psi_L\left(\frac{\Delta_L}{L}\right)^{1/m_-}\ \ge\ \psi_L,
\end{equation}
and $w_{0,2}=\underline{w}_2$ if and only if $L=\Delta_L$.
\end{thm}

Theorem~\ref{theorem wealth threshold} highlights how institutional safety nets shape post-retirement spending near the borrowing limit $\underline{w}_2$ in \eqref{eq:BC1}: when $L>\Delta_L$, the agent sets private consumption to zero on $[\underline w_2,w_{0,2}]$. At $L=\Delta_L$, this interval reduces to the borrowing limit itself. The constant $\psi_L=\kappa^{1-\gamma}L^{-\gamma}=U_2'(0)$ is the marginal utility of effective consumption when the agent consumes the support $L$ only. The agent consumes nothing exactly when the marginal value of wealth is at least $\psi_L$, because an additional unit of private consumption is then worth less than saving it. The construction requires $\psi_\nu\ge\psi_L$, i.e., $L\ge\Delta_L$; this explains the role of $\Delta_L$ as the minimum support for a zero-consumption region. The case $L<\Delta_L$, in which no zero-consumption region is expected, requires a different construction and is not treated in this paper.

\begin{cor}[Closed-form post-retirement policies]\label{optimal post-retirement policies}
Under condition \eqref{eq:post_case1_condition_intro}, the optimal post-retirement consumption and risky investment policies are given in closed form across the two regions by
\begin{equation*}
	c_2^*(w)
	=
	\begin{cases}
		0, & \psi_L\leq \psi(w)\leq \psi_\nu \quad (\underline{w}_2 \le w \le w_{0,2}), \\[0.8em]
		A\psi(w)^{-1/\gamma}-L, & 0<\psi(w)<\psi_L \quad (w > w_{0,2}),
	\end{cases}
\end{equation*}
and
\begin{equation*}
	\pi_2^*(w)
	=
	\begin{cases}
		-\dfrac{\theta}{\sigma}\left(m_+C_1\psi(w)^{m_+}+m_-C_2\psi(w)^{m_-}\right), & \psi_L\leq \psi(w)\leq \psi_\nu, \\[1.2em]
		-\dfrac{\theta}{\sigma}\left(m_+D_1\psi(w)^{m_+}-\dfrac{A}{\gamma K}\psi(w)^{-1/\gamma}\right), & 0<\psi(w)<\psi_L,
	\end{cases}
\end{equation*}
where $C_1, C_2$ are as in Theorem~\ref{theorem wealth threshold}, $K>0$ is the Merton constant defined in Section~\ref{sec:model}, and
\begin{equation*}
    D_1=C_1 - \frac{(\gamma m_-+1)L}{\gamma^2m_+(m_+-m_-)K} \psi_L^{-m_+} > 0.
\end{equation*}
\end{cor}

The construction above is verified directly. The proof uses the dual objects $\mathcal{W}_2$ and $\Vt_2$ introduced in Section~\ref{subsec:dual} and is given in Appendix~\ref{app:proof_post_verification}.

\begin{pro}[Verification of the post-retirement solution]\label{prop:post_verification}
Under condition \eqref{eq:post_case1_condition_intro} and $K>0$, the function $V_2$ given by \eqref{eq:post_value_zero_region_derivation}--\eqref{eq:post_value_positive_region_derivation} in the Appendix is the post-retirement value function, and the feedback controls of Corollary~\ref{optimal post-retirement policies}, $c_2^*=(A\psi^{-1/\gamma}-L)^+$ and $\pi_2^*=-\frac{\theta}{\sigma}\psi\mathcal{W}_2'(\psi)$ with $\psi=V_2'(W_t)$, are optimal.
\end{pro}


\section{Pre-Retirement Solutions} \label{sec:pre}

We now solve the agent's pre-retirement problem by formulating the voluntary retirement decision as an optimal stopping problem with a free boundary. Throughout this section, we maintain the generous-support condition $L\geq\Delta_{L}$ in \eqref{eq:post_case1_condition_intro}, so that the post-retirement value function $V_2(w)$ has zero private consumption on $[\underline w_2,w_{0,2}]$, a singleton at $L=\Delta_L$ (Theorem~\ref{theorem wealth threshold}).

The pre-retirement problem is an optimal stopping problem with the post-retirement value function as the reward. The following proposition justifies this reduction using only the results of Section~\ref{sec:post}.

\begin{pro}[Reduction to the stopping problem]\label{prop:reduction}
Let $V$ be the value function \eqref{value_function}, and let $V_1$ be the pre-retirement value function \eqref{pre-retirement value function}. Then the following statements hold for every $w>\underline{w}_1$.
\begin{enumerate}
\item[(a)] If $V_1(w)<\infty$, then $V(w)=V_1(w)$.
\item[(b)] Suppose that $V_1(w)<\infty$ and that $(\bc^*,\bpi^*,\tau^*)\in\mathcal{A}_1(w)$ is optimal for $V_1(w)$. On $\{\tau^*<\infty\}$, extend it after $\tau^*$ by the post-retirement feedback controls $(c_2^*,\pi_2^*)$ of Proposition~\ref{prop:post_verification}, started at $W_{\tau^*}$. Then the resulting policy belongs to $\mathcal{A}(w)$ and is optimal for \eqref{value_function}.
\end{enumerate}
\end{pro}

In the continuation region, where the agent chooses to remain in the labor force, the pre-retirement value function $V_1(w)$ in \eqref{pre-retirement value function} satisfies the Hamilton--Jacobi--Bellman (HJB) equation:
\begin{equation} \label{eq:pre_HJB}
	\rho V_1(w)=\sup_{c\ge0,\,\pi\in\mathbb{R}} \left[ \left(rw+Y_1+\pi(\mu-r)-c\right)V_1'(w) + \frac{1}{2}\sigma^2\pi^2V_1''(w) + \frac{c^{1-\gamma}}{1-\gamma} \right], \qquad w\in\mathcal C.
\end{equation}
Assuming $V_1$ is strictly increasing and strictly concave, the first-order conditions yield the feedback controls for pre-retirement consumption and risky investment:
\begin{equation} \label{eq:pre_c_star_general}
	c_1^*(w)=\left(V_1'(w)\right)^{-1/\gamma} \qquad \text{and} \qquad \pi_1^*(w)=-\frac{\theta}{\sigma} \frac{V_1'(w)}{V_1''(w)}.
\end{equation}
Substituting \eqref{eq:pre_c_star_general} back into \eqref{eq:pre_HJB} reduces the HJB equation to
\begin{equation} \label{eq:pre_HJB_reduced}
	\rho V_1(w)=(rw+Y_1)V_1'(w)- \frac{1}{2}\theta^2 \frac{\left(V_1'(w)\right)^2}{V_1''(w)} + \frac{\gamma}{1-\gamma} \left(V_1'(w)\right)^{-\frac{1-\gamma}{\gamma}}.
\end{equation}

Because post-retirement borrowing constraints are tighter ($\underline{w}_2 > \underline{w}_1$), retirement is infeasible when $w < \underline{w}_2$, forcing the agent to continue working. At or above $\underline{w}_2$, the agent weighs the value of continuing in the labor market against the immediate retirement payoff $V_2(w)$. For a sufficiently smooth function $f$ satisfying $f''<0<f'$, define the pre-retirement Bellman operator
\begin{equation}\label{eq:pre-retirement-HJB-operator}
\mathcal{H}_1 f(w)
:=\sup_{c\ge0,\,\pi\in\mathbb{R}}
\left\{\bigl(rw+Y_1+\pi(\mu-r)-c\bigr)f'(w)
+\frac{1}{2}\sigma^2\pi^2 f''(w)+\frac{c^{1-\gamma}}{1-\gamma}\right\},
\end{equation}
which, by the first-order conditions, equals
\begin{equation}\label{eq:pre-retirement-HJB-operator-optimized}
\mathcal{H}_1 f(w)=(rw+Y_1)f'(w)-\frac{1}{2}\theta^2\frac{\bigl(f'(w)\bigr)^2}{f''(w)}+\frac{\gamma}{1-\gamma}\bigl(f'(w)\bigr)^{-\frac{1-\gamma}{\gamma}}.
\end{equation}
Formally, the trade-off between working and retiring leads to the variational inequality \citep{Bensoussan1982}
\begin{equation} \label{Variational_inequality}
\begin{aligned}
\mathcal{H}_{1}V_{1}(w)-\rho V_{1}(w)&=0, & &\underline{w}_{1}<w<\underline{w}_{2},\\
\max \big\{ \mathcal{H}_1 V_1(w) -\rho V_1(w),\, V_2(w)-V_1(w) \big\}&=0, & &w>\underline{w}_{2}.
\end{aligned}
\end{equation}
We do not assume \eqref{Variational_inequality}: Theorem~\ref{thm:main} below constructs $V_1$ by duality and shows directly that it satisfies \eqref{Variational_inequality} classically away from the free boundaries.

Accordingly, the state space partitions into the continuation region $\mathcal{C}$ and the retirement region $\mathcal{R}$:
\begin{align}
    \mathcal{C} & := (\underline{w}_1,\infty)\setminus\mathcal{R}, \label{Continuation_region} \\
    \mathcal{R} & := \{w\ge\underline{w}_2:V_1(w)=V_2(w)\}, \label{Retirement_region}
\end{align}
and the optimal retirement time is the first hitting time of $\mathcal{R}$:
\begin{equation} \label{Optimal_retirement_time}
	\tau^*=\inf\{t\geq 0 : W_t\in \mathcal{R}\},\quad \inf\varnothing:=\infty.
\end{equation}

Finally, to establish an analytical benchmark, let $V^{\mathrm{NR}}(w)$ denote the value function of a hypothetical agent who is permanently restricted from retiring and receives the perpetual labor income $Y_1$. Following the standard Merton framework, this never-retiring value function and its convex conjugate (dual) function $\Vt^{\mathrm{NR}}(\phi):=\sup_{w>\underline{w}_1}[V^{\mathrm{NR}}(w)-\phi w]$ are explicitly given by
\begin{equation}\label{eq:db-never-retire-value}
 V^{\mathrm{NR}}(w) =\frac{1}{(1-\gamma)K^{\gamma}} \left(w+\frac{Y_1}{r}\right)^{1-\gamma}
 \qquad \text{and} \qquad
 \Vt^{\mathrm{NR}}(\phi)=\frac{\gamma}{(1-\gamma)K}\phi^{1-\frac1\gamma}+\frac{Y_1}{r}\phi.
\end{equation}

\subsection{Dual variables and local stopping incentives} \label{subsec:dual}

To analyze the continuation region $\mathcal{C}$ in \eqref{Continuation_region} and the retirement region $\mathcal{R}$ in \eqref{Retirement_region} analytically, we transition to the dual state space parameterized by the marginal value of wealth $\phi > 0$. For any twice-differentiable function $f$ on $(0,\infty)$, define the dual differential operator
\begin{equation}\label{eq:Lt}
  \Lt f(\phi):=\tfrac12\theta^2\phi^2 f''(\phi)+(\rho-r)\phi f'(\phi)-\rho f(\phi),
\end{equation}
which corresponds to the infinitesimal generator of $e^{-\rho t}f(\Phi_t)$ under the dual state-price process $\mathrm{d}\Phi_t=\Phi_t[(\rho-r)\mathrm{d}t-\theta \mathrm{d}\beta_t]$. Here, $\beta$ is the Brownian motion in \eqref{eq:wealth1}. Notice that $\Lt\phi^n=Q(n)\phi^n$ for any $n\in\mathbb{R}$, where $Q(n)=\frac12\theta^2n(n-1)+(\rho-r)n-\rho=CE(n-1)$ and $CE(\cdot)$ is the characteristic polynomial in \eqref{eq:characteristic_eq}. We also define the dual pre-retirement utility function
\[
  \Ut(\phi):=\sup_{c\ge0}\Big[\frac{c^{1-\gamma}}{1-\gamma}-c\phi\Big]=\frac{\gamma}{1-\gamma}\phi^{1-\frac1\gamma}.
\]

Next, we represent the post-retirement wealth level and value function in dual coordinates. Let $\mathcal{W}_2(\psi)$ denote the generalized inverse of the post-retirement marginal value function $\psi(w)=V_2'(w)$:
\[
  \mathcal{W}_2(\psi):=\begin{cases}
  W_+(\psi)=D_1\psi^{m_+}+\dfrac{A}{K}\psi^{-1/\gamma}-\dfrac{L+Y_2}{r}, & 0<\psi<\psi_{L},\\[1.0em]
  W_0(\psi)=C_1\psi^{m_+}+C_2\psi^{m_-}-\dfrac{Y_2}{r}, & \psi_{L}\le\psi\le\psi_\nu,\\[1.0em]
  \underline{w}_2, & \psi>\psi_\nu,
  \end{cases}
\]
and define the post-retirement dual value function $\Vt_2(\phi):=\sup_{w\ge\underline{w}_2}[V_2(w)-\phi w]$ explicitly by
\begin{equation}\label{eq:Vt2}
  \Vt_2(\phi) :=
  \begin{cases}
    -\dfrac{D_1}{m_++1}\phi^{m_++1}+\dfrac{\gamma A}{(1-\gamma)K}\phi^{1-\frac1\gamma}
      +\dfrac{L+Y_2}{r}\phi, & 0<\phi<\psi_{L},\\[2ex]
    -\dfrac{C_1}{m_++1}\phi^{m_++1}-\dfrac{C_2}{m_-+1}\phi^{m_-+1}
      +\dfrac{(\kappa L)^{1-\gamma}}{\rho(1-\gamma)}+\dfrac{Y_2}{r}\phi,
      & \psi_{L}\le\phi\le\psi_\nu,\\[2ex]
    V_2(\underline{w}_2)-\phi\, \underline{w}_2, & \phi>\psi_\nu,
  \end{cases}
\end{equation}
where $\rho V_2(\underline{w}_2)=(1-\nu)Y_2\psi_\nu+(\kappa L)^{1-\gamma}/(1-\gamma)$ from \eqref{eq:post_HJB_after_pi} evaluated at $w=\underline{w}_2$.

Using $\mathcal{W}_2(\phi)$, we can explicitly evaluate the Bellman residual $(\mathcal{H}_1V_2-\rho V_2)$ along the post-retirement wealth schedule, which measures the instantaneous incentive to delay retirement.

\begin{lem}[Instantaneous gain from postponing retirement]\label{lem:S_meaning}
For $0<\phi<\psi_{\nu}$, the pre-retirement Bellman residual evaluated at $w=\mathcal{W}_2(\phi)$ is given by
\begin{equation*}
\big(\mathcal{H}_1V_2-\rho V_2\big)\big(\mathcal{W}_2(\phi)\big)=\left\{
\begin{aligned}
&\phi\Big(Y_{1}-Y_{2}-L-\dfrac{\gamma(A-1)}{1-\gamma}\phi^{-1/\gamma}\Big), & &0<\phi<\psi_{L},\\[0.5em]
&(Y_{1}-Y_{2})\phi+\Ut(\phi)-\dfrac{(\kappa L)^{1-\gamma}}{1-\gamma}, & &\psi_{L}\leq\phi<\psi_{\nu}.
\end{aligned}\right.
\end{equation*}
\end{lem}

Economically, $\big(\mathcal{H}_1V_2-\rho V_2\big)\big(\mathcal{W}_2(\phi)\big)$ represents the net instantaneous gain per unit of time from delaying retirement slightly rather than stopping immediately at wealth $\mathcal{W}_2(\phi)$. When this residual is strictly positive, continuing to work strictly dominates immediate retirement in the short run; conversely, a negative residual favors immediate retirement locally, and a zero residual is inconclusive. Denote this residual by $S(\phi)$ for $0<\phi<\psi_\nu$. For $\phi\ge\psi_\nu$, use its dual extension $S(\phi)=(Y_1-\nu Y_2)\phi+\Ut(\phi)-\rho V_2(\underline w_2)$, as in \eqref{eq:S}. For $\phi>\psi_\nu$, the inverse wealth schedule is constant at $\underline w_2$; these $\phi$ are not distinct marginal values attained at interior wealth levels.

\begin{lem}[Structure of local stopping incentives]\label{lem:S_sign}
Let $\mathcal N:=\{\phi>0:S(\phi)<0\}$. This set is either empty or an interval $(u_1,u_2)$ with $0\le u_1<u_2<\infty$. The negative Bellman residual at interior post-retirement wealth corresponds to $\mathcal N\cap(0,\psi_\nu)$. Specifically:
\begin{enumerate}
\item[(a)] \textbf{Enhanced post-retirement life ($\kappa>1$):} $\mathcal N=(0,u_2)$ for some $u_2>0$.
\item[(b)] \textbf{Neutral post-retirement life ($\kappa=1$):} $\mathcal N$ is nonempty if and only if $Y_1<L+Y_2$, in which case $\mathcal N=(0,u_2)$.
\item[(c)] \textbf{Degraded post-retirement life ($0<\kappa<1$):} If $\mathcal N$ is nonempty, then $u_1>0$.
\end{enumerate}
\end{lem}

\subsection{Inverse marginal-value transformation and boundary residuals} \label{subsec:residual}

To solve the free-boundary problem analytically, we transform the HJB equation into dual coordinates using the marginal value of wealth $\phi(w):=V_1'(w)$. Let $X(\phi)$ denote its inverse function, satisfying $X(\phi(w))=w$. Differentiating \eqref{eq:pre_HJB_reduced} with respect to $w$ and applying standard inverse-function identities yields a linear second-order ODE for $X(\phi)$:
\begin{equation} \label{eq:X_ODE_pre}
	\frac{1}{2}\theta^2\phi^2X''(\phi) + (\rho-r+\theta^2)\phi X'(\phi) -rX(\phi) + \phi^{-1/\gamma}-Y_1=0.
\end{equation}

Depending on whether the agent continues working at low or high wealth levels, the general solution to \eqref{eq:X_ODE_pre} splits into two continuation branches:
\begin{itemize}
\item \textbf{Low-wealth continuation branch ($\phi>\phi_L$):} Imposing the natural borrowing boundary condition $X(\phi)\to \underline{w}_1$ as $\phi\to\infty$ eliminates the $m_+$ term, yielding
\begin{equation}\label{eq:db-low-inverse}
 X_L(\phi) = B_L\phi^{m_-}+\frac{1}{K}\phi^{-1/\gamma}-\frac{Y_1}{r},
\end{equation}
with the corresponding parametric value function
\begin{equation}\label{eq:db-low-value}
V_{1,L}(\phi) = \frac{m_-}{m_-+1}B_L\phi^{m_-+1} +\frac{1}{(1-\gamma)K} \phi^{-(1-\gamma)/\gamma}.
\end{equation}

\item \textbf{High-wealth continuation branch ($0<\phi<\phi_U$):} Imposing the standard transversality growth condition as $\phi\downarrow 0$ eliminates the $m_-$ term, yielding
\begin{equation}\label{eq:db-high-inverse}
 X_U(\phi) = B_U\phi^{m_+}+\frac{1}{K}\phi^{-1/\gamma}-\frac{Y_1}{r},
\end{equation}
with the corresponding parametric value function
\begin{equation}\label{eq:db-high-value}
V_{1,U}(\phi) = \frac{m_+}{m_++1}B_U\phi^{m_++1} +\frac{1}{(1-\gamma)K} \phi^{-(1-\gamma)/\gamma}.
\end{equation}
\end{itemize}
Here, $\phi_L$ and $\phi_U$ represent the marginal values associated with the lower and upper wealth boundaries, respectively, while $B_L$ and $B_U$ are free-boundary constants.

At an interior wealth boundary, value matching and smooth pasting require equal values and equal marginal values, respectively. At a contact with the borrowing limit, the working-side slope can instead be a supporting slope of the retirement value at that limit. In both cases, matching wealth at a candidate slope $\phi>0$ gives the coefficient for $m\in\{m_-,m_+\}$:
\begin{equation}\label{eq:Bmatch}
  B_m(\phi):=\Big(\mathcal{W}_2(\phi)-\frac1K\phi^{-1/\gamma}+\frac{Y_1}{r}\Big)\phi^{-m},
\end{equation}
which ensures that the continuation branch ($m=m_-$ for a lower boundary; $m=m_+$ for an upper boundary) takes the value $\mathcal{W}_2(\phi)$ at $\phi$; the derivatives of the two inverse schedules need not coincide. It then remains to check value matching. The \emph{boundary residual} defined below measures this remaining gap in value; hence, candidate free boundaries correspond precisely to its zeros.

\begin{dfn}[Boundary residuals]\label{def:F}
For $m\in\{m_-,m_+\}$ and $\phi>0$, define the boundary residual function by
\begin{equation}\label{eq:Fdef}
  F_m(\phi):=\frac1\phi\Big(V_2(\mathcal{W}_2(\phi))-\frac{m}{m+1}B_m(\phi)\,\phi^{m+1}-\frac{1}{(1-\gamma)K}\phi^{-\frac{1-\gamma}{\gamma}}\Big).
\end{equation}
\end{dfn}

Using the dual identities $V_2(\mathcal{W}_2(\phi))=\Vt_2(\phi)+\phi \mathcal{W}_2(\phi)$ and $\Vt^{\mathrm{NR}\prime}(\phi)-\Vt_2'(\phi)=B_m(\phi)\phi^m$, the boundary residual simplifies directly in terms of the dual value functions:
\begin{equation}\label{eq:Fh}
  F_m(\phi)=-\frac1\phi\Big(\widetilde{V}^{\mathrm{NR}}(\phi)-\widetilde{V}_{2}(\phi)-\frac{\phi\{\widetilde{V}^{\mathrm{NR}\prime}(\phi)-\widetilde{V}_{2}'(\phi)\}}{m+1}\Big).
\end{equation}
Substituting \eqref{eq:Vt2} into \eqref{eq:Fh} yields explicit piecewise expressions for $F_{m_-}$ and $F_{m_+}$:
\begin{equation}\label{eq:Fminus}
F_{m_{-}}(\phi)=
\begin{cases}
\tfrac{m_+-m_-}{(m_++1)(m_-+1)}D_1\phi^{m_+}+\tfrac{(\gamma m_-+1)(A-1)}{(1-\gamma)(m_-+1)K}\phi^{-\frac1\gamma}-\tfrac{m_-(Y_1-L-Y_2)}{r(m_-+1)}, & 0<\phi<\psi_L,\\[1.2em]
\tfrac{m_+-m_-}{(m_++1)(m_-+1)}C_1\phi^{m_+}-\tfrac{\gamma m_-+1}{(1-\gamma)(m_-+1)K}\phi^{-\frac1\gamma}-\tfrac{m_-(Y_1-Y_2)}{r(m_-+1)}+\tfrac{(\kappa L)^{1-\gamma}}{\rho(1-\gamma)\phi}, & \psi_L\le\phi\le\psi_\nu,\\[1.2em]
-\tfrac{\gamma m_-+1}{(1-\gamma)(m_-+1)K}\phi^{-\frac1\gamma}-\tfrac{m_-(Y_1-\nu Y_2)}{r(m_-+1)}+\tfrac{V_2(\underline{w}_2)}{\phi}, & \phi>\psi_\nu,
\end{cases}
\end{equation}
and
\begin{equation}\label{eq:Fplus}
F_{m_{+}}(\phi)=
\begin{cases}
\tfrac{(\gamma m_++1)(A-1)}{(1-\gamma)(m_++1)K}\phi^{-\frac1\gamma}-\tfrac{m_+(Y_1-L-Y_2)}{r(m_++1)}, & 0<\phi<\psi_L,\\[1.2em]
-\tfrac{\gamma m_++1}{(1-\gamma)(m_++1)K}\phi^{-\frac1\gamma}-\tfrac{m_+-m_-}{(m_++1)(m_-+1)}C_2\phi^{m_-}-\tfrac{m_+(Y_1-Y_2)}{r(m_++1)}+\tfrac{(\kappa L)^{1-\gamma}}{\rho(1-\gamma)\phi}, & \psi_L\le\phi\le\psi_\nu,\\[1.2em]
-\tfrac{\gamma m_++1}{(1-\gamma)(m_++1)K}\phi^{-\frac1\gamma}-\tfrac{m_+(Y_1-\nu Y_2)}{r(m_++1)}+\tfrac{V_2(\underline{w}_2)}{\phi}, & \phi>\psi_\nu.
\end{cases}
\end{equation}
Because $\Vt_2\in C^2(0,\infty)$ (Lemma~\ref{lem:Vt2}(a)), \eqref{eq:Fh} shows that both $F_{m_-}$ and $F_{m_+}$ are continuous on $(0,\infty)$.

The next proposition links the boundary residuals to the flow gain $S$ of Lemma~\ref{lem:S_meaning}, extended to $\phi\ge\psi_\nu$ as in \eqref{eq:S}. Let $h(\phi):=\Vt^{\mathrm{NR}}(\phi)-\Vt_2(\phi)$ as in \eqref{eq:h}.

\begin{pro}[Residuals are weighted tail integrals of the flow gain]\label{prop:Fint}
For every $\phi>0$,
\begin{align}
  F_{m_{-}}(\phi)&=-\frac{m_+m_-}{r(m_-+1)}\;\phi^{m_+}\int_\phi^\infty s^{-m_+-2}S(s)\,\mathrm{d}s,\label{eq:Fmint}\\
  F_{m_{+}}(\phi)&=\phantom{-}\frac{m_+m_-}{r(m_++1)}\;\phi^{m_-}\int_0^\phi s^{-m_--2}S(s)\,\mathrm{d}s,\label{eq:Fpint}\\
  h(\phi)&=-\frac{m_+m_-}{r(m_+-m_-)}\Big[\phi^{m_++1}\int_\phi^\infty s^{-m_+-2}S(s)\,\mathrm{d}s+\phi^{m_-+1}\int_0^\phi s^{-m_--2}S(s)\,\mathrm{d}s\Big],\label{eq:Green}
\end{align}
and all integrals above converge. In particular,
\begin{equation*}
\begin{aligned}
  F_{m_{-}}(\phi)>0 &\iff I_-(\phi):=\int_\phi^\infty s^{-m_+-2}S(s)\,\mathrm{d}s<0,\\
  F_{m_{+}}(\phi)>0 &\iff I_+(\phi):=\int_0^\phi s^{-m_--2}S(s)\,\mathrm{d}s<0.
\end{aligned}
\end{equation*}
\end{pro}

These identities are \eqref{eq:Fmint}--\eqref{eq:Green}, proved in Step~1 of Appendix~\ref{app:proof_S_sign}. They show that the root selection in Definition~\ref{def:phiLU} below is determined by the signs of the tail integrals of $S$, so no additional root-selection criterion is needed.

We now use the boundary residuals $F_{m_-}$ and $F_{m_+}$ to identify the optimal retirement region. By Definition~\ref{def:F}, having $F_m(\phi)>0$ means that at wealth $\mathcal{W}_2(\phi)$, immediate retirement yields a strictly higher value than continuing along the working branch constructed at that candidate slope. The inverse schedule is strictly decreasing up to $\psi_\nu$ and constant above it.

When the post-retirement quality of life degrades ($0<\kappa<1$), the boundary equations $F_{m_\pm}(\phi)=0$ can be non-monotone and admit multiple roots. Definition~\ref{def:phiLU} selects the supremum of the positive set for the low-wealth residual and the infimum for the high-wealth residual; a positive residual at one trial slope is not itself an optimal stopping rule.

\begin{dfn}[Marginal-value threshold selection]\label{def:phiLU}
The marginal-value thresholds associated with the lower and upper wealth boundaries are defined by
\[
\begin{aligned}
  \phi_L&:=\sup\{\phi>0:F_{m_{-}}(\phi)>0\}\quad (\text{with }\phi_L:=0\text{ if the set is empty}),\\
  \phi_U&:=\inf\{\phi>0:F_{m_{+}}(\phi)>0\}\quad (\text{with }\phi_U:=\infty\text{ if the set is empty}).
\end{aligned}
\]
\end{dfn}

The following lemma summarizes the structural properties of $\phi_L$ and $\phi_U$ and establishes a simple criterion for when voluntary retirement is optimal.

\begin{lem}[Properties of the marginal-value thresholds]\label{lem:phiLU}
The thresholds $\phi_U$ and $\phi_L$ satisfy the following properties:
\begin{enumerate}
\item[(a)] \textbf{Existence of retirement:} The following three statements are equivalent:
  \begin{enumerate}
  \item[(i)] $\phi_U<\phi_L$;
  \item[(ii)] there exists $\phi>0$ such that both $F_{m_{-}}(\phi)>0$ and $F_{m_{+}}(\phi)>0$;
  \item[(iii)] there exists $\phi>0$ such that $\widetilde{V}^{\mathrm{NR}}(\phi)<\widetilde{V}_{2}(\phi)$ (i.e., retirement beats perpetual work in dual value).
  \end{enumerate}

\item[(b)] \textbf{Ordering and sign behavior ($\phi_U<\phi_L$):} When retirement is optimal, the thresholds satisfy $u_1\le\phi_U<\phi_L<u_2$, with $\phi_U=0$ if $u_1=0$ and $u_1<\phi_U$ otherwise. Moreover, $F_{m_{-}}(\phi_L)=0$, $F_{m_{-}}<0$ on $(\phi_L,\infty)$, and $F_{m_{-}}>0$ on $(\phi_U,\phi_L)$; symmetrically, if $\phi_U>0$, then $F_{m_{+}}(\phi_U)=0$, $F_{m_{+}}<0$ on $(0,\phi_U)$, and $F_{m_{+}}>0$ on $(\phi_U,\phi_L]$. In addition, $\widetilde{V}^{\mathrm{NR}}(\phi_L)<\widetilde{V}_{2}(\phi_L)$, and $\widetilde{V}^{\mathrm{NR}}(\phi_U)<\widetilde{V}_{2}(\phi_U)$ whenever $\phi_U>0$.

\item[(c)] \textbf{No-retirement condition ($\phi_L\le\phi_U$):} If $\phi_L\le\phi_U$, then $\widetilde{V}^{\mathrm{NR}}(\phi)\ge\widetilde{V}_{2}(\phi)$ for all $\phi>0$, with equality holding for at most one $\phi>0$.
\end{enumerate}
\end{lem}

Combining the dual characterization and boundary residual analysis established above, we now present our main theoretical result, which completely characterizes the optimal retirement region and the pre-retirement value function.

\begin{thm}[Characterization of the retirement region]\label{thm:main}
Let $\kappa>0$, and let $\phi_L,\phi_U$ be the marginal-value thresholds defined in Definition~\ref{def:phiLU}.
\begin{enumerate}
\item[(a)] \textbf{Never-retire regime ($\phi_L\le\phi_U$).} 
Voluntary retirement is never strictly optimal. Specifically, $V_1(w)=V^{\mathrm{NR}}(w)$ for all $w\in(\underline{w}_1,\infty)$, and the agent follows the standard Merton policies $c^*=K(w+Y_1/r)$ and $\pi^*=\frac{\theta}{\gamma\sigma}(w+Y_1/r)$ indefinitely. Moreover, $V^{\mathrm{NR}}(w)\ge V_2(w)$ on $[\underline{w}_2,\infty)$, with equality holding at most at a single wealth level. Consequently, $\mathcal{R}=\emptyset$ when $\phi_L<\phi_U$; if $\phi_L=\phi_U$, $\mathcal{R}$ contains at most one point where the agent is indifferent between retiring and continuing, so perpetual work remains optimal.

\item[(b)] \textbf{Voluntary retirement regime ($\phi_U<\phi_L$).} 
Define the lower and upper retirement wealth boundaries by
\begin{equation*}
  w_L:=\mathcal{W}_2(\phi_L),\qquad w_U:=\mathcal{W}_2(\phi_U)\quad (\text{with the convention } w_U:=\infty\text{ if }\phi_U=0),
\end{equation*}
which satisfy $\underline{w}_2\le w_L\le w_U\le\infty$. Then:
\begin{enumerate}
\item[(i)] \textbf{Retirement region:} Voluntary retirement is optimal exclusively on the closed interval $\mathcal{R}=[w_L,w_U]\cap\mathbb{R}$, whereas continuing to work is strictly optimal outside this region ($V_1>V_2$ on $[\underline{w}_2,\infty)\setminus\mathcal{R}$).

\item[(ii)] \textbf{Value function decomposition:} In the low-wealth working region $(\underline{w}_1,w_L)$, the pre-retirement value function is parametrically given by $V_1(w)=V_{1,L}(\phi)$ with $w=X_L(\phi)$ for $\phi\in(\phi_L,\infty)$. It coincides with the post-retirement value function $V_2(w)$ on $\mathcal{R}$, and equals $V_1(w)=V_{1,U}(\phi)$ with $w=X_U(\phi)$ for $\phi\in(0,\phi_U)$ in the high-wealth working region $(w_U,\infty)$. The associated free-boundary coefficients are
\begin{equation}\label{eq:BLU}
  \begin{aligned}
  B_L&=B_{m_-}(\phi_L)=(m_-+1)\{\widetilde{V}^{\mathrm{NR}}(\phi_L)-\widetilde{V}_{2}(\phi_L)\}\phi_L^{-m_--1}>0,\\
  B_U&=B_{m_+}(\phi_U)=(m_++1)\{\widetilde{V}^{\mathrm{NR}}(\phi_U)-\widetilde{V}_{2}(\phi_U)\}\phi_U^{-m_+-1}<0\quad(\phi_U>0),
  \end{aligned}
\end{equation}
with $X_L'<0$ on $(\phi_L,\infty)$ and $X_U'<0$ on $(0,\phi_U)$.

\item[(iii)] \textbf{Optimal controls:} The stopping time $\tau^*=\inf\{t\ge 0:W_t\in\mathcal{R}\}$ in \eqref{Optimal_retirement_time}, paired with the feedback controls in \eqref{eq:pre_c_star_general} on the continuation region $\mathcal{C}$, is optimal for the pre-retirement problem \eqref{pre-retirement value function}.

\item[(iv)] \textbf{Smoothness and concavity:} $V_1$ is strictly concave on $(\underline{w}_1,\infty)$, $C^1$ on $(\underline{w}_1,\infty)$, and $C^2$ away from $\{w_L,w_U\}$. An exception arises when $\phi_L>\psi_\nu$: in this case, the lower threshold hits the borrowing limit ($w_L=\underline{w}_2$), and $V_1$ exhibits a concave kink at $\underline{w}_2$ satisfying $V_1'(\underline{w}_2-)=\phi_L>\max(\phi_U,\psi_\nu)=V_1'(\underline{w}_2+)$.

\item[(v)] \textbf{Strict local dominance of stopping:} Inside the retirement region (excluding $\underline{w}_2$), immediate retirement strictly dominates infinitesimal continuation: $\rho V_2(w)-\mathcal{H}_1V_2(w)>0$ for every $w\in\mathcal{R}\cap(\underline{w}_2,\infty)$, including the free boundaries.
\end{enumerate}
\end{enumerate}
In both regimes, $V_1$ satisfies the variational inequality \eqref{Variational_inequality} classically away from the free boundaries.
\end{thm}

The proof of Theorem~\ref{thm:main} concerns the stopping problem \eqref{pre-retirement value function} only and does not use Proposition~\ref{prop:reduction}. It also requires no transversality or uniform-integrability condition beyond the admissibility constraints: the homogeneous terms $\phi^{m_\pm+1}$ are bounded on the dual continuation pieces where they appear (Lemma~\ref{lem:cand}(d)). Combining Theorem~\ref{thm:main} with Proposition~\ref{prop:reduction}, we obtain the solution of the original problem.

\begin{cor}[Solution of the original problem]\label{cor:original}
For every $w>\underline{w}_1$, $V(w)=V_1(w)$, and the policy in Theorem~\ref{thm:main}(b)(iii) (or the never-retiring policy in Theorem~\ref{thm:main}(a)), followed after $\tau^*$ by the post-retirement feedback controls $(c_2^*,\pi_2^*)$ of Corollary~\ref{optimal post-retirement policies}, is optimal for the original problem \eqref{value_function}.
\end{cor}

The explicit representation in Theorem~\ref{thm:main}(b)(ii) also pins down the shape of pre-retirement policies below the lower retirement boundary.

\begin{cor}[Pre-retirement policies at low wealth]\label{cor:low_policy}
Suppose that $\phi_U<\phi_L$. On the low-wealth continuation region $(\underline{w}_1,w_L)$, the optimal pre-retirement consumption and risky investment
\[
c_1^*(w)=\phi^{-1/\gamma},\qquad \pi_1^*(w)=\frac{\theta}{\sigma}\Big(-m_-B_L\phi^{m_-}+\frac{1}{\gamma K}\phi^{-1/\gamma}\Big),\qquad w=X_L(\phi),
\]
are strictly increasing in wealth, and $c_1^*(w)\to0$, $\pi_1^*(w)\to0$ as $w\downarrow\underline{w}_1$. In particular, the agent does not reduce risky exposure as wealth rises toward $w_L$; any reduction in risky investment associated with retirement occurs discretely at the boundary.
\end{cor}

As noted in Section~\ref{sec:model}, the integrability requirement imposed on $\mathcal{A}(w)$ is not needed in \eqref{pre-retirement value function}; restricting to policies that satisfy it does not change $V_1$, since the optimal policy in Theorem~\ref{thm:main}(b)(iii) satisfies it: its payoff equals $V_1(w)$, which is finite. Proofs of Corollaries~\ref{cor:original} and \ref{cor:low_policy} are given in Appendix~\ref{app:proof_cor_original}.

Theorem~\ref{thm:main} offers clear economic intuition regarding both where retirement occurs and why the agent chooses to stop working at those specific thresholds.

First, the location of the retirement boundary $w_L$ (and symmetrically $w_U$) reveals the agent's immediate financial condition upon exiting the labor market. Because wealth and marginal value move in opposite directions, on the strictly decreasing part of the inverse wealth schedule, a higher marginal-value threshold $\phi_L$ corresponds to retiring at a lower wealth level:
\begin{itemize}
\item \textbf{Comfortable retirement ($w_L > w_{0,2}$, or $\phi_L < \psi_L$):} The agent accumulates enough wealth to fund strictly positive consumption immediately upon retirement.
\item \textbf{Support-reliant retirement ($\psi_L\le\phi_L\le\psi_\nu$}): The agent retires close to the borrowing limit, optimally choosing zero consumption at first and relying entirely on the government support $L$ for utility.
\item \textbf{Immediate corner retirement ($\phi_L>\psi_\nu$, hence $w_L=\underline w_2$}): The desire to retire is so strong that the agent exits the labor force the very instant post-retirement borrowing constraints permit, creating a concave kink in the value function at $\underline{w}_2$. (If $\phi_U \ge \psi_\nu$ as well, the entire retirement region collapses to the single boundary point $\mathcal{R} = \{\underline{w}_2\}$.)
\end{itemize}
The same classification applies to $w_U$ and $\phi_U$. In each case, $\phi_L$ and $\phi_U$ solve the corresponding line of \eqref{eq:Fminus}--\eqref{eq:Fplus}.

Second, Theorem~\ref{thm:main} highlights a fundamental tension between \emph{short-term gains} and \emph{long-term option value}. Recall from Lemma~\ref{lem:S_meaning} that $\mathcal{H}_1V_2(w)-\rho V_2(w)$ measures the instantaneous gain from delaying retirement for just a brief moment. While $\mathcal{H}_1V_2(w)-\rho V_2(w)<0$ at interior wealth in the retirement region confirms that retiring beats working in the immediate short run, having $\mathcal{H}_1V_2(w)-\rho V_2(w)<0$ alone is not enough to trigger retirement. Even when working yields a short-term loss ($\mathcal{H}_1V_2(w)-\rho V_2(w)< 0$), the agent may still choose to stay in the labor market to preserve the option of retiring later under much better wealth conditions.

The two boundary residuals aggregate this local gain over different ranges of the marginal value of wealth, as shown in Proposition~\ref{prop:Fint}.The distinction also has a portfolio implication. At a finite interior entry boundary $w_b>\underline w_2$, let $\phi_b=V_2'(w_b)$ and let $\pi_{\mathrm{work}}(w_b)$ be the limit of the risky holding from the working side. Taking the continuation-side limit in the working Bellman equation and subtracting $\mathcal H_1 V_2(w_b)$ gives
\[
\pi_{\mathrm{work}}(w_b)-\pi_2^*(w_b)=\frac{2[\rho V_2(w_b)-\mathcal H_1 V_2(w_b)]}{\sigma\theta\phi_b}>0.
\]
Indeed, by Theorem~\ref{thm:main}(b)(ii), $V_1''$ has a one-sided limit $V_1''(w_b\pm)=1/X'(\phi_b)<0$ from the working side, where $X\in\{X_L,X_U\}$. Letting $w\to w_b$ in $\rho V_1=\mathcal{H}_1V_1$ and using $V_1(w_b)=V_2(w_b)$ and $V_1'(w_b)=V_2'(w_b)=\phi_b$, \eqref{eq:pre-retirement-HJB-operator-optimized} gives $\rho V_2(w_b)-\mathcal{H}_1V_2(w_b)=\frac12\theta^2\phi_b^2\big(1/V_2''(w_b)-1/V_1''(w_b\pm)\big)$, while $\pi_i=-\frac{\theta}{\sigma}\phi_b/V_i''$. The strict sign follows from Theorem~\ref{thm:main}(b)(v). Thus entry from either side reduces risky holdings; the argument does not cover a kink at the borrowing limit.


\subsection{Retirement decisions under enhanced post-retirement quality of life} \label{subsec:kappa_gt_one}

We now specialize the general characterization in Theorem~\ref{thm:main} to the regime where the post-retirement quality of life is enhanced relative to the working phase ($\kappa>1$), and to the neutral case $\kappa=1$. We first identify the level of labor income that decides whether the agent retires with strictly positive consumption. Let
\begin{equation} \label{eq:Y1_tilde_def}
	\begin{split}
		\widetilde{Y}_1:= 
		& Y_2\left\{1-\frac{1-\nu}{m_++1} \left(\frac{\Delta_L}{L}\right)^{-m_+/m_-}\right\}  \\
	 & \quad+L\left\{1-\frac{r(\gamma m_-+1)\left\{1-\gamma-\gamma^2m_+(m_++1)(A-1)\kappa^{-\frac{1-\gamma}{\gamma}}\right\}}{(1-\gamma)\gamma^2m_+m_-(m_++1)K}\right\}.
	\end{split}
\end{equation}

\begin{lem}[Residual at the consumption threshold]\label{lem:F_at_psiL}
For every $\kappa>0$,
\begin{equation}\label{eq:F_at_psiL}
F_{m_-}(\psi_L)=-\frac{m_-}{r(m_-+1)}\big(Y_1-\widetilde{Y}_1\big).
\end{equation}
In particular, since $\frac{m_-}{r(m_-+1)}>0$, $F_{m_-}(\psi_L)<0$ if and only if $Y_1>\widetilde{Y}_1$.
\end{lem}

The following theorem shows that the retirement region is always an upper half-line when $\kappa\ge1$, determines in which post-retirement consumption region the threshold lies, and provides closed-form expressions in the canonical case in which the agent retires with strictly positive consumption, $w^*>w_{0,2}$, or equivalently $V_2'(w^*)=\phi^*<\psi_L$.

\begin{thm}[Single retirement threshold for $\kappa\ge1$]\label{theorem kappa>1}
Suppose that $\kappa>1$, or that $\kappa=1$ and $Y_1<L+Y_2$. Then $F_{m_-}$ has exactly one zero $\phi^*\in(0,\infty)$, with $F_{m_-}>0$ on $(0,\phi^*)$ and $F_{m_-}<0$ on $(\phi^*,\infty)$, and $\phi_U=0<\phi_L=\phi^*$. Consequently,
\[
\mathcal{R}=[w^*,\infty),\qquad w^*:=\mathcal{W}_2(\phi^*),\qquad \tau^*=\inf\{t\ge0:W_t\ge w^*\},
\]
$V_1(w)=V_{1,L}(\phi)$ with $w=X_L(\phi)$ on $(\underline{w}_1,w^*)$, and $V_1=V_2$ on $[w^*,\infty)$. Moreover:
\begin{enumerate}
\item[(i)] \textbf{Interior threshold.} $\phi^*<\psi_L$ if and only if
	\begin{equation}\label{ref:kappa_greater_than_1}
		Y_1>\widetilde{Y}_1.
	\end{equation}
In this case, $\phi^*$ is the unique root in $(0,\psi_L)$ of the scalar free-boundary equation
\begin{equation} \label{eq:free_boundary_phi}
	\frac{m_+-m_-}{(m_++1)(m_-+1)}D_1(\phi^*)^{m_+}+\frac{(\gamma m_-+1)(A-1)}{(1-\gamma)(m_-+1)K}(\phi^*)^{-\frac{1}{\gamma}}-\frac{m_-\big(Y_1-(L+Y_2)\big)}{r(m_-+1)}=0,
\end{equation}
whose left-hand side, the first line of \eqref{eq:Fminus}, is strictly decreasing in $\phi^*$, and the retirement wealth threshold is given in closed form by
\begin{equation*}
	w^*=D_1(\phi^*)^{m_+} + \frac{A}{K}(\phi^*)^{-1/\gamma} - \frac{L+Y_2}{r} > w_{0,2}.
\end{equation*}
In the continuation region $(\underline{w}_1,w^*)$, the pre-retirement value function is parametrically given by
\begin{equation}\label{pre-retirement value function solution}
	V_1(w)=\frac{m_-}{m_-+1}B\phi^{m_-+1} + \frac{1}{(1-\gamma)K} \phi^{-\frac{1-\gamma}{\gamma}},
\end{equation}
where wealth $w$ and marginal value $\phi=V_1'(w) \in (\phi^*,\infty)$ are linked via
\begin{equation} \label{eq:X_solution_pre}
	w=B\phi^{m_-} + \frac{1}{K}\phi^{-1/\gamma} - \frac{Y_1}{r},
\end{equation}
with the free-boundary constant
\begin{equation}\label{eq:BB}
	B=B_L=\left[ D_1(\phi^*)^{m_+} + \frac{A-1}{K}(\phi^*)^{-1/\gamma} + \frac{Y_1-(L+Y_2)}{r} \right] (\phi^*)^{-m_-} > 0.
\end{equation}
\item[(ii)] \textbf{Support-reliant threshold.} $\psi_L\le\phi^*\le\psi_\nu$ if and only if $Y_1\le\widetilde{Y}_1$ and $F_{m_-}(\psi_\nu)\le0$. In this case $w^*=W_0(\phi^*)\in[\underline{w}_2,w_{0,2}]$, and $\phi^*$ solves the second line of \eqref{eq:Fminus}.
\item[(iii)] \textbf{Corner threshold.} $\phi^*>\psi_\nu$ if and only if $F_{m_-}(\psi_\nu)>0$, where
\begin{equation*}
F_{m_-}(\psi_\nu)=-\frac{\gamma m_-+1}{(1-\gamma)(m_-+1)K}\psi_\nu^{-\frac1\gamma}-\frac{m_-(Y_1-\nu Y_2)}{r(m_-+1)}+\frac{1}{\rho}\Big((1-\nu)Y_2+\frac{(\kappa L)^{1-\gamma}}{(1-\gamma)\psi_\nu}\Big).
\end{equation*}
In this case $w^*=\underline{w}_2$: the agent retires as soon as retirement becomes feasible, and $V_1'(\underline{w}_2-)=\phi^*>\psi_\nu$.
\end{enumerate}
If $\kappa=1$ and $Y_1\ge L+Y_2$, never retiring is optimal. In the continuation region, $\phi=\phi(w)$ is determined by $w=X_L(\phi)$ with $B_L>0$, and the optimal feedback controls are
\[
c_1^*(w)=\phi(w)^{-1/\gamma},\qquad \pi_1^*(w)=-\frac{\theta}{\sigma}\phi X_L'(\phi)=-\frac{\theta}{\sigma}\Big(B_Lm_-\phi^{m_-}-\frac{1}{\gamma K}\phi^{-1/\gamma}\Big)>0.
\]
For $w\ge w^*$ the agent retires immediately and follows the post-retirement policies of Corollary~\ref{optimal post-retirement policies}.
\end{thm}

No additional condition needs to be verified in Theorem~\ref{theorem kappa>1}: the monotonicity $X_L'<0$ is automatic because $m_-B_L<0$; the dominance $V_1\ge V_2$ below $w^*$ and the stopping-region inequality $\rho V_2\ge\mathcal{H}_1V_2$ on $[w^*,\infty)$ are part of Theorem~\ref{thm:main}; and an upper retirement boundary is impossible because $\phi_U=0$. The proofs of Lemma~\ref{lem:F_at_psiL} and Theorem~\ref{theorem kappa>1} are given in Appendix~\ref{app:proof_kappa_ge_one}.

Figure~\ref{fig:wealth_phi_correspondence} summarizes the wealth thresholds and their respective marginal-value labels when $\kappa>1$ in case (i) of Theorem~\ref{theorem kappa>1}, illustrating the semi-infinite retirement region $[w^*,\infty)$ above the positive-consumption threshold $w_{0,2}$.

\begin{figure}[h!]
\centering
\begin{tikzpicture}[
    x=1.05cm,
    y=1.25cm,
    axis/.style={black, line width=0.75pt},
    tick/.style={black, line width=0.75pt},
    guide/.style={black!45, dashed, dash pattern=on 4pt off 4pt, line width=0.55pt},
    boundary/.style={black, dashed, dash pattern=on 4pt off 3pt, line width=0.75pt},
    every node/.style={font=\large}
]

\def\xleft{0.00}
\def\xwone{0.85}
\def\xwtwo{3.10}
\def\xwzero{5.50}
\def\xwstar{8.20}
\def\xright{10.50}

\def\yphi{0.00}
\def\yw{1.55}

\fill[blue!10] (\xwstar,\yphi) rectangle (\xright,\yw);

\draw[axis] (\xleft,\yw) -- (\xright,\yw);
\draw[axis] (\xleft,\yphi) -- (\xright,\yphi);

\draw[guide]    (\xwone,\yphi)  -- (\xwone,\yw);
\draw[guide]    (\xwtwo,\yphi)  -- (\xwtwo,\yw);
\draw[guide]    (\xwzero,\yphi) -- (\xwzero,\yw);
\draw[boundary] (\xwstar,\yphi) -- (\xwstar,\yw);

\foreach \x in {\xwone,\xwtwo,\xwzero,\xwstar}{
    \draw[tick] (\x,{\yw-0.08}) -- (\x,{\yw+0.08});
    \draw[tick] (\x,{\yphi-0.08}) -- (\x,{\yphi+0.08});
}

\node[anchor=east] at (-0.32,\yw) {$w$};
\node[anchor=south] at ({\xwone-0.10},{\yw+0.25}) {$\underline{w}_1=-\dfrac{Y_1}{r}$};
\node[anchor=south] at ({\xwtwo+0.10},{\yw+0.25}) {$\underline{w}_2=-\nu\dfrac{Y_2}{r}$};
\node[anchor=south] at (\xwzero,{\yw+0.25}) {$w_{0,2}$};
\node[anchor=south] at (\xwstar,{\yw+0.25}) {$w^*$};
\node[anchor=west] at ({\xright+0.22},\yw) {$+\infty$};

\node[anchor=east] at (-0.32,\yphi) {$\phi,\psi$};
\node[anchor=north] at (\xwone,{\yphi-0.25}) {$+\infty$};
\node[anchor=north] at (\xwtwo,{\yphi-0.25}) {$\psi_\nu$};
\node[anchor=north] at (\xwzero,{\yphi-0.25}) {$\psi_L=\kappa^{1-\gamma}L^{-\gamma}$};
\node[anchor=north] at (\xwstar,{\yphi-0.25}) {$\phi^*$};
\node[anchor=west] at ({\xright+0.22},\yphi) {$0$};

\node[text=black!60, font=\normalsize] at (4.35,{0.5*(\yw+\yphi)}) {Continuation};
\node[font=\normalsize] at ({0.5*(\xwstar+\xright)},{0.5*(\yw+\yphi)}) {Retirement};

\end{tikzpicture}
\caption{Wealth and marginal-value thresholds when $\kappa>1$. The labels $\psi_\nu=V_2^{\prime}(\underline w_2+)$ and $\psi_L=V_2^{\prime}(w_{0,2})$ are post-retirement marginal values; $\phi^*$ is the common marginal value at entry.}
\label{fig:wealth_phi_correspondence}
\end{figure}
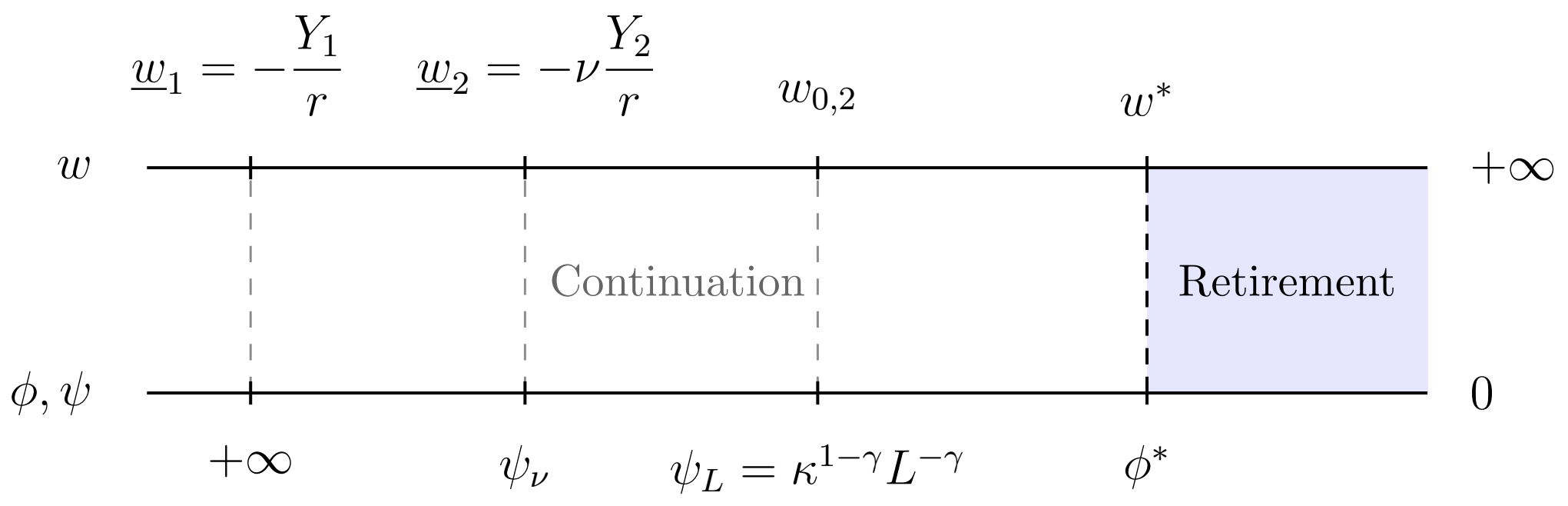

\subsection{Retirement decisions under degraded post-retirement quality of life} \label{subsec:kappa_lt_one}

When $0<\kappa<1$, retirement discounts effective consumption, but support $L$ can still make retirement attractive at intermediate wealth. We distinguish between never retiring, retiring at the borrowing limit, and retiring on an interval with two interior boundaries.

First, we establish that the classical semi-infinite retirement region breaks down completely whenever $0<\kappa<1$: high-wealth agents strictly prefer continuing to work at their current wealth to retiring immediately.

\begin{pro}[Failure of semi-infinite retirement regions]\label{prop:db-high-wealth-continuation}
Suppose that $0<\kappa<1$.Then the retirement region $\mathcal{R}$ is bounded above; in particular, it cannot be of the form $[w^{*},\infty)$.
\end{pro}

Moreover, if pre-retirement labor income exceeds post-retirement resources by a sufficiently large margin, working forever yields at least as much value as any policy with a retirement decision.

\begin{pro}[No-retirement condition]\label{gpsdb:no-retirement-income}
Suppose that $0<\kappa<1$. If
\begin{equation}\label{gpsdb:no-retirement-condition}
Y_1-Y_2\geq\kappa L,
\end{equation}
then never retiring is optimal: $V_1(w)=V^{\mathrm{NR}}(w)$ for $w>\underline w_1$. In particular, immediate retirement cannot be optimal on an interior wealth interval of positive length. In fact, $\mathcal{R}$ contains at most one point.
\end{pro}
By Proposition~\ref{gpsdb:no-retirement-income}, the condition $\kappa L>Y_1-Y_2$ is necessary for retirement to be optimal at some wealth level. When $0<\kappa<1$, it implies $L+Y_2>Y_1$ and is strictly stronger. For $\kappa=1$, the same conclusion follows from Theorem~\ref{theorem kappa>1}, since $Y_1-Y_2\ge L$ is then equivalent to $Y_1\ge L+Y_2$.

Consistent with Theorem~\ref{thm:main}, Proposition~\ref{prop:db-high-wealth-continuation} implies that every retirement region in this regime has a finite upper end; Corollary~\ref{cor:configurations} below lists all possible forms. In the double-free-boundary case, the state space splits into a bounded retirement corridor flanked by two continuation regions:
\begin{equation}\label{eq:db-continuation-and-retirement-regions}
 \mathcal{C}=(\underline{w}_1,w_L)\cup(w_U,\infty),
 \qquad
 \mathcal{R}=[w_L,w_U].
\end{equation}

The following corollary lists all possible configurations of the retirement region and gives a necessary and sufficient condition for retirement in terms of the two residuals.

\begin{cor}[Exhaustive configurations for $0<\kappa<1$]\label{cor:configurations}
Let $0<\kappa<1$. Then $\phi_U<\phi_L$, i.e., retirement is strictly optimal at some wealth level, if and only if both of the following hold:
\begin{enumerate}
\item[(i)] $F_{m_+}$ takes a positive value, so that $\phi_U\in(0,\infty)$ is the smallest zero of $F_{m_+}$;
\item[(ii)] $F_{m_-}(\phi_U)>0$.
\end{enumerate}
In that case, $\phi_L$ is the unique zero of $F_{m_-}$ in $(\phi_U,\infty)$ and $\mathcal{R}=[w_L,w_U]$ with $w_U<\infty$. Moreover, $\mathcal{R}$ takes exactly one of the following forms: $\mathcal{R}$ has at most one point (no retirement; $\phi_L\le\phi_U$, and $\mathcal{R}=\emptyset$ unless $\phi_L=\phi_U$); $\mathcal{R}=[\underline{w}_2,w_U]$ (corner-lower-boundary case; $\phi_L\ge\psi_\nu>\phi_U$); $\mathcal{R}=[w_L,w_U]$ with $\underline{w}_2<w_L<w_U<\infty$ (double-free-boundary case; $\phi_L<\psi_\nu$); or the degenerate form $\mathcal{R}=\{\underline{w}_2\}$, which occurs when $\phi_L>\phi_U\ge\psi_\nu$.
\end{cor}

In the corner-lower-boundary case with $\phi_L>\psi_\nu$, the low-wealth branch $X_L$ with $B_L$ from \eqref{eq:BLU} meets $V_2$ at $\underline{w}_2$ with a concave kink (Theorem~\ref{thm:main}(b)(iv)), and $\phi_L$ solves the third line of \eqref{eq:Fminus}. If $\phi_L=\psi_\nu$, the branch meets $V_2$ smoothly at $\underline{w}_2$. The proof is given in Appendix~\ref{app:proof_configurations}.

In the interior double-free-boundary case, both retirement thresholds lie strictly above the post-retirement zero-consumption level $w_{0,2}$, i.e.,
\begin{equation}\label{eq:db-boundaries-positive-consumption}
 0<\phi_U<\phi_L<\psi_L,
\end{equation}
so that $w_L=W_+(\phi_L)$ and $w_U=W_+(\phi_U)$ with $W_+(\psi)=D_1\psi^{m_+}+\frac{A}{K}\psi^{-1/\gamma}-\frac{L+Y_2}{r}$. On this branch, the boundary equations $F_{m_-}(\phi_L)=0$ and $F_{m_+}(\phi_U)=0$ reduce by the first lines of \eqref{eq:Fminus}--\eqref{eq:Fplus} to
\begin{equation}\label{eq:db-lower-boundary-equation}
 \frac{m_+-m_-}{(m_++1)(m_-+1)}D_1\phi_L^{m_+} +\frac{(\gamma m_-+1)(A-1)}{(1-\gamma)(m_-+1)K}\phi_L^{-1/\gamma} -\frac{m_-\{Y_1-(L+Y_2)\}}{r(m_-+1)}=0
\end{equation}
and
\begin{equation}\label{eq:db-F-plus}
 \frac{(\gamma m_++1)(A-1)}{(1-\gamma)(m_++1)K}\phi_U^{-1/\gamma} -\frac{m_+\{Y_1-(L+Y_2)\}}{r(m_++1)} = 0.
\end{equation}

For $0<\kappa<1$, we have $(A-1)/(1-\gamma)<0$. Consequently, whenever the consumption-equivalent support $L$ exceeds the cash income loss $Y_1-Y_2$,
\begin{equation}\label{eq:db-upper-root-income-condition}
 L+Y_2>Y_1,
\end{equation}
the positive-consumption formula \eqref{eq:db-F-plus}, viewed as an algebraic equation on $(0,\infty)$, admits a unique closed-form positive root:
\begin{equation}\label{eq:db-explicit-upper-root}
 \phi_U^\circ
 =\left[ \frac{r(\gamma m_++1)(1-A)} {m_+(1-\gamma)K(L+Y_2-Y_1) }\right]^{\gamma}.
\end{equation}
The next lemma tests whether this algebraic root lies on the branch on which the formula was derived.

\begin{lem}[Branch test for the upper-boundary root]\label{lem:db-upper-boundary-admissibility}
Suppose that $0<\kappa<1$ and $L+Y_2>Y_1$, and let $\phi_U^\circ$ be given by \eqref{eq:db-explicit-upper-root}. Then the algebraic root satisfies $\phi_U^\circ<\psi_L$ if and only if
\begin{equation}\label{eq:db-upper-boundary-parameter-condition}
 \frac{r(\gamma m_++1)(1-A)L} {m_+(1-\gamma)K(L+Y_2-Y_1) }<A.
\end{equation}
At equality the algebraic root is $\psi_L$. If the inequality is reversed strictly, the root lies outside this branch; any actual boundary must then be determined using the full residual \eqref{eq:Fplus} and Definition~\ref{def:phiLU}.
\end{lem}

Two boundaries strictly above $w_{0,2}$ can therefore be tested without a two-dimensional search: compute $\phi_U^\circ$ in closed form and check the remaining conditions of the following corollary. By Lemma~\ref{lem:F_at_psiL}, the condition $F_{m_-}(\psi_L)<0$, which places the lower threshold before $\psi_L$, is equivalent to $Y_1>\widetilde{Y}_1$.

\begin{cor}[Explicit criterion for two interior free boundaries]\label{cor:explicit_criterion}
Suppose that $0<\kappa<1$. The following statements are equivalent:
\begin{enumerate}
\item[(i)] there exist $0<\phi_1<\phi_2<\psi_L$ with $F_{m_+}(\phi_1)=0$ and $F_{m_-}(\phi_2)=0$;
\item[(ii)] the following four conditions hold:
\begin{equation}\label{eq:four_conditions}
L+Y_2>Y_1,\qquad \phi_U^\circ<\psi_L,\qquad F_{m_-}(\phi_U^\circ)>0,\qquad Y_1>\widetilde{Y}_1.
\end{equation}
\end{enumerate}
In this case, $\phi_U=\phi_1=\phi_U^\circ$ and $\phi_L=\phi_2$, where $\phi_U,\phi_L$ are as in Definition~\ref{def:phiLU}, and $\phi_L$ is the unique root of \eqref{eq:db-lower-boundary-equation} in $(\phi_U^\circ,\psi_L)$. Moreover, $w_L=W_+(\phi_L)$ and $w_U=W_+(\phi_U^\circ)$ with $w_{0,2}<w_L<w_U<\infty$, $\mathcal{R}=[w_L,w_U]$ and $\mathcal{C}=(\underline{w}_1,w_L)\cup(w_U,\infty)$; $V_1$ is given by Theorem~\ref{thm:main}(b)(ii) with
\begin{equation}\label{eq:db-coefficients}
\begin{aligned}
 B_L&=\Big[ D_1\phi_L^{m_+}+\frac{A-1}{K}\phi_L^{-1/\gamma} +\frac{Y_1-(L+Y_2)}{r} \Big]\phi_L^{-m_-},\\
 B_U&=\Big[ D_1\phi_U^{m_+}+\frac{A-1}{K}\phi_U^{-1/\gamma} +\frac{Y_1-(L+Y_2)}{r} \Big]\phi_U^{-m_+};
\end{aligned}
\end{equation}
and $\tau^*=\inf\{t\ge0:W_t\in[w_L,w_U]\}$ is optimal.
\end{cor}

The proof is given in Appendix~\ref{app:proof_explicit_criterion}.

\begin{rem}[Interpretation]\label{rem:interpretation}
At low wealth, labor income and pre-retirement borrowing capacity make continued work valuable. At intermediate wealth, the income support $L$ can make retirement optimal. At high wealth, $L$ becomes negligible, while $\kappa<1$ makes post-retirement utility smaller. In terms of Theorem~\ref{thm:main}, $S$ is positive at both ends and negative in between, and $\mathcal{R}$ corresponds to the sub-interval of $\{S<0\}$ on which both tail integrals of $S$ in Proposition~\ref{prop:Fint} are negative.
\end{rem}

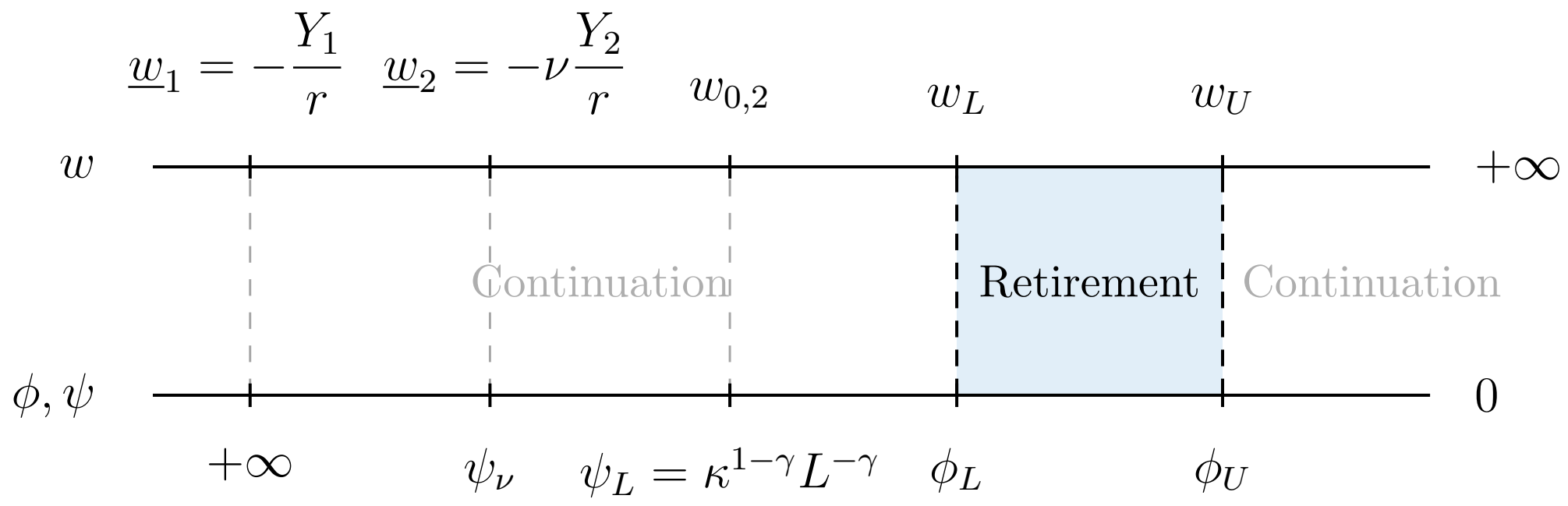
\begin{figure}[h!]
\centering
\begin{tikzpicture}[
    x=1.10cm,
    y=1.25cm,
    axis/.style={black, line width=0.75pt},
    tick/.style={black, line width=0.75pt},
    guide/.style={guidegray, dashed, dash pattern=on 4pt off 4pt,
                  line width=0.55pt},
    boundary/.style={black, dashed, dash pattern=on 4pt off 3pt,
                     line width=0.75pt},
    every node/.style={font=\large}
]

\def\xleft{0.00}
\def\xwone{0.75}
\def\xwtwo{2.60}
\def\xwzero{4.45}
\def\xwL{6.20}
\def\xwU{8.25}
\def\xright{9.85}

\def\yphi{0.00}
\def\yw{1.55}

\fill[retirementblue]
    (\xwL,\yphi) rectangle (\xwU,\yw);

\draw[axis] (\xleft,\yw) -- (\xright,\yw);
\draw[axis] (\xleft,\yphi) -- (\xright,\yphi);

\draw[guide]    (\xwone,\yphi) -- (\xwone,\yw);
\draw[guide]    (\xwtwo,\yphi) -- (\xwtwo,\yw);
\draw[guide]    (\xwzero,\yphi) -- (\xwzero,\yw);
\draw[boundary] (\xwL,\yphi)   -- (\xwL,\yw);
\draw[boundary] (\xwU,\yphi)   -- (\xwU,\yw);

\foreach \x in {\xwone,\xwtwo,\xwzero,\xwL,\xwU}{
    \draw[tick] (\x,{\yw-0.08}) -- (\x,{\yw+0.08});
    \draw[tick] (\x,{\yphi-0.08}) -- (\x,{\yphi+0.08});
}

\node[anchor=east] at (-0.32,\yw) {$w$};
\node[anchor=south] at ({\xwone-0.10},{\yw+0.25})
    {$\underline{w}_1=-\dfrac{Y_1}{r}$};
\node[anchor=south] at ({\xwtwo+0.12},{\yw+0.25})
    {$\underline{w}_2=-\nu\dfrac{Y_2}{r}$};
\node[anchor=south] at (\xwzero,{\yw+0.25}) {$w_{0,2}$};
\node[anchor=south] at (\xwL,{\yw+0.25}) {$w_L$};
\node[anchor=south] at (\xwU,{\yw+0.25}) {$w_U$};
\node[anchor=west] at ({\xright+0.22},\yw) {$+\infty$};

\node[anchor=east] at (-0.32,\yphi) {$\phi,\psi$};
\node[anchor=north] at (\xwone,{\yphi-0.25}) {$+\infty$};
\node[anchor=north] at (\xwtwo,{\yphi-0.25}) {$\psi_\nu$};
\node[anchor=north] at (\xwzero,{\yphi-0.25})
    {$\psi_L=\kappa^{1-\gamma}L^{-\gamma}$};
\node[anchor=north] at (\xwL,{\yphi-0.25}) {$\phi_L$};
\node[anchor=north] at (\xwU,{\yphi-0.25}) {$\phi_U$};
\node[anchor=west] at ({\xright+0.22},\yphi) {$0$};

\node[text=continuationgray, font=\normalsize]
    at (3.45,{0.5*(\yw+\yphi)})
    {Continuation};
\node[font=\normalsize]
    at ({0.5*(\xwL+\xwU)},{0.5*(\yw+\yphi)})
    {Retirement};
\node[text=continuationgray, font=\normalsize]
    at (9.4,{0.5*(\yw+\yphi)})
    {Continuation};

\end{tikzpicture}
\caption{Wealth and marginal-value thresholds when $0<\kappa<1$. The labels $\psi_\nu=V_2^{\prime}(\underline w_2+)$ and $\psi_L=V_2^{\prime}(w_{0,2})$ refer to retirement preferences; the working and retirement marginal values agree at the interior entry boundaries shown.}
\label{fig:wealth_phi_correspondence 0<kappa<1}
\end{figure}

Figure~\ref{fig:wealth_phi_correspondence 0<kappa<1} illustrates the wealth thresholds and their respective marginal-value labels in the two-interior-boundary case, where voluntary retirement is confined to the bounded intermediate wealth corridor $[w_L,w_U]$.

\subsection{Numerical Results} \label{subsec:numerical}

This section illustrates the geometric configurations of the optimal retirement boundaries and the corresponding consumption and asset allocation policies. In each comparative-static exercise, we vary a single target parameter while holding all other baseline parameters fixed.

For the first regime where the post-retirement quality of life is enhanced relative to the working phase ($\kappa>1$), the baseline parameter vector is specified as
\[
(r,\mu,\sigma,\rho,\gamma,Y_1,Y_2,\nu,L,\kappa)
=(0.01,0.05,0.20,0.05,2,1,0.10,0.20,1,1.20).
\]
Under this parameterization, the system yields a single lower retirement boundary $w_L=w^*=120.0218$, an unbounded upper threshold $w_U=\infty$, and a zero-consumption threshold $w_{0,2}=-1.8036$. Here, $K=0.035$, $\Delta_L=0.4253<L$, and $Y_1=1>\widetilde{Y}_1=0.4240$, so that by Theorem~\ref{theorem kappa>1}(i) the retirement threshold lies in the positive-consumption region.

Conversely, for the second regime where the post-retirement quality of life is degraded ($0<\kappa<1$), we employ the parameter configuration
\[
(r,\mu,\sigma,\rho,\gamma,Y_1,Y_2,\nu,L,\kappa)
=(0.03,0.07,0.20,0.07,2,1.30,0.15,0.50,1.80,0.86).
\]
This setup generates two distinct marginal-value roots and compact wealth boundaries given by
\[
\phi_U=0.07955<\phi_L=0.11644<\psi_L=0.35889,
\qquad
(w_L,w_U)=(10.9819,20.2005),
\]
with a post-retirement zero-consumption threshold of $w_{0,2}=-1.9936$. Both illustrative parameterizations satisfy $K>0$ and $L>\Delta_L$ in \eqref{eq:post_case1_condition_intro}. For the second parameterization, $K=0.055$ and $\Delta_L=0.4405<L$, and the four conditions \eqref{eq:four_conditions} of Corollary~\ref{cor:explicit_criterion} hold: $L+Y_2-Y_1=0.65>0$, $\phi_U^\circ=0.07955<\psi_L$, $F_{m_-}(\phi_U^\circ)=1.0295>0$, and $Y_1=1.30>\widetilde{Y}_1=1.1154$.

\begin{figure}[h!]
\centering
\subfigure[$\kappa$]{
\includegraphics[width=.45\columnwidth]{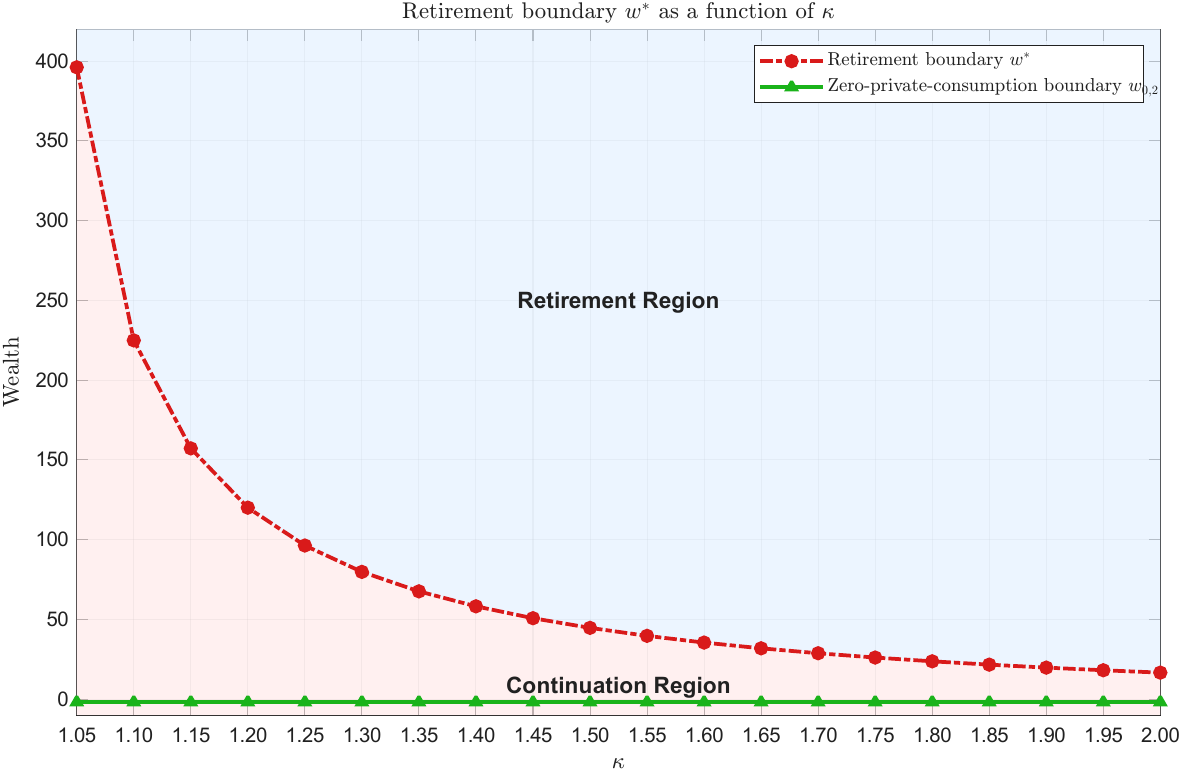}
\label{fig:kappa_gt_one_kappa}
}
\quad
\subfigure[$Y_2$]{
\includegraphics[width=.45\columnwidth]{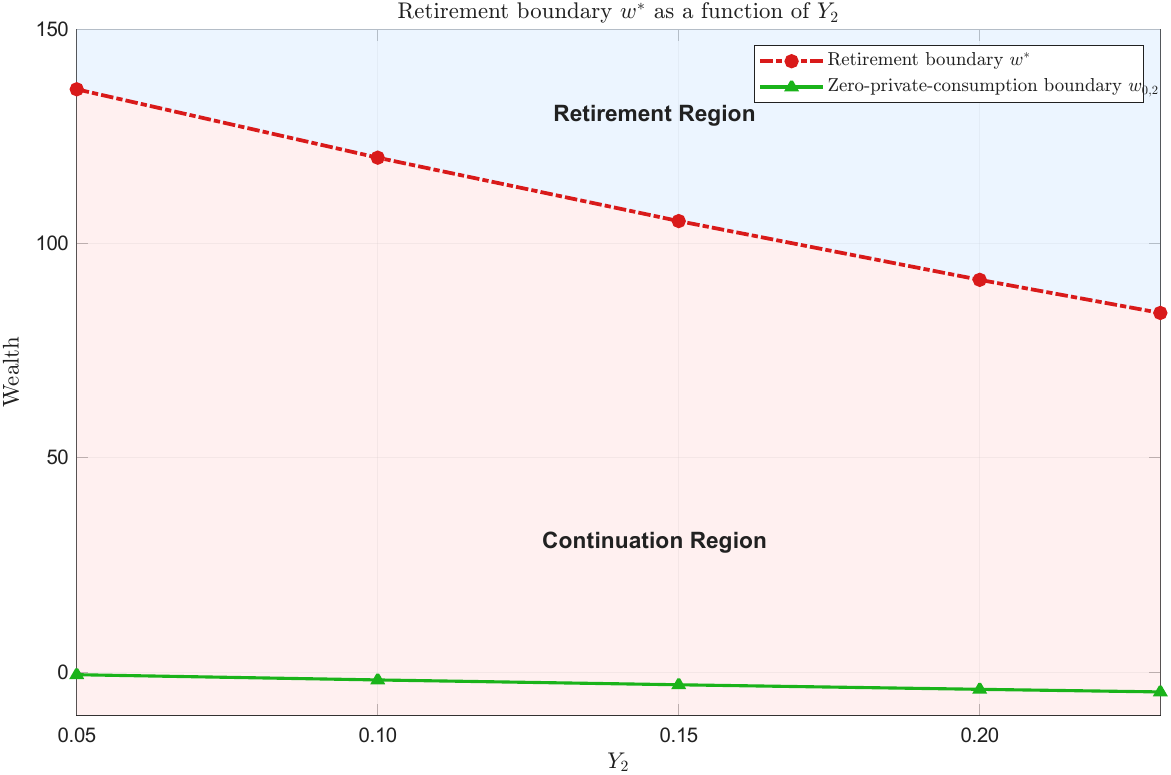}
\label{fig:kappa_gt_one_Y2}
}
\subfigure[$\nu$]{
\includegraphics[width=.45\columnwidth]{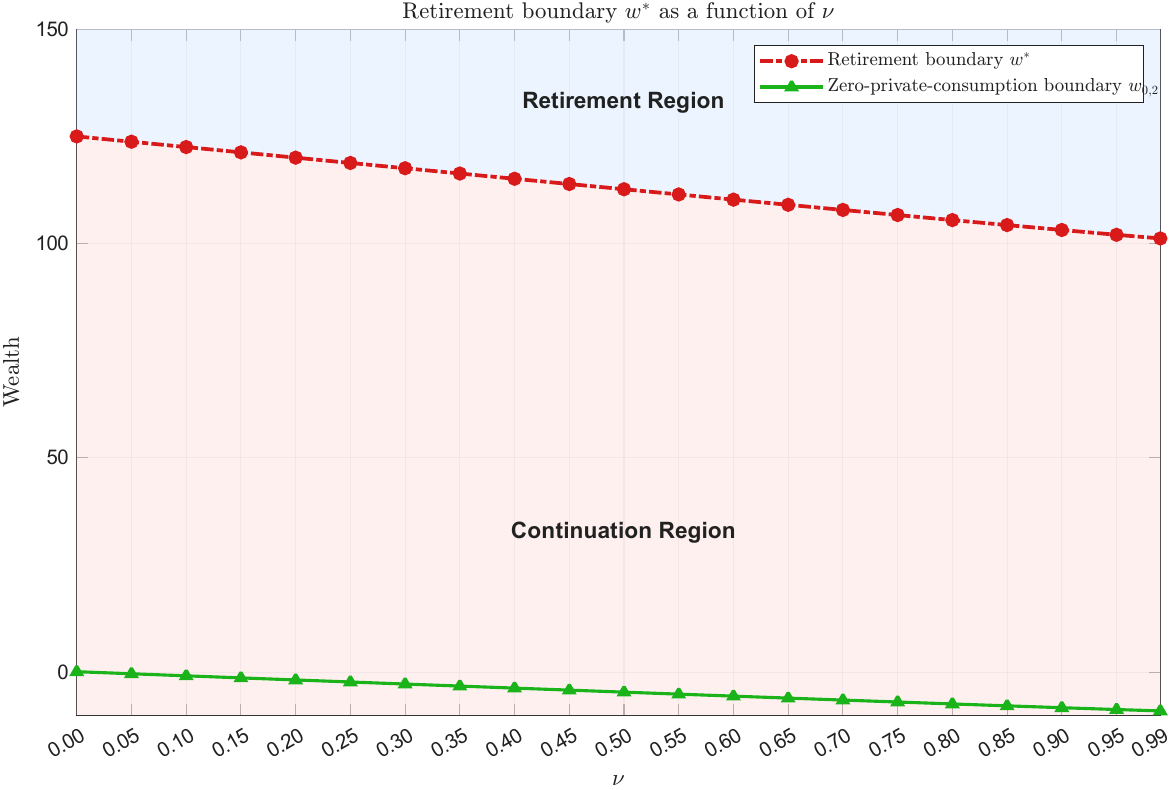}
\label{fig:kappa_gt_one_nu}
}
\quad
\subfigure[$L$]{
\includegraphics[width=.45\columnwidth]{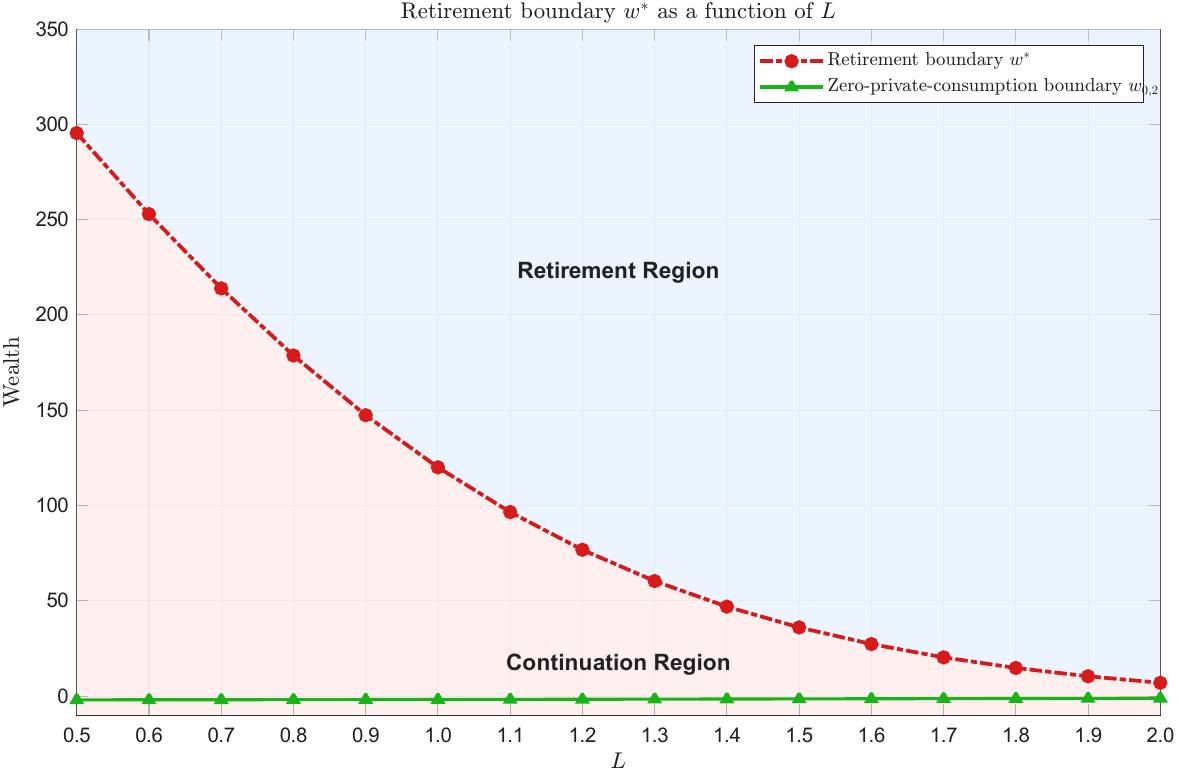}
\label{fig:kappa_gt_one_L}
}
\caption{The retirement boundary $w_L=w^*$ and the zero-consumption boundary $w_{0,2}$ when $\kappa>1$ (with $w_U=\infty$).}
\label{fig:kappa_gt_one}
\end{figure}

Figure~\ref{fig:kappa_gt_one} presents the single-boundary comparative statics for the first regime where the post-retirement quality of life is enhanced ($\kappa>1$). In this environment, the retirement region is the semi-infinite interval $[w_L,\infty)$, and the red curve in Figure~\ref{fig:kappa_gt_one} represents the unique optimal retirement boundary. Over the displayed ranges, $w_L$ decreases monotonically with $\kappa$, $Y_2$, $\nu$, and $L$. These shifts expand the set of wealth levels at which the agent retires immediately; however, they do not by themselves rank random retirement times, because the wealth trajectory also depends on the endogenous pre-retirement controls. Finally, the green curve $w_{0,2}$ separates the regions of zero and positive private consumption after retirement.

\begin{figure}[h!]
\centering
\subfigure[$\kappa$]{
\includegraphics[width=.45\columnwidth]{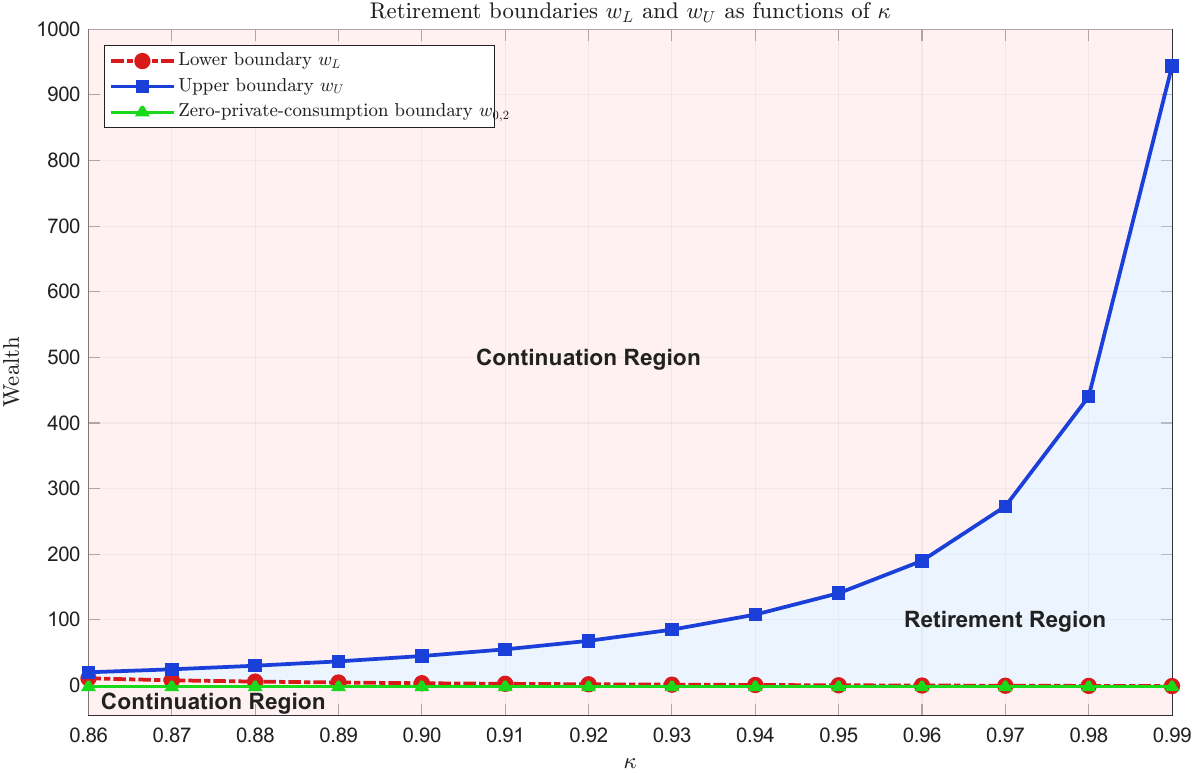}
\label{fig:kappa_lt_one_kappa}
}
\quad
\subfigure[$Y_2$]{
\includegraphics[width=.45\columnwidth]{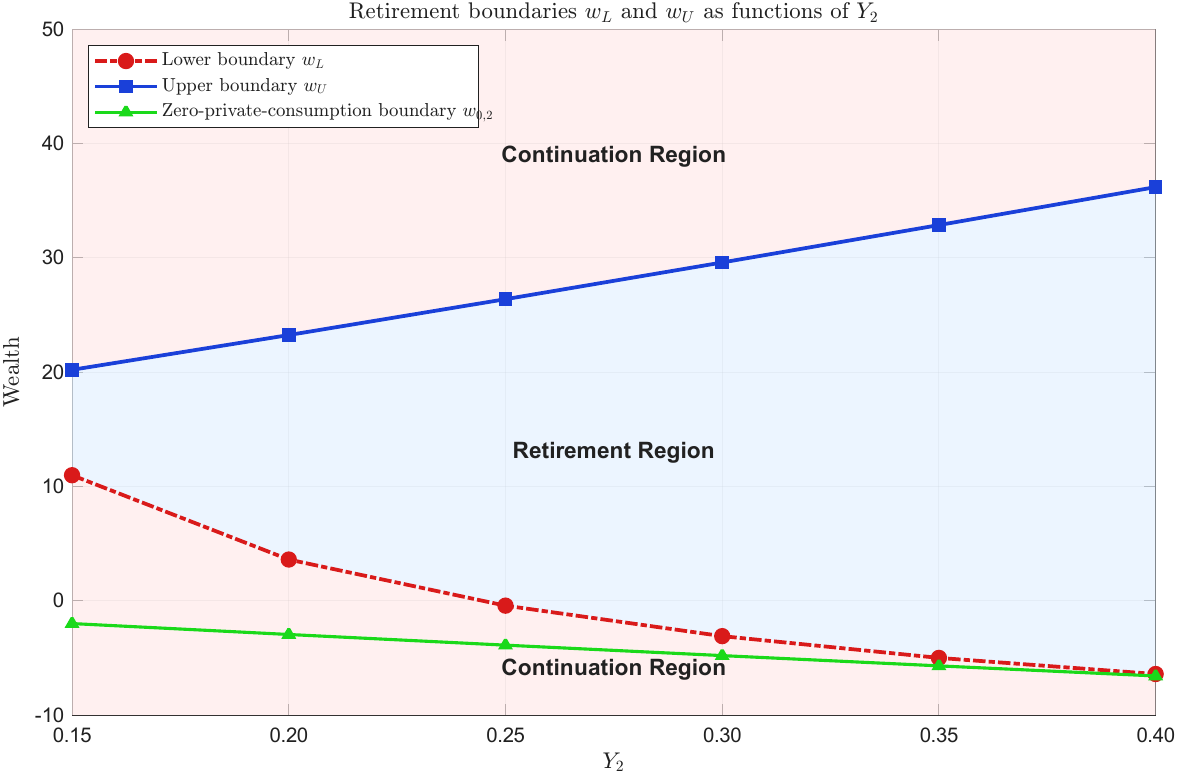}
\label{fig:kappa_lt_one_Y2}
}
\subfigure[$\nu$]{
\includegraphics[width=.45\columnwidth]{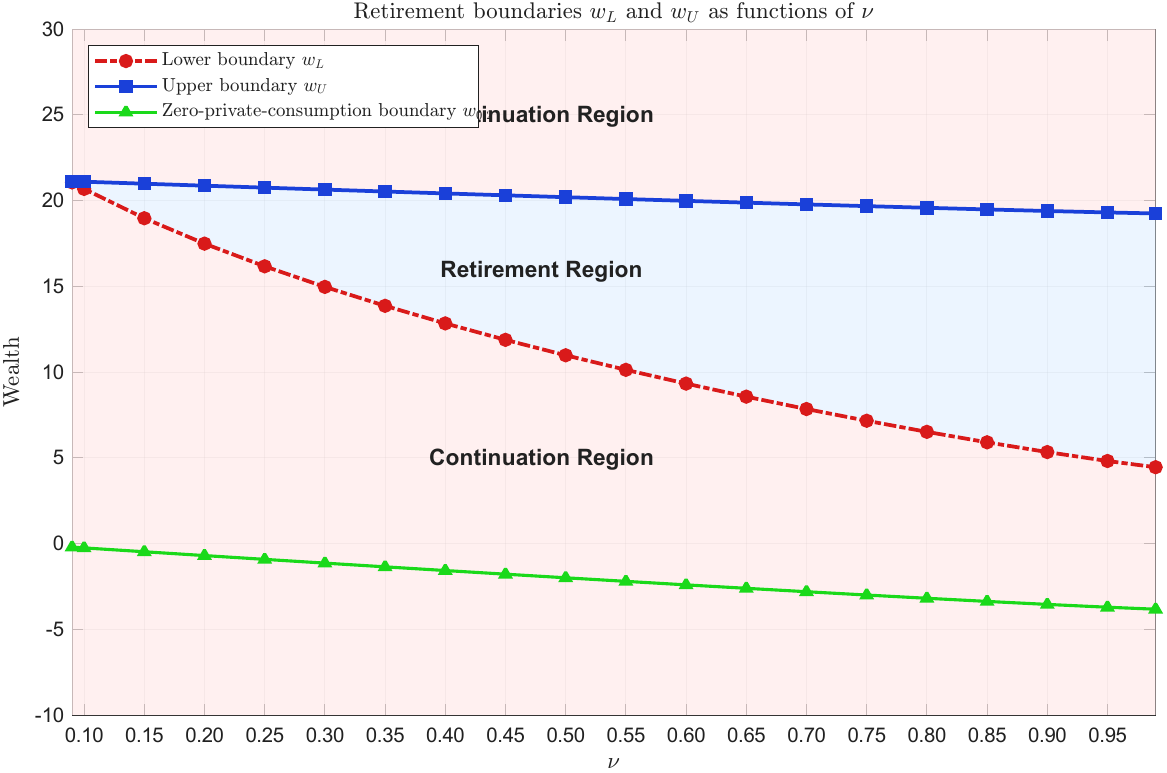}
\label{fig:kappa_lt_one_nu}
}
\quad
\subfigure[$L$]{
\includegraphics[width=.45\columnwidth]{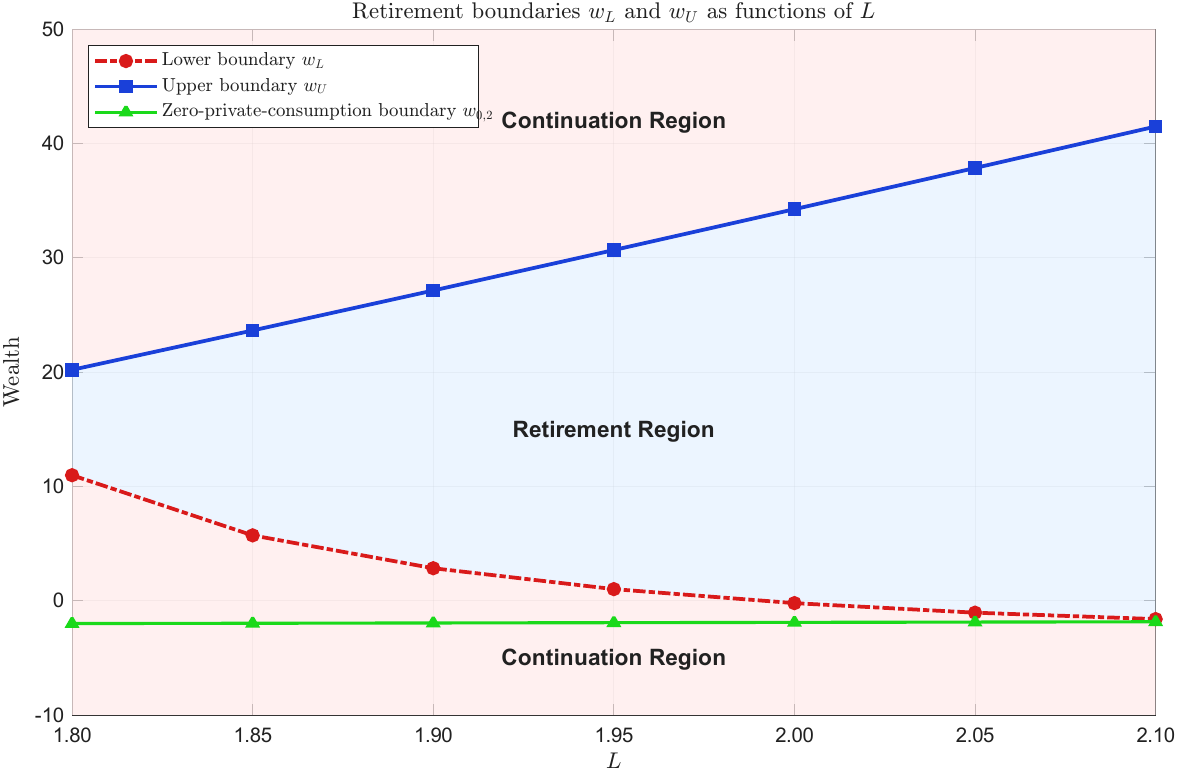}
\label{fig:kappa_lt_one_L}
}
\caption{The retirement boundaries $w_L$ and $w_U$ and the zero-consumption boundary $w_{0,2}$ when $0<\kappa<1$.}
\label{fig:kappa_lt_one}
\end{figure}

Figure~\ref{fig:kappa_lt_one} presents the two-boundary comparative statics for the second regime where the post-retirement quality of life degrades ($0<\kappa<1$). Here, voluntary retirement is optimal exclusively within the compact interval $w\in[w_L,w_U]$. Over the displayed ranges, increases in $\kappa$, $Y_2$, or $L$ lower $w_L$ and raise $w_U$, thereby expanding the retirement region on both ends.\footnote{In the limiting case where $\kappa$ approaches unity from below, the high-wealth continuation motive vanishes and $w_U$ diverges to infinity, recovering the single-boundary retirement structure.} By contrast, an increase in the pledgeability parameter $\nu$ shifts both thresholds downward: although the overall retirement interval widens because $w_L$ falls more sharply than $w_U$, some high-wealth workers near the upper boundary who previously retired now choose to continue working.

\begin{rem}[On the magnitude of income support]\label{rem:calibration}
Two theoretical conditions require the support parameter $L$ to be relatively large compared with the income drop at retirement. First, condition \eqref{eq:post_case1_condition_intro} requires $L\ge\Delta_L=-(\gamma m_-+1)(1-\nu)Y_2$, where $-(\gamma m_-+1)=5.87$ in the second baseline. Second, Proposition~\ref{gpsdb:no-retirement-income} (for $0 \lt \kappa \lt 1$) together with Lemma~\ref{lem:S_sign}(b) and Theorem~\ref{thm:main} (for $\kappa=1$) implies that $\kappa L \gt Y_1-Y_2$ is a necessary condition for voluntary retirement whenever $\kappa\le 1$. Because $L$ captures the consumption-equivalent value of in-kind transfers (such as healthcare or housing assistance), it need not be small relative to cash earnings for financially vulnerable households. Nevertheless, the existence of a bounded retirement interval does not depend on $L$ exceeding pre-retirement labor income $Y_1$: for example, setting $(Y_1,Y_2,\nu,L,\kappa)=(1,0.5,0.7,0.9,0.9)$ while keeping all other parameters at their second-baseline values satisfies all four conditions in \eqref{eq:four_conditions} and yields $(w_L,w_U)=(-10.4023,19.8314)$. The complementary regime $L \lt \Delta_L$, in which no zero-consumption region arises, is left for future research.
\end{rem}

\begin{figure}[t!]
\centering
\subfigure[Consumption]{
\includegraphics[width=.45\columnwidth]{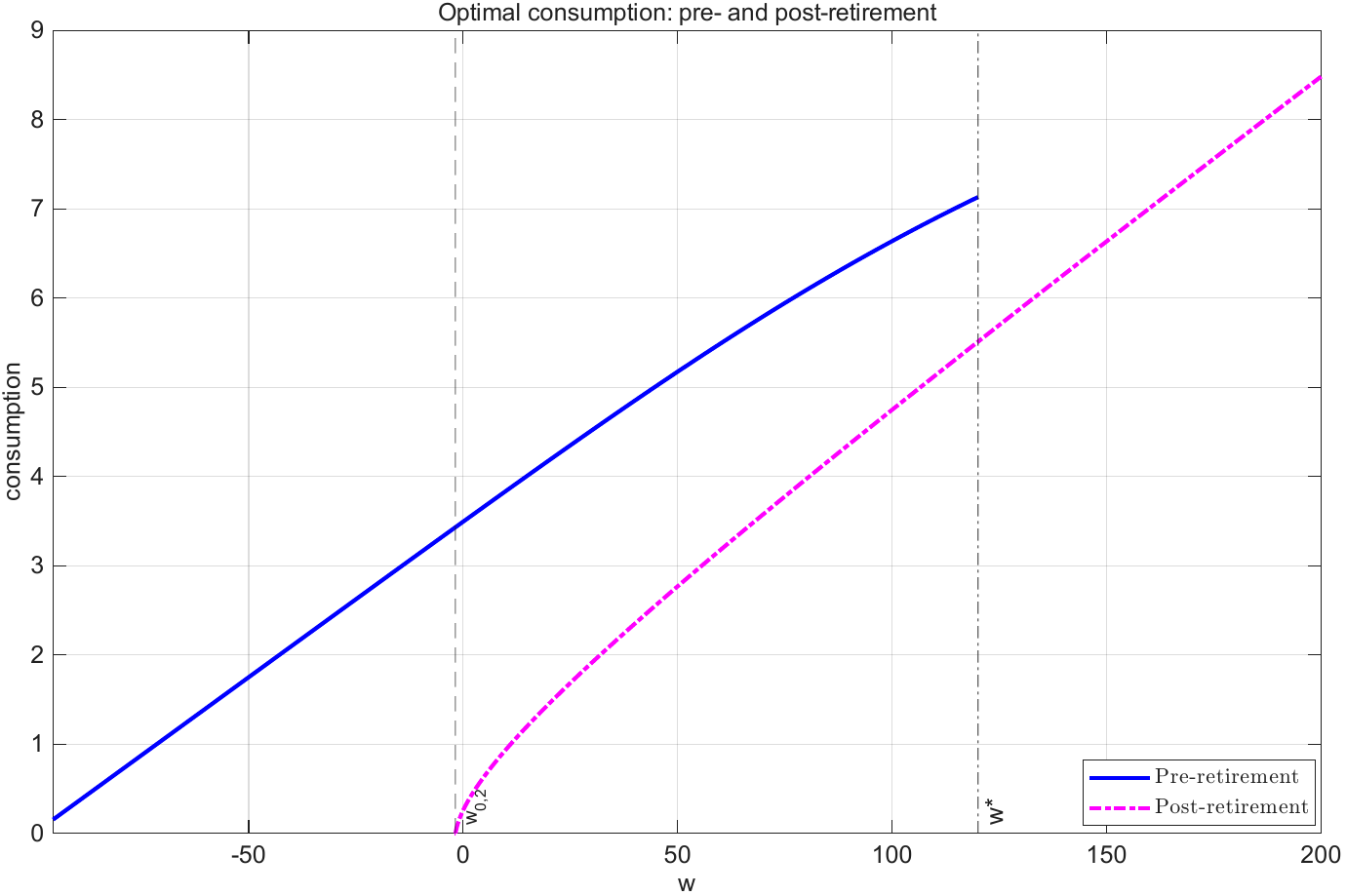}
\label{fig:kappa_gt_one_consumption}
}
\quad
\subfigure[Risky investment]{
\includegraphics[width=.45\columnwidth]{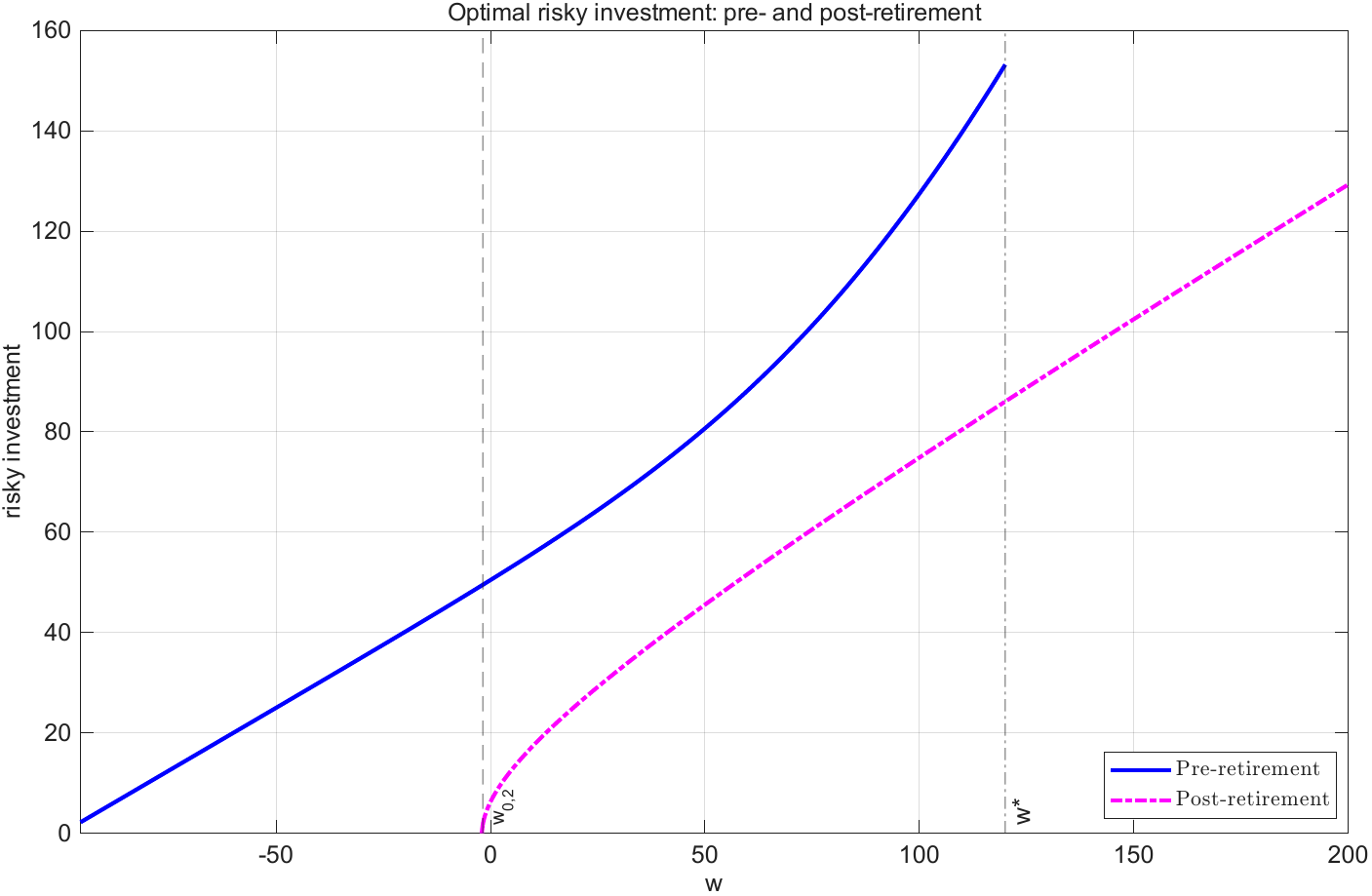}
\label{fig:kappa_gt_one_portfolio}
}
\caption{Optimal consumption and risky investment policies when $\kappa>1$.}
\label{fig:kappa_gt_one2}
\end{figure}

\begin{figure}[t!]
\centering
\subfigure[Consumption]{
\includegraphics[width=.45\columnwidth]{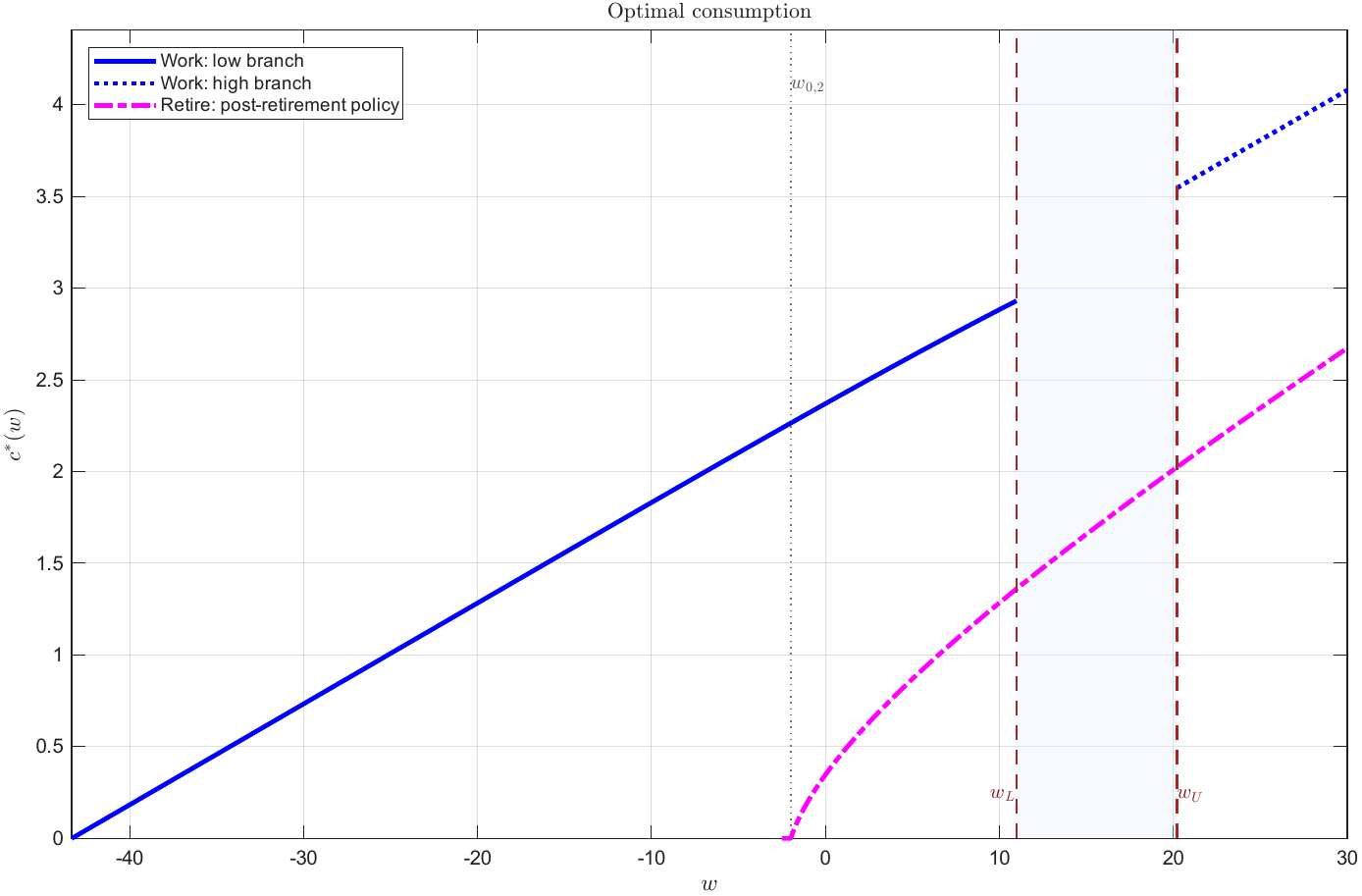}
\label{fig:kappa_lt_one_consumption}
}
\quad
\subfigure[Risky investment]{
\includegraphics[width=.45\columnwidth]{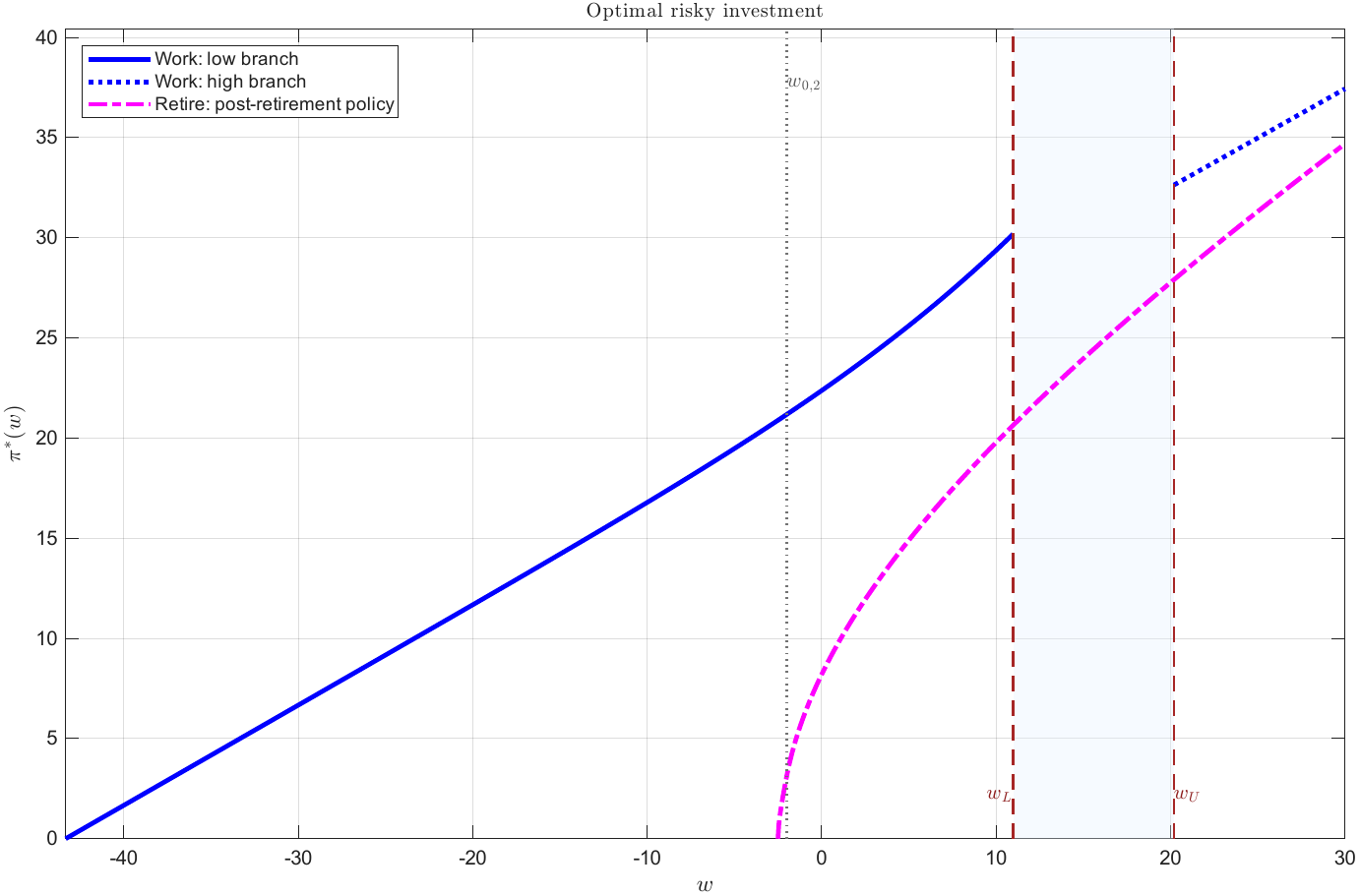}
\label{fig:kappa_lt_one_portfolio}
}
\caption{Optimal consumption and risky investment policies when $0<\kappa<1$.}
\label{fig:kappa_lt_one2}
\end{figure}

Figures~\ref{fig:kappa_gt_one2} and \ref{fig:kappa_lt_one2} illustrate the optimal consumption and risky investment policies for the regimes where the post-retirement quality of life is enhanced ($\kappa > 1$) and degraded ($0 < \kappa < 1$), respectively. The blue curves depict pre-retirement working policies, whereas the pink curves depict policies for agents who have already retired. Because retirement is irreversible, a retired agent remains out of the labor force even if post-retirement wealth subsequently exits the entry region. Private consumption increases with wealth within each regime; below $w_{0,2}$, it equals zero after retirement because effective consumption is supplied entirely by the support component $L$. When the post-retirement quality of life improves ($\kappa > 1$), the agent remains in the labor market as long as her wealth is below $w_{L}$ and transitions to voluntary retirement immediately upon reaching this threshold. Risky investment drops discretely at entry, as it does at every interior entry boundary by the portfolio comparison following Theorem~\ref{thm:main}. Private consumption also drops discretely at entry, because at the boundary $c_1^*=(\phi^*)^{-1/\gamma}$ strictly exceeds $c_2^*=\max\{A(\phi^*)^{-1/\gamma}-L,\,0\}$ whenever $A \le 1$ and $L > 0$, which holds here since $\kappa > 1$ and $\gamma > 1$ imply $A = \kappa^{(1-\gamma)/\gamma} < 1$.

In Figure~\ref{fig:kappa_lt_one2}, risky investment increases with wealth throughout the low-wealth working region (consistent with Corollary~\ref{cor:low_policy}). A worker whose wealth rises to $w_L$ discretely reduces risky holdings from $30.19$ to $20.64$ upon entering retirement. Likewise, a high-wealth worker whose wealth falls to $w_U$ from above reduces risky holdings from $32.64$ to $27.91$ upon retirement. Thus, the apparent upward jump at $w_U$ when reading the graph from left to right simply reflects the reverse of the actual direction of retirement entry. Note that these figures represent dollar holdings rather than portfolio shares.


\section{Conclusion} \label{sec:conclusion}

This paper provides a comprehensive characterization of a continuous-time optimal consumption, investment, and retirement timing problem under joint credit constraints and post-retirement preference shifts. We introduce post-retirement preferences featuring a consumption-equivalent support component. Crucially, this support component directly augments effective consumption without entering the wealth dynamics as cash income. This formulation allows us to cleanly disentangle the welfare-enhancing utility effect of institutional support programs from the standard income effects generated by retirement pension benefits.

We derive closed-form characterizations for the post-retirement value function and the associated optimal policies. When the income support level is sufficiently large relative to effective post-retirement borrowing capacity, a zero-consumption region emerges near the borrowing boundary. Within this region, private spending is zero, whereas effective consumption funded by income support remains strictly positive and yields finite flow utility; at the critical support threshold, this region collapses to the single borrowing-limit point. The complementary case of a smaller support level, in which no zero-consumption region arises, is left for future research.

The post-retirement solution is verified directly, and the original life-cycle problem reduces to an optimal stopping problem whose termination payoff is the post-retirement value function. We transform the free-boundary conditions into scalar equations for the marginal value of wealth at retirement and show that their residuals represent weighted tail integrals of the instantaneous gain from postponing retirement. This representation yields a complete characterization of the retirement region for every level of post-retirement quality of life without requiring additional verification conditions. When post-retirement quality of life improves or remains unchanged, the retirement region is an upper half-line of wealth above a critical threshold, which may lie in the positive-consumption region, in the zero-consumption region, or at the borrowing limit (with the sole exception that retirement is never optimal when quality of life is unchanged and pre-retirement labor income is at least as large as the sum of pension income and income support). When post-retirement quality of life degrades, retirement is optimal only on a bounded wealth interval or never, and the economically relevant roots of the possibly non-monotone boundary equations are uniquely selected by the signs of the tail integrals.

The retirement decision exhibits two economically distinct entry directions. A worker can retire either after accumulating wealth up to a lower threshold or, when retirement is optimal only on a bounded interval, after wealth falls to an upper threshold. In the latter regime, stronger income support in our baseline example expands the retirement interval at both ends, whereas greater pension pledgeability shifts both boundaries downward. Moreover, ranking random retirement times across parameter regimes requires tracking the optimally controlled wealth trajectories from a common initial wealth level.

Risky investment rises with wealth along the low-wealth working branch but drops discretely upon retirement at either interior boundary. A conventional single-target-wealth rule would miss both the continued labor force participation of households above the upper threshold and their discrete reduction in risky asset holdings when wealth declines to that threshold. A natural next question for future research is to quantify the welfare cost of following such a naive single-threshold rule for households starting on either side of the retirement interval.

\newpage

\clearpage
\newpage

\appendix

\begin{center}
\Large\textbf{Appendix}
\end{center}

{\small

\section{Proof of Theorem~\ref{theorem wealth threshold}} \label{app:proof_post_threshold}

Recall from Section~\ref{sec:post} that $\psi(w):=V_2'(w)$ denotes the marginal value of post-retirement wealth. Because $V_2(w)$ is strictly concave, $\psi(w)$ is strictly decreasing in $w$. From the first-order condition \eqref{FOC post-retirement consumption}, the post-retirement consumption regimes are characterized in terms of $\psi(w)$ and $\psi_L:=\kappa^{1-\gamma}L^{-\gamma}$ by
\begin{equation*}
	c_2^*(w)=0 \quad\Longleftrightarrow\quad \psi(w)\geq \psi_L,
	\qquad\text{and}\qquad
	c_2^*(w)>0 \quad\Longleftrightarrow\quad 0<\psi(w)<\psi_L.
\end{equation*}
Since $\psi(w)$ is strictly decreasing, optimal post-retirement consumption $c_2^*(w)$ increases strictly with wealth once $w$ exceeds the zero-consumption threshold.

To solve the post-retirement HJB equation \eqref{eq:post_HJB_after_pi}, we differentiate it with respect to $w$, yielding
\begin{equation} \label{eq:post_derivative_HJB}
	0=(r-\rho-\theta^2)\psi(w) + (rw+Y_2)\psi'(w) + \frac{1}{2}\theta^2\frac{\psi(w)^2\psi''(w)}{\big(\psi'(w)\big)^2} + \mathcal{M}'\big(\psi(w)\big)\psi'(w).
\end{equation}
Let $W(\psi)$ denote the inverse function of $\psi(w)$, satisfying
\begin{equation}\label{dual function W}
	W(\psi(w))=w,
\end{equation}
which implies the inverse-function identities
\begin{equation} \label{eq:inverse_first}
	W'(\psi(w))\psi'(w)=1
	\qquad\text{and}\qquad
	W''(\psi(w))(\psi'(w))^2+W'(\psi(w))\psi''(w)=0.
\end{equation}
Substituting \eqref{dual function W} and \eqref{eq:inverse_first} into \eqref{eq:post_derivative_HJB} linearizes the problem into the following dual ODE:
\begin{equation}\label{eq:post_dual_ODE}
	\frac{1}{2}\theta^2\psi^2W''(\psi) + (\rho-r+\theta^2)\psi W'(\psi) -rW(\psi) -Y_2 -\mathcal{M}'(\psi)=0.
\end{equation}

We analyze this ODE separately across the two consumption regions:
\begin{itemize}
\item \textbf{Zero-consumption region ($\psi\geq\psi_L$):} Let $W_0(\psi)$ denote the inverse marginal value function in this region. Since $\mathcal{M}'(\psi)=0$ by \eqref{eq:M_solution}, $W_0(\psi)$ satisfies
\begin{equation} \label{eq:post_W0_ODE}
	\frac{1}{2}\theta^2\psi^2W_0''(\psi) + (\rho-r+\theta^2)\psi W_0'(\psi) -rW_0(\psi) -Y_2=0,
\end{equation}
whose general solution is
\begin{equation*}
	W_0(\psi)=C_1\psi^{m_+}+C_2\psi^{m_-}-\frac{Y_2}{r}.
\end{equation*}

\item \textbf{Positive-consumption region ($0<\psi<\psi_L$):} Let $W_+(\psi)$ denote the inverse marginal value function in this region. Differentiating \eqref{eq:M_solution} yields $\mathcal{M}'(\psi)=-A\psi^{-1/\gamma}+L$, so $W_+(\psi)$ satisfies
\begin{equation} \label{eq:post_Wp_ODE}
	\frac{1}{2}\theta^2\psi^2W_+''(\psi)+(\rho-r+\theta^2)\psi W_+'(\psi)-rW_+(\psi)+A\psi^{-1/\gamma}-(L+Y_2)=0.
\end{equation}
Imposing the standard transversality growth condition as $w\to\infty$ (equivalently, $\psi\downarrow 0$) eliminates the $m_- < 0$ homogeneous term, yielding the general solution
\begin{equation}\label{Wealth_post}
	W_+(\psi)=D_1\psi^{m_+}+\frac{A}{K}\psi^{-1/\gamma}-\frac{L+Y_2}{r},
\end{equation}
\end{itemize}
where $C_1$, $C_2$, and $D_1$ are constants to be determined from the boundary and smooth-pasting conditions. More precisely, the general solution of \eqref{eq:post_Wp_ODE} is $D_1\psi^{m_+}+D_2\psi^{m_-}+\frac{A}{K}\psi^{-1/\gamma}-\frac{L+Y_2}{r}$, since $CE(-1/\gamma)=-K$. As $\psi\downarrow0$, i.e., as $w\to\infty$, the term $D_2\psi^{m_-}$ dominates the Merton term $\frac AK\psi^{-1/\gamma}$ because $m_-<-1/\gamma$, so that the solution would not converge to the Merton solution for large wealth unless $D_2=0$. The choice $D_2=0$ is justified a posteriori by the verification in Proposition~\ref{prop:post_verification}.

Under condition \eqref{eq:post_case1_condition_intro} ($L\geq\Delta_L$), a non-empty zero-consumption region emerges near the borrowing limit $\underline{w}_2$. Let $w_{0,2}\ge\underline{w}_2$ denote the wealth threshold satisfying
\begin{equation}\label{borrowing limit psi}
	\psi(w_{0,2})=\psi_L.
\end{equation}
Since $\psi(w)$ is strictly decreasing in $w$, the agent optimally chooses zero consumption on $\underline{w}_2\leq w\leq w_{0,2}$ and maintains strictly positive consumption for $w>w_{0,2}$. Let
\begin{equation*}
	\psi_\nu:=\psi\left(-\nu\frac{Y_2}{r}\right)=\psi\left(\underline{w}_2\right)
\end{equation*}
denote the marginal value of wealth at the post-retirement borrowing limit $\underline{w}_2$. By \eqref{dual function W}, the borrowing constraint \eqref{eq:BC1} implies $W_0(\psi_\nu)=-\nu\frac{Y_2}{r}$. Moreover, since $\underline{w}_2$ lies in the zero-consumption region and represents a reflecting lower boundary where risky exposure vanishes, we have $W_0'(\psi_\nu)=0$. Indeed, the wealth process stays above $\underline{w}_2$ only if the risky investment vanishes there, i.e., $\pi_2^*(\underline{w}_2)=-\frac\theta\sigma\psi_\nu W_0'(\psi_\nu)=0$ by \eqref{FOC_pi} and \eqref{eq:inverse_first}; equivalently, $V_2''(w)\to-\infty$ as $w\downarrow\underline{w}_2$. At the borrowing limit, the agent consumes nothing and the wealth drift is $r\underline{w}_2+Y_2=(1-\nu)Y_2>0$, so that wealth is pushed back above $\underline{w}_2$. Solving these two boundary conditions yields
\begin{equation} \label{eq:C1_post}
	C_1=-\frac{m_-(1-\nu)Y_2}{(m_+-m_-)r}\psi_\nu^{-m_+}>0
\end{equation}
and
\begin{equation} \label{eq:C2_post_boundary}
	C_2=\frac{m_+(1-\nu)Y_2}{(m_+-m_-)r}\psi_\nu^{-m_-}>0.
\end{equation}

Using \eqref{eq:post_HJB_after_pi}, we recover the post-retirement value function in the zero-consumption region ($\underline{w}_2\leq w\leq w_{0,2}$) as
\begin{align}
	V_2(w)&=\frac{1}{\rho} \left[ (rw+Y_2)\psi(w) - \frac{1}{2}\theta^2 \frac{\psi(w)^2}{\psi'(w)} + \mathcal{M}\big(\psi(w)\big) \right] \nonumber \\
&=\frac{1}{\rho}	\left[(rw+Y_2)\psi	-\frac{1}{2}\theta^2\psi^2W_0'(\psi)+\frac{(\kappa L)^{1-\gamma}}{1-\gamma}\right] \nonumber   \\
&=\frac{1}{\rho}	\left[-\frac{1}{2}\theta^2\psi^2W_0'(\psi)+r\psi W_0(\psi)	+Y_2\psi
		+\frac{(\kappa L)^{1-\gamma}}{1-\gamma}	\right] \nonumber  \\
&=\frac{r-\frac{1}{2}\theta^2m_+}{\rho}C_1\psi^{m_++1}+	\frac{r-\frac{1}{2}\theta^2m_-}{\rho}C_2\psi^{m_-+1}+\frac{(\kappa L)^{1-\gamma}}{\rho(1-\gamma)} \nonumber \\
&=\frac{m_+}{m_++1}C_1\psi^{m_++1}	+\frac{m_-}{m_-+1}C_2\psi^{m_-+1}+	\frac{(\kappa L)^{1-\gamma}}{\rho(1-\gamma)}. 	\label{eq:post_value_zero_region_derivation}
\end{align}
Similarly, in the positive-consumption region ($w>w_{0,2}$), we recover $V_2(w)$ as
\begin{align}
	V_2(w)
	 & =	\frac{1}{\rho}	\left[(rw+Y_2)\psi-\frac{1}{2}\theta^2\psi^2W_+'(\psi)+		\frac{\gamma}{1-\gamma}A\psi^{-\frac{1-\gamma}{\gamma}}+L\psi	\right] \nonumber \\
	 & =	\frac{1}{\rho}	\left[-\frac{1}{2}\theta^2\psi^2W_+'(\psi)	+r\psi W_+(\psi)+\frac{\gamma}{1-\gamma}A\psi^{-\frac{1-\gamma}{\gamma}}+(L+Y_2)\psi	\right] \nonumber \\
	 & = \frac{r-\frac{1}{2}\theta^2 m_+}{\rho}D_1\psi^{m_++1}+\dfrac{\gamma A}{\rho(1-\gamma)K}\left[\frac{1-\gamma}{2\gamma^2}\theta^2+\frac{r(1-\gamma)}{\gamma}+K	\right]\psi^{-\frac{1-\gamma}{\gamma}}\nonumber \\
	 & =	\frac{m_+}{m_++1}D_1\psi^{m_++1}	+	\frac{A}{(1-\gamma)K}	\psi^{-\frac{1-\gamma}{\gamma}},	\label{eq:post_value_positive_region_derivation}
\end{align}
where the final equality follows from the Merton constant definition
\begin{equation} \label{eq:Merton_K}
K:=r + \frac{\rho - r}{\gamma} + \frac{\gamma - 1}{2 \gamma^{2}} \theta^{2} > 0
\end{equation}
together with the characteristic root identities
\begin{equation} \label{eq:root_identity_r_rho}
	r=-\frac{m_+m_-\theta^2}{2}, \qquad \rho=-\frac{(m_++1)(m_-+1)\theta^2}{2}.
\end{equation}

Continuity of the value function across the threshold $w_{0,2}$ requires $V_2(w_{0,2}-)=V_2(w_{0,2}+)$. Equating \eqref{eq:post_value_zero_region_derivation} and \eqref{eq:post_value_positive_region_derivation} at $\psi=\psi_L$ and using $A\psi_L^{-\frac{1-\gamma}{\gamma}}=(\kappa L)^{1-\gamma}$ yields
\begin{align}
	0
	 & =	\frac{m_+}{m_++1}(C_1-D_1)\psi_L^{m_++1}+\frac{m_-}{m_-+1}C_2\psi_L^{m_-+1} -\frac{A}{(1-\gamma)K}\psi_L^{-\frac{1-\gamma}{\gamma}}+\frac{(\kappa L)^{1-\gamma}}{\rho(1-\gamma)} \nonumber \\
	 & =	\frac{m_+}{m_++1}(C_1-D_1)\psi_L^{m_++1}+\frac{m_-}{m_-+1}C_2\psi_L^{m_-+1}+\frac{1}{1-\gamma}\left(\frac{1}{\rho}-\frac{1}{K}\right)(\kappa L)^{1-\gamma}. \label{eq:continuity_case1}
\end{align}
Furthermore, smooth pasting of the dual function $W$ at $\psi_L$ imposes
\begin{equation} \label{eq:W_C1}
	W_0'(\psi_L)=W_+'(\psi_L).
\end{equation}
Multiplying both sides of \eqref{eq:W_C1} by $\psi_L^2$ gives
\begin{equation} \label{eq:smooth_pasting_case1}
	0 = m_+(C_1-D_1)\psi_L^{m_++1} + m_-C_2\psi_L^{m_-+1} + \frac{1}{\gamma K}(\kappa L)^{1-\gamma}.
\end{equation}
Notice that \eqref{eq:continuity_case1} and \eqref{eq:W_C1}, together with \eqref{eq:post_HJB_after_pi} written as $\rho V_2=(rW+Y_2)\psi-\frac12\theta^2\psi^2W'(\psi)+\mathcal{M}(\psi)$ and the continuity of $\mathcal{M}$, imply $W_0(\psi_L)=W_+(\psi_L)$. Hence, the dual function $W$ is continuously differentiable at $\psi_L$.

Solving the linear system \eqref{eq:continuity_case1} and \eqref{eq:smooth_pasting_case1} with \eqref{eq:root_identity_r_rho} and the identity
\begin{equation} \label{eq:root_identity_K}
	K=-\frac{(\gamma m_++1)(\gamma m_-+1)\theta^2}{2\gamma^2}
\end{equation}
yields
\begin{equation} \label{eq:D1_case1_solved}
	D_1=C_1 - \frac{(\gamma m_-+1)L}{\gamma^2m_+(m_+-m_-)K} \psi_L^{-m_+}>0
\end{equation}
and
\begin{equation} \label{eq:C2_case1_solved}
	C_2=- \frac{(\gamma m_++1)L}{\gamma^2m_-(m_+-m_-)K} \psi_L^{-m_-}.
\end{equation}
Equating \eqref{eq:C2_case1_solved} with \eqref{eq:C2_post_boundary} and solving for $\psi_\nu$ gives
\begin{equation} \label{eq:psi_nu_case1_solved}
	\psi_\nu=\psi_L\left( \frac{-(\gamma m_-+1)(1-\nu)Y_2}{L} \right)^{1/m_-}=\psi_L\left( \frac{\Delta_L}{L} \right)^{1/m_-}.
\end{equation}
Since $m_- < 0$, condition \eqref{eq:post_case1_condition_intro} ($L \geq \Delta_L$) ensures that $\psi_\nu \geq \psi_L$, confirming that $w_{0,2} = W_0(\psi_L) \ge W_0(\psi_\nu) = \underline{w}_2$ and that the zero-consumption region $[\underline{w}_2, w_{0,2}]$ is non-empty. Finally, evaluating $W_0(\psi_L)$ yields the explicit expression for $w_{0,2}$ in Theorem~\ref{theorem wealth threshold}. This completes the proof.~~~\textbf{Q.E.D.}

\section{Monotonicity and concavity of the post-retirement value function $V_2(w)$} \label{app:V2_concavity}

\begin{lem}\label{lem:V2concave}
The post-retirement value function $V_2(w)$ given by \eqref{eq:post_value_zero_region_derivation}--\eqref{eq:post_value_positive_region_derivation} is strictly increasing and strictly concave on $(\underline{w}_2,\infty)$.
\end{lem}

\noindent\textbf{Proof.} Since $CE(0)=-r<0$, $CE(-1)=-\rho<0$, and $CE(\cdot)$ is a strictly convex quadratic polynomial, its real roots satisfy $m_-<-1<0<m_+$. The positivity of $C_1$ and $C_2$ follows immediately from \eqref{eq:C1_post}--\eqref{eq:C2_post_boundary} because $m_-<0<m_+$ and $\nu<1$. Moreover, since $\gamma m_-+1<0$, $m_+>0$, and $K>0$, equation \eqref{eq:D1_case1_solved} implies $D_1>C_1>0$.

Next, differentiating $\psi W_0'(\psi)$ yields
\[
  \big(\psi {W}_0'(\psi)\big)'=\big(m_+C_1\psi^{m_+}+m_-C_2\psi^{m_-}\big)'=m_+^2C_1\psi^{m_+-1}+m_-^2C_2\psi^{m_--1}>0.
\]
Combined with the boundary condition $\psi_\nu {W}_0'(\psi_\nu)=0$, it follows that $\psi {W}_0'(\psi)<0$, and hence ${W}_0'(\psi)<0$, on $[\psi_{L},\psi_\nu)$. Similarly,
\[
  \big(\psi {W}_+'(\psi)\big)'=\Big(m_+D_1\psi^{m_+}-\frac{A}{\gamma K}\psi^{-1/\gamma}\Big)'=m_+^2D_1\psi^{m_+-1}+\frac{A}{\gamma^2K}\psi^{-1/\gamma-1}>0.
\]
Using the smooth-pasting condition \eqref{eq:W_C1}, we have $\psi_{L}{W}_+'(\psi_{L})=\psi_{L}{W}_0'(\psi_{L})\le0$, which implies ${W}_+'(\psi)<0$ on $(0,\psi_{L})$.

Because ${W}_0(\psi_{L})=w_{0,2}={W}_+(\psi_{L})$ and ${W}_0'(\psi_{L})={W}_+'(\psi_{L})$, the piecewise function ${W}(\psi)$ defined by ${W}_+$ on $(0,\psi_{L})$ and by ${W}_0$ on $[\psi_{L},\psi_\nu]$ is continuously differentiable and strictly decreasing on $(0,\psi_\nu]$, with limits ${W}(0+)=\infty$ and ${W}(\psi_\nu)=\underline{w}_2$. Consequently, its inverse $\psi={W}^{-1}$ maps $[\underline{w}_2,\infty)$ bijectively onto $(0,\psi_\nu]$. Differentiating \eqref{eq:post_value_zero_region_derivation}--\eqref{eq:post_value_positive_region_derivation} with respect to $\psi$ gives
\begin{align*}
  \frac{\mathrm d}{\mathrm d\psi}\Big(\frac{m_+}{m_++1}C_1\psi^{m_++1}+\frac{m_-}{m_-+1}C_2\psi^{m_-+1}\Big) &= \psi {W}_0'(\psi), \\
  \frac{\mathrm d}{\mathrm d\psi}\Big(\frac{m_+}{m_++1}D_1\psi^{m_++1}+\frac{A}{(1-\gamma)K}\psi^{-\frac{1-\gamma}{\gamma}}\Big) &= \psi {W}_+'(\psi).
\end{align*}
Applying the chain rule along $w={W}(\psi)$ therefore yields
\[
  V_2'(w)=\frac{\psi {W}'(\psi)}{{W}'(\psi)}=\psi>0
  \qquad\text{and}\qquad
  V_2''(w)=\frac{1}{{W}'(\psi)}<0
\]
for all $w\in(\underline{w}_2,\infty)$.~~~\textbf{Q.E.D.}

\section{Proof of Corollary~\ref{optimal post-retirement policies}} \label{app:proof_post_policies}

By Lemma~\ref{lem:V2concave}, ${W}_0'(\psi)<0$ on $[\psi_{L},\psi_\nu)$ and ${W}_+'(\psi)<0$ on $(0,\psi_{L})$, ensuring that $V_2(w)$ is strictly increasing and strictly concave on $(\underline{w}_2,\infty)$ and that $\psi(w)=V_2'(w)$ is the inverse of the dual function ${W}$ in \eqref{dual function W}. Consequently, the supremum in the HJB equation \eqref{eq:Bellman1} is uniquely attained by the first-order conditions \eqref{FOC_pi} and \eqref{FOC post-retirement consumption}. 

Specifically, optimal post-retirement consumption is given by $c_2^*(w)=(A\psi(w)^{-1/\gamma}-L)^+$, which vanishes if and only if $\psi(w)\geq\psi_{L}$ (corresponding to the region $\psi_L\le\psi(w)\le\psi_\nu$). Moreover, applying the inverse-function identity $\psi'(w)=1/{W}'(\psi)$ to \eqref{FOC_pi} yields the optimal risky investment policy
\[
\pi_2^*(w)=-\frac{\theta}{\sigma}\frac{V_2'(w)}{V_2''(w)}=-\frac{\theta}{\sigma}\psi(w){W}'(\psi(w)).
\]
Substituting $\psi{W}_0'(\psi)=m_+C_1\psi^{m_+}+m_-C_2\psi^{m_-}$ on $[\psi_L,\psi_\nu]$ and $\psi{W}_+'(\psi)=m_+D_1\psi^{m_+}-\frac{A}{\gamma K}\psi^{-1/\gamma}$ on $(0,\psi_L)$ produces the closed-form expressions stated in Corollary~\ref{optimal post-retirement policies}. Finally, the positivity of $C_1$, $C_2$, and $D_1$ follows directly from Lemma~\ref{lem:V2concave}.~~~\textbf{Q.E.D.}

\section{Auxiliary lemmas for the post-retirement verification} \label{app:post_aux}

The dual objects $\mathcal{W}_2$, $\Vt_2$ and the operator $\Lt$ are those of Section~\ref{subsec:dual}. The following lemma, moved here from the proof of Theorem~\ref{thm:main}, collects their properties.

\begin{lem}[Properties of the dual post-retirement value]\label{lem:Vt2}
The dual function $\Vt_2$ in \eqref{eq:Vt2} satisfies:
\begin{enumerate}
\item[(a)] $\Vt_2\in C^2(0,\infty)$, $\Vt_2'=-\mathcal{W}_2$, and $\Vt_2$ is convex on $(0,\infty)$.
\item[(b)] $\Vt_2(\phi)=\sup_{w\ge \underline{w}_2}[V_2(w)-\phi w]$ for every $\phi>0$, with the supremum attained at $w=\mathcal{W}_2(\phi)$.
\item[(c)] $\Lt\Vt_2(\phi)=-N_2(\phi)$ on $(0,\infty)\setminus\{\psi_\nu\}$, where
\begin{equation*}
  N_2(\phi):= \begin{cases}
              Y_2\phi+\mathcal{M}(\phi), & \quad 0<\phi\le\psi_\nu, \\
              \nu Y_2\phi+\rho V_2(\underline{w}_2), & \quad \phi>\psi_\nu,
              \end{cases}
\end{equation*}
and $\mathcal{M}$ is given in \eqref{eq:M_solution}.
\end{enumerate}
\end{lem}

\begin{proof}
(a) Differentiating each branch of \eqref{eq:Vt2} directly yields $\Vt_2'=-\mathcal{W}_2$ via \eqref{Wealth_post}. From Appendix~\ref{app:V2_concavity}, $W_0(\psi_L)=W_+(\psi_L)$, $W_0'(\psi_L)=W_+'(\psi_L)$, $W_0(\psi_\nu)=\underline{w}_2$, and $W_0'(\psi_\nu)=0$, which implies $\mathcal{W}_2\in C^1(0,\infty)$: its derivative also matches the zero derivative of the constant extension at $\psi_\nu$. Continuity of $\Vt_2$ at $\psi_\nu$ follows from \eqref{eq:post_value_zero_region_derivation} evaluated at $\psi=\psi_\nu$. Convexity of $\Vt_2$ then follows from $\Vt_2''=-\mathcal{W}_2'\ge 0$ and Lemma~\ref{lem:V2concave}.

\noindent (b) By Lemma~\ref{lem:V2concave}, for $0<\phi\le\psi_\nu$, the first-order condition $V_2'(w)=\phi$ admits the unique solution $w=\mathcal{W}_2(\phi)$, yielding the maximum value $V_2(\mathcal{W}_2(\phi))-\phi\mathcal{W}_2(\phi)=\Vt_2(\phi)$ by \eqref{eq:post_value_zero_region_derivation}--\eqref{eq:post_value_positive_region_derivation}. For $\phi>\psi_\nu$, we have $V_2'(w)-\phi<0$ for all $w\ge \underline{w}_2$, so the supremum is attained at the boundary $w=\underline{w}_2$.

\noindent (c) Applying $\Lt$ term by term using $\Lt\phi^n=Q(n)\phi^n$ and $\Lt c=-\rho c$ for constants gives: on $(0,\psi_L)$, $\Lt\Vt_2(\phi)=-(L+Y_2)\phi-\frac{\gamma A}{1-\gamma}\phi^{1-1/\gamma}=-Y_2\phi-\mathcal{M}(\phi)$; on $(\psi_L,\psi_\nu)$, $\Lt\Vt_2(\phi)=-Y_2\phi-\frac{(\kappa L)^{1-\gamma}}{1-\gamma}=-Y_2\phi-\mathcal{M}(\phi)$; and on $(\psi_\nu,\infty)$, $\Lt\Vt_2(\phi)=-\rho V_2(\underline{w}_2)+r\underline{w}_2\phi=-\rho V_2(\underline{w}_2)-\nu Y_2\phi$.
\end{proof}

\begin{lem}[Bounds of $V_2$]\label{lem:V2bounds}
Let $V_2$ be given by \eqref{eq:post_value_zero_region_derivation}--\eqref{eq:post_value_positive_region_derivation}. Then
\begin{enumerate}
\item[(i)] $\rho V_2(\underline{w}_2)=(1-\nu)Y_2\psi_\nu+\dfrac{(\kappa L)^{1-\gamma}}{1-\gamma}$.
\item[(ii)] If $\gamma>1$, then $\dfrac{(\kappa L)^{1-\gamma}}{\rho(1-\gamma)}<V_2(w)<0$ for all $w\ge\underline{w}_2$.
\item[(iii)] If $\gamma<1$, then $0<V_2(w)\le f(w)$ for all $w\ge\underline{w}_2$, where $f(w):=\dfrac{\kappa^{1-\gamma}}{(1-\gamma)K^\gamma}\Big(w+\dfrac{L+Y_2}{r}\Big)^{1-\gamma}$.
\end{enumerate}
\end{lem}

\begin{proof}
(i) Formula \eqref{eq:post_value_zero_region_derivation} is \eqref{eq:post_HJB_after_pi} in the dual variable and extends continuously to $\psi=\psi_\nu$. Evaluating it with $W_0(\psi_\nu)=\underline{w}_2$, $r\underline{w}_2+Y_2=(1-\nu)Y_2$, $W_0'(\psi_\nu)=0$, and $\mathcal{M}(\psi_\nu)=(\kappa L)^{1-\gamma}/(1-\gamma)$ (because $\psi_\nu\ge\psi_L$) gives the claim.

\noindent (ii) $V_2$ is increasing by Lemma~\ref{lem:V2concave}, so the lower bound follows from (i). As $w\to\infty$, $\psi\downarrow0$ and both terms of \eqref{eq:post_value_positive_region_derivation} tend to $0$, because $m_++1>0$ and $\frac{\gamma-1}{\gamma}>0$. Since $V_2$ is strictly increasing, $V_2<0$.

\noindent (iii) Positivity follows from (i) and monotonicity. For the upper bound, let $\tilde f(\phi):=\frac{\gamma A}{(1-\gamma)K}\phi^{1-1/\gamma}+\frac{L+Y_2}{r}\phi$ be the convex conjugate of $f$, and let $q:=\Vt_2-\tilde f\in C^2(0,\infty)$. By \eqref{eq:Q} and Lemma~\ref{lem:Vt2}(c), $\Lt q=\mathcal{M}_f-\mathcal{M}\ge0$ on $(0,\psi_\nu)$, where $\mathcal{M}_f(\phi):=\frac{\gamma A}{1-\gamma}\phi^{1-1/\gamma}+L\phi=\sup_{c>-L}[U_2(c)-c\phi]\ge\mathcal{M}(\phi)$. On $(\psi_\nu,\infty)$, $\Lt q=(1-\nu)Y_2\phi+\mathcal{M}_f(\phi)-\rho V_2(\underline{w}_2)$; this expression is nonnegative at $\psi_\nu$ by (i) and increasing, since its derivative is $(1-\nu)Y_2+L-A\phi^{-1/\gamma}>0$ for $\phi>\psi_L$. Moreover, $q=-\frac{D_1}{m_++1}\phi^{m_++1}<0$ on $(0,\psi_L)$, $q(0+)=0$, and $\lim_{\phi\to\infty}q(\phi)=-\infty$. A positive interior maximum of $q$ at some $\phi_0$ would give $q'(\phi_0)=0$, $q''(\phi_0)\le0$, and hence $\Lt q(\phi_0)\le-\rho q(\phi_0)<0$, a contradiction. Hence $\Vt_2\le\tilde f$, and $V_2(w)\le\inf_{\phi>0}[\Vt_2(\phi)+\phi w]\le\inf_{\phi>0}[\tilde f(\phi)+\phi w]=f(w)$.
\end{proof}

\begin{lem}[Transversality bound]\label{lem:transversality}
Let $\gamma<1$ and let $(\bc,\bpi)\in\mathcal{A}_2(w)$ with $W_0=w\ge\underline{w}_2$. Then $\E[e^{-\rho T}f(W_T)]\le f(w)e^{-\gamma KT}$, where $f$ is defined in Lemma~\ref{lem:V2bounds}(iii).
\end{lem}

\begin{proof}
Let $\xi_t:=\exp\{-(r+\frac12\theta^2)t-\theta\beta_t\}$ be the state-price density and $\bar W_t:=W_t+(L+Y_2)/r$, which is positive because $W_t\ge\underline{w}_2>-(L+Y_2)/r$. Since $\mathrm{d}\bar W_t=(r\bar W_t-L+\pi_t(\mu-r)-c_t)\mathrm{d}t+\sigma\pi_t\mathrm{d}\beta_t$ and $c_t+L\ge0$, the process $\xi_t\bar W_t+\int_0^t\xi_s(c_s+L)\mathrm{d}s$ is a nonnegative local martingale and thus a supermartingale, so $\E[\xi_T\bar W_T]\le\bar W_0$. Writing $e^{-\rho T}\bar W_T^{1-\gamma}=(\xi_T\bar W_T)^{1-\gamma}e^{-\rho T}\xi_T^{\gamma-1}$ and applying H\"older's inequality with exponents $\frac1{1-\gamma}$ and $\frac1\gamma$, we obtain
\[
\E[e^{-\rho T}\bar W_T^{1-\gamma}]\le\big(\E[\xi_T\bar W_T]\big)^{1-\gamma}\big(\E[e^{-\rho T/\gamma}\xi_T^{(\gamma-1)/\gamma}]\big)^{\gamma}\le\bar W_0^{1-\gamma}e^{-\gamma KT},
\]
where we used $\E[e^{-\rho T/\gamma}\xi_T^{(\gamma-1)/\gamma}]=e^{-KT}$. Since $f(W_T)=\frac{\kappa^{1-\gamma}}{(1-\gamma)K^\gamma}\bar W_T^{1-\gamma}$, the claim follows.
\end{proof}

\section{Proof of Proposition~\ref{prop:post_verification}} \label{app:proof_post_verification}

\paragraph{Step 1 (Upper bound).} Since $\mathcal{W}_2$ is continuously differentiable with $\mathcal{W}_2'<0$ on $(0,\psi_\nu)$ (Lemmas~\ref{lem:V2concave} and \ref{lem:Vt2}(a)), its inverse $V_2'$ is continuously differentiable on $(\underline{w}_2,\infty)$, i.e., $V_2\in C^2((\underline{w}_2,\infty))$, and $V_2$ satisfies the HJB equation \eqref{eq:Bellman1} for every $w>\underline{w}_2$ by construction. Since $V_2''(w)\to-\infty$ as $w\downarrow\underline{w}_2$, we shift the wealth slightly. Fix $w\ge\underline{w}_2$, $(\bc,\bpi)\in\mathcal{A}_2(w)$, and $\varepsilon>0$, and set $Z_t:=W_t+\varepsilon$. Then $Z_t\ge\underline{w}_2+\varepsilon$ and
\[
\mathrm{d}Z_t=[rZ_t-r\varepsilon+Y_2+\pi_t(\mu-r)-c_t]\mathrm{d}t+\sigma\pi_t\mathrm{d}\beta_t.
\]
Applying It\^o's formula to $e^{-\rho t}V_2(Z_t)$ and using \eqref{eq:Bellman1} at $Z_t$ and $V_2'>0$,
\[
e^{-\rho t}V_2(Z_t)-V_2(Z_0)\le-\int_0^te^{-\rho s}U_2(c_s)\mathrm{d}s+\int_0^te^{-\rho s}\sigma\pi_sV_2'(Z_s)\mathrm{d}\beta_s.
\]
Let $\tau_n:=n\wedge\inf\{t\ge0:\int_0^t\pi_s^2\mathrm{d}s\ge n\}$. Since $0<V_2'\le\psi_\nu$, the stopped stochastic integral is a martingale, and hence
\[
V_2(w+\varepsilon)\ge\E^w\Big[\int_0^{\tau_n}e^{-\rho t}U_2(c_t)\mathrm{d}t\Big]+\E^w\big[e^{-\rho\tau_n}V_2(Z_{\tau_n})\big].
\]
If $\gamma<1$, then $U_2\ge0$ and $V_2>0$ by Lemma~\ref{lem:V2bounds}(iii); dropping the nonnegative terminal term and letting $n\to\infty$ by monotone convergence gives $V_2(w+\varepsilon)\ge\E^w[\int_0^\infty e^{-\rho t}U_2(c_t)\mathrm{d}t]$. If $\gamma>1$, then $U_2$ and $V_2$ are bounded and negative by Lemma~\ref{lem:V2bounds}(ii), and dominated convergence gives the same inequality. Letting $\varepsilon\downarrow0$ and using the continuity of $V_2$, we obtain $V_2(w)\ge\E^w[\int_0^\infty e^{-\rho t}U_2(c_t)\mathrm{d}t]$ for every admissible $(\bc,\bpi)$.

\paragraph{Step 2 (Optimal wealth).} Fix $w\ge\underline{w}_2$ and let $\Phi_t:=\psi_0\exp\{(\rho-r-\frac12\theta^2)t-\theta\beta_t\}$ with $\psi_0:=V_2'(w)\in(0,\psi_\nu]$. Since the borrowing constraint $W_t\ge\underline{w}_2$ corresponds to the upper bound $\psi\le\psi_\nu$ of the marginal value, we reflect $\Phi$ downward at $\psi_\nu$:
\[
\ell_t:=\sup_{0\le s\le t}(\log\Phi_s-\log\psi_\nu)^+,\qquad \psi_t:=\Phi_te^{-\ell_t}.
\]
Then $\ell$ is continuous and nondecreasing with $\ell_0=0$, it increases only on $\{t:\psi_t=\psi_\nu\}$, $\psi_t\le\psi_\nu$, and $\mathrm{d}\psi_t=\psi_t[(\rho-r)\mathrm{d}t-\theta\mathrm{d}\beta_t-\mathrm{d}\ell_t]$. Set $W_t^*:=\mathcal{W}_2(\psi_t)$, $c_{2,t}^*:=(A\psi_t^{-1/\gamma}-L)^+$, and $\pi_{2,t}^*:=-\frac\theta\sigma\psi_t\mathcal{W}_2'(\psi_t)$. Since $\mathcal{W}_2\in C^1$ and $\mathcal{W}_2''$ is bounded on compact subsets of $(0,\psi_\nu]$ with a jump only at $\psi_L$, It\^o's formula gives
\[
\mathrm{d}W_t^*=\big[(\rho-r)\psi_t\mathcal{W}_2'(\psi_t)+\tfrac12\theta^2\psi_t^2\mathcal{W}_2''(\psi_t)\big]\mathrm{d}t-\theta\psi_t\mathcal{W}_2'(\psi_t)\mathrm{d}\beta_t-\psi_t\mathcal{W}_2'(\psi_t)\mathrm{d}\ell_t.
\]
The reflection term vanishes because $\ell$ increases only when $\psi_t=\psi_\nu$ and $\mathcal{W}_2'(\psi_\nu)=0$. By the dual ODE \eqref{eq:post_dual_ODE} and $\mathcal{M}'(\psi)=-c_2^*$, $(\rho-r)\psi\mathcal{W}_2'+\frac12\theta^2\psi^2\mathcal{W}_2''=r\mathcal{W}_2+Y_2-c_2^*-\theta^2\psi\mathcal{W}_2'$, and $-\theta^2\psi\mathcal{W}_2'=\pi_2^*(\mu-r)$, $-\theta\psi\mathcal{W}_2'=\sigma\pi_2^*$. Hence
\[
\mathrm{d}W_t^*=\big[rW_t^*+Y_2+\pi_{2,t}^*(\mu-r)-c_{2,t}^*\big]\mathrm{d}t+\sigma\pi_{2,t}^*\mathrm{d}\beta_t,\qquad W_0^*=\mathcal{W}_2(\psi_0)=w,
\]
so that $W^*$ is the wealth process generated by $(c_2^*,\pi_2^*)$, and $W_t^*\ge\underline{w}_2$ because $\psi_t\le\psi_\nu$. Since $\mathcal{W}_2$ is the inverse of $V_2'$ on $(0,\psi_\nu]$, $\psi_t=V_2'(W_t^*)$, so that $(c_2^*,\pi_2^*)$ coincide with the feedback controls of Corollary~\ref{optimal post-retirement policies} evaluated at $W_t^*$. In particular, $(c_2^*,\pi_2^*)$ is admissible.

\paragraph{Step 3 (Attainment).} Write $e^{-\rho t}V_2(W_t^*)=e^{-\rho t}\Vh_2(\psi_t)$ with $\Vh_2(\psi):=V_2(\mathcal{W}_2(\psi))=\Vt_2(\psi)+\psi\mathcal{W}_2(\psi)$. By the chain rule, $\Vh_2'(\psi)=\psi\mathcal{W}_2'(\psi)$, which is continuous on $(0,\psi_\nu]$ and absolutely continuous; hence It\^o's formula for $C^1$ functions with absolutely continuous derivative applies, and the jump of $\mathcal{W}_2''$ at $\psi_L$ is irrelevant because $\{t:\psi_t=\psi_L\}$ has Lebesgue measure zero. The reflection term is $-\Vh_2'(\psi_\nu)\psi_\nu\mathrm{d}\ell_t=-\psi_\nu^2\mathcal{W}_2'(\psi_\nu)\mathrm{d}\ell_t=0$. Moreover, on $(0,\psi_\nu)$, a direct computation using \eqref{eq:post_dual_ODE} gives $\Lt(\psi\mathcal{W}_2(\psi))=\psi(Y_2+\mathcal{M}'(\psi))$, so that by Lemma~\ref{lem:Vt2}(c),
\[
\begin{aligned}
\Lt\Vh_2(\psi)&=-Y_2\psi-\mathcal{M}(\psi)+\psi\big(Y_2+\mathcal{M}'(\psi)\big)\\
&=\psi\mathcal{M}'(\psi)-\mathcal{M}(\psi)=-U_2\big(c_2^*(\psi)\big),
\end{aligned}
\]
where the last equality follows from $\mathcal{M}(\psi)=U_2(c_2^*(\psi))-c_2^*(\psi)\psi$ and $\mathcal{M}'(\psi)=-c_2^*(\psi)$. Since $\{t:\psi_t=\psi_\nu\}$ has Lebesgue measure zero, It\^o's formula yields
\[
e^{-\rho T}V_2(W_T^*)+\int_0^Te^{-\rho t}U_2(c_{2,t}^*)\mathrm{d}t=V_2(w)+\int_0^Te^{-\rho t}\sigma\psi_t\pi_{2,t}^*\mathrm{d}\beta_t.
\]
The stochastic integral is a true martingale, because $0<\psi\le\psi_\nu$, $\pi_2^*$ is a finite sum of real powers of $\psi$ by Corollary~\ref{optimal post-retirement policies}, and $\E[\sup_{t\in[0,T]}\psi_t^{-p}]<\infty$ for every $p>0$ (since $\psi_t^{-1}\le\Phi_t^{-1}\max(1,\sup_{s\le t}\Phi_s/\psi_\nu)$ and $\Phi$ is a geometric Brownian motion). Letting $T\to\infty$, the terminal term vanishes by boundedness if $\gamma>1$, and by Lemmas~\ref{lem:V2bounds}(iii) and \ref{lem:transversality} if $\gamma<1$; the time integral converges by monotone convergence. Therefore $V_2(w)=\E^w[\int_0^\infty e^{-\rho t}U_2(c_{2,t}^*)\mathrm{d}t]$, which is finite. Combined with Step 1, $V_2$ is the post-retirement value function and $(c_2^*,\pi_2^*)$ is optimal.~~~\textbf{Q.E.D.}

\section{Proof of Proposition~\ref{prop:reduction}} \label{app:proof_reduction}

Write $J$ for the payoff in \eqref{value_function} and $J_1$ for the payoff in \eqref{pre-retirement value function}. Since $U_1$, $U_2$ and $V_2$ all have the sign of $1-\gamma$ (Lemma~\ref{lem:V2bounds}), both payoffs are well defined for every policy, and conditional expectations of the integrals below may be taken termwise.

\paragraph{Step 1 (Conditional upper bound after retirement).} Let $(\bc,\bpi,\tau)\in\mathcal{A}(w)$. On $\{\tau<\infty\}$, $W_\tau\ge\underline{w}_2$ by \eqref{eq:BC1}, and after $\tau$ the wealth follows the post-retirement dynamics. Put $\tau_n:=(\tau+n)\wedge\inf\{t\ge\tau:\int_\tau^t\pi_s^2\mathrm{d}s\ge n\}$, fix $\varepsilon>0$, and set $Z_t:=W_t+\varepsilon$ for $t\ge\tau$. The argument in Step 1 of the proof of Proposition~\ref{prop:post_verification}, applied to $e^{-\rho t}V_2(Z_t)$ on $[\tau,\tau_n]$, gives
\[
e^{-\rho\tau_n}V_2(Z_{\tau_n})+\int_\tau^{\tau_n}e^{-\rho s}U_2(c_s)\mathrm{d}s\le e^{-\rho\tau}V_2(W_\tau+\varepsilon)+\int_\tau^{\tau_n}e^{-\rho s}V_2'(Z_s)\sigma\pi_s\mathrm{d}\beta_s.
\]
The stochastic integral has zero $\mathcal{F}_\tau$-conditional expectation, since $0<V_2'\le\psi_\nu$ and $\int_\tau^{\tau_n}\pi_s^2\mathrm{d}s\le n$. Letting $n\to\infty$ as in Proposition~\ref{prop:post_verification} (conditional monotone convergence for $\gamma<1$ after dropping the nonnegative terminal term; conditional dominated convergence for $\gamma>1$, where $V_2$ and $U_2$ are bounded) and then $\varepsilon\downarrow0$ yields
\begin{equation}\label{eq:cond_upper}
\E\Big[\int_\tau^\infty e^{-\rho s}U_2(c_s)\mathrm{d}s\,\Big|\,\mathcal{F}_\tau\Big]\le e^{-\rho\tau}V_2(W_\tau)\qquad\text{on }\{\tau<\infty\}.
\end{equation}

\paragraph{Step 2 ($V\le V_1$).} By \eqref{eq:BC1}, $W_t\ge\underline{w}_1$ for $t<\tau$ and $W_\tau\ge\underline{w}_2>\underline{w}_1$ on $\{\tau<\infty\}$, so the restriction of $(\bc,\bpi,\tau)$ to $[0,\tau]$ belongs to $\mathcal{A}_1(w)$. Taking expectations in \eqref{eq:cond_upper}, we obtain $J(w;\bc,\bpi,\tau)\le J_1(w;\bc,\bpi,\tau)\le V_1(w)$. Taking the supremum over $\mathcal{A}(w)$ gives $V\le V_1$.

\paragraph{Step 3 (Proof of (a)).} Suppose that $V_1(w)<\infty$, and let $(\bc,\bpi,\tau)\in\mathcal{A}_1(w)$ with $J_1(w;\bc,\bpi,\tau)>-\infty$, so that $J_1(w;\bc,\bpi,\tau)$ is finite. On $\{\tau<\infty\}$, $\beta'_s:=\beta_{\tau+s}-\beta_\tau$ is a Brownian motion independent of $\mathcal{F}_\tau$. Let $\psi$ be the reflected dual process of Step 2 of the proof of Proposition~\ref{prop:post_verification}, driven by $\beta'$ and started at $\psi_0=V_2'(W_\tau)\in(0,\psi_\nu]$; for $s\ge0$ set $W_{\tau+s}:=\mathcal{W}_2(\psi_s)$ and use $(c_2^*,\pi_2^*)$ evaluated at $\psi_s$. The concatenated policy $(\hat\bc,\hat\bpi,\tau)$ is progressively measurable, satisfies \eqref{eq:wealth1} with income $Y_2$ after $\tau$, and satisfies $W_t\ge\underline{w}_2$ for $t\ge\tau$. Step 3 of the proof of Proposition~\ref{prop:post_verification}, together with the independence of $\beta'$ from $\mathcal{F}_\tau$, gives equality in \eqref{eq:cond_upper}. Taking expectations, $J(w;\hat\bc,\hat\bpi,\tau)=J_1(w;\bc,\bpi,\tau)$. Since $U_1$, $U_2$ and $V_2$ have a constant sign, the expected discounted absolute utility of the concatenated policy equals $|J_1(w;\bc,\bpi,\tau)|<\infty$, so the policy belongs to $\mathcal{A}(w)$, and $V(w)\ge J_1(w;\bc,\bpi,\tau)$. Since the never-retiring Merton policy $c_t=K(W_t+Y_1/r)$, $\pi_t=\frac{\theta}{\gamma\sigma}(W_t+Y_1/r)$ belongs to $\mathcal{A}_1(w)$ and has a finite payoff, policies with $J_1=-\infty$ do not affect the supremum in \eqref{pre-retirement value function}. Taking the supremum gives $V(w)\ge V_1(w)$, and together with Step 2, $V(w)=V_1(w)$.

\paragraph{Step 4 (Proof of (b)).} Applying Step 3 to $(\bc^*,\bpi^*,\tau^*)$, whose payoff $J_1(w;\bc^*,\bpi^*,\tau^*)=V_1(w)$ is finite, the concatenated policy belongs to $\mathcal{A}(w)$ and its payoff equals $V_1(w)=V(w)$.~~~\textbf{Q.E.D.}

\section{Proof of Lemma \ref{lem:S_meaning}}

By \eqref{eq:characteristic_eq}, \eqref{eq:Merton_K}, and \eqref{eq:root_identity_r_rho}, we know that
\begin{equation}\label{eq:Q}
  Q(m_\pm+1)=0,\quad Q(0)=-\rho,\quad Q(1)=-r,\quad Q\big(1-\tfrac1\gamma\big)=-K,\quad
  r=-\tfrac12\theta^2m_+m_- .
\end{equation}
Applying $\Lt$ term by term to $\Vt^{\mathrm{NR}}$ and using $Q(1)=-r$ and $Q(1-\frac1\gamma)=-K$ in \eqref{eq:Q}, we obtain $\Lt\Vt^{\mathrm{NR}}=-Y_1\phi-\frac{\gamma}{1-\gamma}\phi^{1-\frac1\gamma}=-Y_1\phi-\Ut$. Apply also $\Lt$ term by term using \eqref{eq:Q} and $\Lt c=-\rho c$ for constants. On $(0,\psi_{L})$, we obtain that $\Lt\Vt_2=-(L+Y_2)\phi-\frac{\gamma A}{1-\gamma}\phi^{1-1/\gamma}=-Y_2\phi-\mathcal{M}(\phi)$; on $[\psi_{L},\psi_\nu]$, we obtain that $-Y_2\phi-\frac{(\kappa L)^{1-\gamma}}{1-\gamma}=-Y_2\phi-\mathcal{M}(\phi)$. Therefore,
$$
\Lt\Vt^{\mathrm{NR}}-\Lt\Vt_2=-\big(\mathcal{H}_{1}V_{2}-\rho V_{2}\big)\big(W_{2}\big).
$$
This is because for $0<\phi<\psi_\nu$ and $w={W}_2(\phi)$, $V_2'(w)=\phi$, $V_2''(w)=1/{W}_2'(\phi)=-1/\Vt_2''(\phi)<0$, $V_2(w)=\Vt_2(\phi)+\phi w$ and $w=-\Vt_2'(\phi)$, and substituting $f = V_{2}$ into \eqref{eq:pre-retirement-HJB-operator-optimized}, we obtain that
\[
  \mathcal H_1V_2(w)-\rho V_2(w)
  =\tfrac12\theta^2\phi^2\Vt_2''+(\rho-r)\phi\Vt_2'-\rho\Vt_2+Y_1\phi+\Ut(\phi)
  =\Lt\Vt_2+Y_1\phi+\Ut,
\]
which is the same as the lemma states.

\section{\texorpdfstring{Proofs of Proposition~\ref{prop:Fint} and Lemma~\ref{lem:S_sign}}{Proofs of Proposition and Lemma}} \label{app:proof_S_sign}

We begin by establishing integral representations for the boundary residuals $F_{m_-}$ and $F_{m_+}$, which will also serve as key building blocks in subsequent proofs. Define the auxiliary functions
\begin{equation}\label{eq:S}
  S(\phi):=\begin{cases}
    \phi\,g(\phi), & 0<\phi<\psi_L,\\[1.2ex]
    (Y_1-Y_2)\phi+\Ut(\phi)-\dfrac{(\kappa L)^{1-\gamma}}{1-\gamma}, & \psi_L\le\phi\le\psi_\nu,\\[1.2ex]
    (Y_1-\nu Y_2)\phi+\Ut(\phi)-\rho V_2(\underline{w}_2), & \phi>\psi_\nu,
  \end{cases}
\end{equation}
where
\begin{equation*}
  g(\phi):=Y_1-Y_2-L-\frac{\gamma(A-1)}{1-\gamma}\phi^{-1/\gamma},\qquad \phi>0,
\end{equation*}
and let
\begin{equation}\label{eq:h}
  h(\phi):=\Vt^{\mathrm{NR}}(\phi)-\Vt_2(\phi).
\end{equation}
Notice from Lemma~\ref{lem:S_meaning} and \eqref{eq:Vt2} that $S(\phi) = \big(\mathcal{H}_1 V_2 - \rho V_2\big)\big(\mathcal{W}_2(\phi)\big)$ for $0<\phi<\psi_\nu$, and $S(\phi)$ extends continuously across $\psi_\nu$ (with an upward kink of size $(1-\nu)Y_2$ in slope). Denote by
\[
  \mathcal{N} := \{\phi > 0 : S(\phi) < 0\}
\]
the negative set of the extended dual residual. On $(0,\psi_\nu)$ it corresponds to a negative instantaneous gain at interior wealth.

\paragraph{Step 1: Integral representations of $F_{m_-}$ and $F_{m_+}$\texorpdfstring{(proof of Proposition~\ref{prop:Fint})}{}.}
We prove the representations \eqref{eq:Fmint}--\eqref{eq:Green} of Proposition~\ref{prop:Fint}, where the improper integrals
\[
  I_-(\phi) := \int_\phi^\infty s^{-m_+-2}S(s)\,\mathrm{d}s \qquad\text{and}\qquad I_+(\phi) := \int_0^\phi s^{-m_--2}S(s)\,\mathrm{d}s
\]
both converge. Indeed, near $s=0$, $|S(s)|\le C(s+s^{1-1/\gamma})$ for some constant $C>0$; since $-m_--1>-1$ and $-m_--1-\frac{1}{\gamma}>-1$ (the latter being equivalent to $K>0$ via \eqref{eq:root_identity_K}), $I_+(\phi)$ converges. Similarly, for sufficiently large $s$, $|S(s)|\le C(1+s)$, so $m_+>0$ ensures the convergence of $I_-(\phi)$.

By the computation in the proof of Lemma~\ref{lem:S_meaning} on $(0,\psi_\nu)$, and by Lemma~\ref{lem:Vt2}(c) on $(\psi_\nu,\infty)$, $\Lt h(\phi) = -S(\phi)$ on $(0,\infty)\setminus\{\psi_\nu\}$; since $h\in C^2(0,\infty)$ by Lemma~\ref{lem:Vt2}(a) and $S$ is continuous, the identity also holds at $\psi_\nu$. Let $m\in\{m_-,m_+\}$ and let $m_{o}$ denote the conjugate root in $\{m_-,m_+\}\setminus\{m\}$. For any $f\in C^2(0,\infty)$, setting $k(\phi):=f(\phi)\phi^{-(m+1)}$ and using $m+m_{o}+2=1-2(\rho-r)/\theta^2$ and $(m+1)(m_{o}+1)=-2\rho/\theta^2$ gives
\[
  \Lt f(\phi)=\tfrac12\theta^2\phi^{m_{o}+2}\big(\phi^{m-m_{o}+1}k'(\phi)\big)',
  \qquad
  f(\phi)-\frac{\phi f'(\phi)}{m+1}=-\frac{\phi^{m+2}k'(\phi)}{m+1}.
\]
Applying these identities to $f=h$ and using $\Lt h=-S$, we see that $G_m(\phi):=\phi^{m-m_{o}+1}k'(\phi)$ satisfies $G_m'(\phi)=-\frac{2}{\theta^2}\phi^{-m_{o}-2}S(\phi)$, while \eqref{eq:Fh} yields $F_m(\phi)=\phi^{m_{o}}G_m(\phi)/(m+1)$.
\begin{itemize}
\item For $m=m_-$ (so $m_o=m_+$), the third branch of \eqref{eq:Fminus} shows that $F_{m_-}(\phi)$ remains bounded as $\phi\to\infty$. Since $m_+>0$, $G_{m_-}(\phi)=(m_-+1)\phi^{-m_+}F_{m_-}(\phi)\to 0$ as $\phi\to\infty$, which integrates to $G_{m_-}(\phi)=\frac{2}{\theta^2}I_-(\phi)$.
\item For $m=m_+$ (so $m_o=m_-$), \eqref{eq:Fplus} implies $F_{m_+}(\phi)=O(\phi^{-1/\gamma}+1)$ as $\phi\downarrow 0$. Since $-m_->\frac{1}{\gamma}$, $G_{m_+}(\phi)=(m_++1)\phi^{-m_-}F_{m_+}(\phi)\to 0$ as $\phi\downarrow 0$, which integrates to $G_{m_+}(\phi)=-\frac{2}{\theta^2}I_+(\phi)$.
\end{itemize}
Substituting $\frac{2}{\theta^2}=-\frac{m_+m_-}{r}$ establishes \eqref{eq:Fmint}--\eqref{eq:Fpint}, and taking the linear combination $(m_-+1)F_{m_-}(\phi)-(m_++1)F_{m_+}(\phi)=(m_+-m_-)h(\phi)/\phi$ via \eqref{eq:Fh} yields \eqref{eq:Green}. Since $m_+m_-<0$, $m_-+1<0$, and $m_++1>0$, it follows immediately that
\begin{equation*}
  F_{m_{-}}(\phi)>0 \iff I_-(\phi)<0 \qquad\text{and}\qquad F_{m_{+}}(\phi)>0 \iff I_+(\phi)<0,
\end{equation*}
with $I_-'(\phi)=-\phi^{-m_+-2}S(\phi)$ and $I_+'(\phi)=\phi^{-m_--2}S(\phi)$.

\paragraph{Step 2: Characterization of $\mathcal{N} = \{S < 0\}$\texorpdfstring{~(proof of Lemma~\ref{lem:S_sign})}{}.}
Observe that the sign of $(A-1)/(1-\gamma)$ coincides with the sign of $\kappa-1$. Hence, on $(0,\psi_L)$, the function $g(\phi)$ is strictly increasing with $g(0+)=-\infty$ when $\kappa>1$, constant at $Y_1-Y_2-L$ when $\kappa=1$, and strictly decreasing with $g(0+)=+\infty$ when $\kappa<1$.

On the upper domain $[\psi_L,\infty)$, $S(\phi)$ is continuous and strictly convex because it consists of piecewise affine terms plus $\Ut(\phi)$, with $\Ut''(\phi)=\frac{1}{\gamma}\phi^{-\frac{1}{\gamma}-1}>0$ and an upward slope jump of $(1-\nu)Y_2>0$ at $\psi_\nu$. Moreover, $0<Y_2<Y_1$ and $\nu<1$ imply
\begin{equation}\label{eq:Y1nu}
  Y_1>\nu Y_2 \qquad\Longleftrightarrow\qquad \underline{w}_1<\underline{w}_2,
\end{equation}
so $\lim_{\phi\to\infty} S(\phi)/\phi = Y_1-\nu Y_2 > 0$ and $\lim_{\phi\to\infty} S(\phi) = +\infty$. Consequently, $\mathcal{N}\cap[\psi_L,\infty)$ is a bounded interval (possibly empty); if it is non-empty, then $S(\phi)>0$ for all $\phi > \sup(\mathcal{N}\cap[\psi_L,\infty))$. Note also that since $\psi_L^{-1/\gamma}=L/A$, the following bound holds for the right-hand derivative at $\psi_L$: $S'(\psi_L+)\ge Y_1-Y_2-L/A$, with equality when $\psi_L<\psi_\nu$ and an additional $(1-\nu)Y_2$ when $\psi_L=\psi_\nu$.

We now examine the three preference regimes:
\begin{enumerate}
\item[(a)] \textbf{Case $\kappa>1$:} Since $\lim_{\phi\downarrow 0} S(\phi)/\phi = -\infty$, we have $S(\phi)<0$ near $\phi=0$. If $g(\psi_L)<0$, monotonicity of $g$ implies $S(\phi)<0$ on $(0,\psi_L]$, and strict convexity on $[\psi_L,\infty)$ yields $\mathcal{N}=(0,u_2)$ for some $u_2>\psi_L$. Conversely, if $g(\psi_L)\ge 0$, then $Y_1-Y_2-L\ge\frac{\gamma(A-1)}{1-\gamma}\frac{L}{A}$, which implies
\[
  S'(\psi_L+)\ge Y_1-Y_2-\frac{L}{A}\ge\frac{L(A-1)}{A}\Big(1+\frac{\gamma}{1-\gamma}\Big)=\frac{L}{A}\,\frac{A-1}{1-\gamma}>0.
\]
By convexity, $S$ is strictly increasing on $[\psi_L,\infty)$ with $S(\phi)\ge S(\psi_L)\ge 0$. Thus, $\mathcal{N}=(0,u_2)$, where $u_2\le\psi_L$ is the unique root of $g(\phi)=0$.

\item[(b)] \textbf{Case $\kappa=1$:} Here $A=1$ and $g(\phi)\equiv Y_1-Y_2-L$. If $Y_1<L+Y_2$, then $S(\phi)<0$ on $(0,\psi_L]$ and convexity on $[\psi_L,\infty)$ yields $\mathcal{N}=(0,u_2)$ with $u_2>\psi_L$. If $Y_1\ge L+Y_2$, then $S(\phi)\ge 0$ on $(0,\psi_L]$ and $S'(\psi_L+)\ge Y_1-Y_2-L\ge 0$, implying $S(\phi)>0$ on $(\psi_L,\infty)$ and $\mathcal{N}=\emptyset$.

\item[(c)] \textbf{Case $0<\kappa<1$:} Since $\lim_{\phi\downarrow 0} S(\phi)/\phi = +\infty$, we have $S(\phi)>0$ near $\phi=0$. If $g(\psi_L)<0$, strict monotonicity of $g$ gives $\mathcal{N}\cap(0,\psi_L)=(u_1,\psi_L)$ where $u_1\in(0,\psi_L)$ is the unique root of $g(\phi)=0$; since $S(\psi_L)<0$, convexity on $[\psi_L,\infty)$ gives $\mathcal{N}=(u_1,u_2)$ with $u_2>\psi_L$. If $g(\psi_L)\ge 0$, then $S(\phi)>0$ on $(0,\psi_L)$ and $S(\psi_L)\ge0$ and $\mathcal{N}\subset(\psi_L,\infty)$ is either an open interval $(u_1,u_2)$ with $u_1\ge\psi_L>0$ or empty.
\end{enumerate}
When $\mathcal N$ is nonempty, these arguments give $S(\phi)>0$ outside $[u_1,u_2]$. When $\mathcal N$ is empty, $S(\phi)\ge0$ for all $\phi>0$.~~~\textbf{Q.E.D.}

\section{Proof of Lemma~\ref{lem:phiLU}} \label{app:proof_phiLU}

Recall from \eqref{eq:h} that $h(\phi) = \Vt^{\mathrm{NR}}(\phi) - \Vt_2(\phi)$, and let $\mathcal{N} = \{\phi > 0 : S(\phi) < 0\}$ as in Appendix~\ref{app:proof_S_sign}. If $\mathcal{N}=\emptyset$, then $S\ge 0$ everywhere, which implies $I_- > 0$ and $I_+ \ge 0$ on $(0,\infty)$; hence $\phi_L=0$, $\phi_U=\infty$, and $h\ge 0$ by \eqref{eq:Green}. 

Suppose now that $\mathcal{N}=(u_1,u_2)\neq\emptyset$ (Lemma~\ref{lem:S_sign}), with $S>0$ outside $[u_1,u_2]$. We first examine the shapes of the tail integrals $I_-$ and $I_+$:
\begin{itemize}
\item \textbf{Behavior of $I_-(\phi)$:} Since $S>0$ on $(u_2,\infty)$, we have $I_- > 0$ on $[u_2,\infty)$. Moreover, because $I_-'(\phi) = -\phi^{-m_+-2}S(\phi)$, $I_-$ is strictly increasing on $(u_1,u_2)$ and strictly decreasing on $(0,u_1)$ whenever $u_1>0$. If $u_1=0$, then $I_-(0+)=-\infty$: near $\phi=0$, $S(s)\le -cs^{1-1/\gamma}$ for $\kappa>1$ and $S(s)=(Y_1-Y_2-L)s<0$ for $\kappa=1$ (for a constant $c>0$), while neither $s^{-m_+-1-1/\gamma}$ nor $s^{-m_+-1}$ is integrable at $0$. Consequently, the negativity set $\{I_-<0\}$ is an open interval $(\phi_L',\phi_L)$. Whenever $\{I_-<0\}\neq\emptyset$, $\phi_L\in(u_1,u_2)$ is the unique zero of $I_-$ in $(u_1,u_2)$, $\phi_L'\in[0,u_1)$, and $I_->0$ on $(\phi_L,\infty)$.

\item \textbf{Behavior of $I_+(\phi)$:} If $u_1>0$, then $S>0$ on $(0,u_1)$ gives $I_+ > 0$ on $(0,u_1]$. Since $I_+'(\phi)=\phi^{-m_--2}S(\phi)$, $I_+$ is strictly decreasing on $(u_1,u_2)$ and strictly increasing on $(u_2,\infty)$, with $\lim_{\phi\to\infty}I_+(\phi)=\infty$ because $S(s)\ge cs$ for large $s$ and $-m_--1>0$. If $u_1=0$, then $I_+ < 0$ on $(0,u_2]$. Thus, whenever $\{I_+<0\}\neq\emptyset$, it forms an open interval $(\phi_U,\phi_U')$ with $\phi_U\in[u_1,u_2)$ and $\phi_U'>u_2$. Specifically, $\phi_U=0$ if $u_1=0$, whereas $\phi_U\in(u_1,u_2)$ is the unique zero of $I_+$ in $(u_1,u_2)$ if $u_1>0$; in both cases, $I_+>0$ on $(0,\phi_U)$.
\end{itemize}

We now establish parts (a)--(c) of Lemma~\ref{lem:phiLU}:
\begin{enumerate}
\item[(a)] If either $\{I_-<0\}$ or $\{I_+<0\}$ is empty, then $\phi_L=0$ or $\phi_U=\infty$, so conditions (i) and (ii) both fail. Otherwise, since $\phi_L'\le u_1\le\phi_U$ and $\phi_L<u_2<\phi_U'$, the intersection $(\phi_L',\phi_L)\cap(\phi_U,\phi_U')$ coincides with $(\phi_U,\phi_L)$, which is non-empty if and only if $\phi_U<\phi_L$. Thus, (i) and (ii) are equivalent. 

If (ii) holds at some $\phi>0$, then $I_-(\phi)<0$ and $I_+(\phi)<0$, so $h(\phi)<0$ directly by \eqref{eq:Green}. Conversely, suppose $h(\phi_0)<0$ for some $\phi_0>0$. If $u_1=0$, then $\phi_U=0$ and $I_-(0+)=-\infty$ ensures $\phi_L>0$, so (i) holds. If $u_1>0$, then $\kappa<1$ (by Lemma~\ref{lem:S_sign}), so $h(\phi)>0$ near $\phi=0$ because its leading term is $\frac{\gamma(1-A)}{(1-\gamma)K}\phi^{1-1/\gamma}$ with a positive coefficient. Since $Y_1>\nu Y_2$ also implies $\lim_{\phi\to\infty}h(\phi)=\infty$, the continuous function $h$ must attain a strictly negative global minimum at some interior point $\phi_m>0$. At this minimum, $h'(\phi_m)=0$, so \eqref{eq:Fh} yields $F_{m_-}(\phi_m)=F_{m_+}(\phi_m)=-h(\phi_m)/\phi_m>0$, establishing (ii).

\item[(b)] As shown above, $\mathcal{N}=\emptyset$ implies $\phi_L=0$ and $\phi_U=\infty$, so $\phi_U<\phi_L$ forces $\mathcal{N}\neq\emptyset$, and the ordering $u_1\le\phi_U<\phi_L<u_2$ follows from the characterization of $\{I_\pm<0\}$. Continuity of $F_{m_\pm}$ gives $F_{m_-}(\phi_L)=0$, as well as $F_{m_+}(\phi_U)=0$ whenever $\phi_U>0$. Observe that the prefactor $-\frac{m_+m_-}{r(m_+-m_-)}$ in \eqref{eq:Green} is strictly positive. Since $I_-(\phi_L)=0$, evaluating \eqref{eq:Green} at $\phi_L$ shows that $h(\phi_L)$ is a positive multiple of $\phi_L^{m_-+1}I_+(\phi_L)$, which is strictly negative because $\phi_L\in(\phi_U,\phi_U')$. Symmetrically, if $\phi_U>0$, then $I_+(\phi_U)=0$ implies that $h(\phi_U)$ is a positive multiple of $\phi_U^{m_++1}I_-(\phi_U)$, which is strictly negative because $\phi_U\in(\phi_L',\phi_L)$.

\item[(c)] Suppose $\phi_L\le\phi_U$. By part (a), $h(\phi)\ge 0$ for all $\phi>0$. Any zero $\phi_1>0$ of $h$ must therefore be a global minimum, so $h'(\phi_1)=0$, which via \eqref{eq:Fh} implies $F_{m_-}(\phi_1)=F_{m_+}(\phi_1)=0$, or equivalently $I_-(\phi_1)=I_+(\phi_1)=0$. From the monotonicity properties established above, if $\{I_-<0\}\neq\emptyset$, the only possible zeros of $I_-$ are $\phi_L'$ and $\phi_L$; if $\{I_-<0\}=\emptyset$, $I_-$ attains its unique global minimum at $u_1$, so its only possible zero is $u_1$ (where $I_+(u_1)>0$). Similarly, if $\{I_+<0\}\neq\emptyset$, the only possible zeros of $I_+$ are $\phi_U$ and $\phi_U'$; if $\{I_+<0\}=\emptyset$, its only possible zero is $u_2$ (where $I_-(u_2)>0$). Since $\phi_L'<u_1\le\phi_U$ (or $\phi_L'=0$) and $\phi_L<u_2<\phi_U'$, any common zero of $I_-$ and $I_+$ must satisfy $\phi_1=\phi_L=\phi_U$. Consequently, $h$ vanishes at most once on $(0,\infty)$.~~~\textbf{Q.E.D.}
\end{enumerate}

\section{Proof of Theorem~\ref{thm:main}} \label{app:proof_thm_main}

We divide the proof of Theorem~\ref{thm:main} into two steps: we first construct a dual candidate value function $\Vt_1$ and establish its regularity and convexity properties (Lemma~\ref{lem:cand}; the properties of $\Vt_2$ are collected in Lemma~\ref{lem:Vt2} of Appendix~\ref{app:post_aux}), and we then verify via dynamic programming and martingale arguments that its convex conjugate coincides with $V_1$ and satisfies the variational inequality \eqref{Variational_inequality}.

\paragraph{Step 1: Construction and properties of the dual candidate $\Vt_1$.}
If $\phi_L\le\phi_U$, set $\Vt_1:=\Vt^{\mathrm{NR}}$. If $\phi_U<\phi_L$, define
\[
  \Vt_1(\phi):=
  \begin{cases}
    \Vt^{\mathrm{NR}}(\phi)-\dfrac{B_U}{m_++1}\phi^{m_++1}, & 0<\phi<\phi_U,\\[1.2ex]
    \Vt_2(\phi), & \phi_U\le\phi\le\phi_L,\\[1.2ex]
    \Vt^{\mathrm{NR}}(\phi)-\dfrac{B_L}{m_-+1}\phi^{m_-+1}, & \phi>\phi_L,
  \end{cases}
\]
where the first branch is omitted when $\phi_U=0$, and $B_L,B_U$ are defined in \eqref{eq:BLU}. Define the excess dual value $\widetilde{\Delta}(\phi):=\Vt_1(\phi)-\Vt_2(\phi)$ for $\phi > 0$. Since $\Vt^{\mathrm{NR}}-\Vt_2=h$, substituting \eqref{eq:BLU} gives
\begin{equation*}
  \widetilde{\Delta}(\phi)=
  \begin{cases}
    h(\phi)-\big(\phi/\phi_U\big)^{m_++1}h(\phi_U), & 0<\phi<\phi_U,\\[0.8ex]
    0, & \phi_U\le\phi\le\phi_L,\\[0.8ex]
    h(\phi)-\big(\phi/\phi_L\big)^{m_-+1}h(\phi_L), & \phi>\phi_L,
  \end{cases}
\end{equation*}
when $\phi_U<\phi_L$, while $\widetilde{\Delta}=h$ when $\phi_L\le\phi_U$. We also define the primal candidate $\Vh_1(w):=\inf_{y>0}[\Vt_1(y)+yw]$ for $w>\underline{w}_1$.

\begin{lem}[Properties of the dual candidate]\label{lem:cand}
The candidate functions $\Vt_1$ and $\widetilde{\Delta}$ satisfy the following properties:
\begin{enumerate}
\item[(a)] $\widetilde{\Delta}\in C^1(0,\infty)$, $\widetilde{\Delta}$ is $C^2$ outside $\{\phi_U,\phi_L\}$, $\widetilde{\Delta}=0$ on $[\phi_U,\phi_L]$, and $\widetilde{\Delta}>0$ elsewhere (when $\phi_L\le\phi_U$, $\widetilde{\Delta}=h\ge 0$).
\item[(b)] When $\phi_U<\phi_L$, the residual $\Lt\Vt_1(\phi)+Y_1\phi+\Ut(\phi)$ is zero on the open continuation branches and equals $S(\phi)<0$ on $(\phi_U,\phi_L)$. These classical identities are not asserted at the joining points. When $\phi_L\le\phi_U$, the residual is zero throughout $(0,\infty)$.
\item[(c)] $-\Vt_1'=X_L$ on $(\phi_L,\infty)$ and $-\Vt_1'=X_U$ on $(0,\phi_U)$ with the coefficients \eqref{eq:BLU}; moreover, $X_L'<0$, $X_U'<0$, and $\Vt_1$ is convex on $(0,\infty)$.
\item[(d)] There exists a constant $C>0$ such that $|\Vt_1(\phi)-Y_1\phi/r|\le C(1+\phi+\phi^{1-1/\gamma})$ for all $\phi>0$.
\end{enumerate}
\end{lem}

\begin{proof}
(a) Write $k_\pm(\phi) := h(\phi)\phi^{-(m_\pm+1)}$. From Appendix~\ref{app:proof_S_sign}, we have $k_\pm'(\phi)=(m_\pm+1)\phi^{-(m_\pm+1)}F_{m_\pm}(\phi)$, so that
\[
  k_-'(\phi)=\tfrac{2}{\theta^2}\phi^{m_+-m_--1}I_-(\phi), \qquad k_+'(\phi)=-\tfrac{2}{\theta^2}\phi^{m_--m_+-1}I_+(\phi).
\]
For $\phi>\phi_L$, $\widetilde{\Delta}(\phi)\phi^{-(m_-+1)}=k_-(\phi)-k_-(\phi_L)>0$ because $I_->0$ on $(\phi_L,\infty)$ by Lemma~\ref{lem:phiLU}(b). Symmetrically, when $\phi_U>0$, for $0<\phi<\phi_U$ we have $\widetilde{\Delta}(\phi)\phi^{-(m_++1)}=k_+(\phi)-k_+(\phi_U)>0$ because $I_+>0$ on $(0,\phi_U)$. Furthermore, the right-hand derivative of $\widetilde{\Delta}$ at $\phi_L$ equals $h'(\phi_L)-(m_-+1)h(\phi_L)/\phi_L=(m_-+1)F_{m_{-}}(\phi_L)=0$ by \eqref{eq:Fh}, and the left-hand derivative at $\phi_U$ vanishes analogously. Thus $\widetilde{\Delta}\in C^1(0,\infty)$.

\noindent (b) Since $\Lt\phi^{m_\pm+1}=0$, we have $\Lt\widetilde{\Delta}=\Lt h=-S$ outside $[\phi_U,\phi_L]$. Using $\Lt\Vt^{\mathrm{NR}}(\phi) = -(Y_1\phi+\Ut(\phi))$ and $\Lt\Vt_2(\phi) = -(Y_1\phi+\Ut(\phi)) + S(\phi)$, it follows that $\Lt\Vt_1(\phi)+Y_1\phi+\Ut(\phi)=\Lt\widetilde{\Delta}(\phi)+S(\phi)$, which vanishes outside $[\phi_U,\phi_L]$ and equals $S(\phi)$ on $(\phi_U,\phi_L)$ when $\phi_U<\phi_L$. Finally, $[\phi_U,\phi_L]\cap(0,\infty)\subset(u_1,u_2)$ by Lemma~\ref{lem:phiLU}(b), ensuring $S<0$ on $[\phi_U,\phi_L]\cap(0,\infty)$.

\noindent (c) By construction and \eqref{eq:db-low-inverse}--\eqref{eq:db-high-inverse}, $-\Vt_1'=X_L$ on $(\phi_L,\infty)$ and $-\Vt_1'=X_U$ on $(0,\phi_U)$. By Lemma~\ref{lem:phiLU}(b), $h(\phi_L)<0$, which implies $B_L>0$ and $m_-B_L<0$; hence $X_L'(\phi)=m_-B_L\phi^{m_--1}-\frac{1}{\gamma K}\phi^{-1/\gamma-1}<0$. Analogously, when $\phi_U>0$, $h(\phi_U)<0$ yields $B_U<0$ and $m_+B_U<0$, so $X_U'(\phi)=m_+B_U\phi^{m_+-1}-\frac{1}{\gamma K}\phi^{-1/\gamma-1}<0$. Thus, $\Vt_1''=-X'>0$ on the continuation branches, while $\Vt_1''=-\mathcal{W}_2'\ge 0$ on $[\phi_U,\phi_L]$. Since $\Vt_1\in C^1(0,\infty)$, $\Vt_1$ is convex. When $\phi_L\le\phi_U$, $-\Vt_1'(\phi)=\frac{1}{K}\phi^{-1/\gamma}-\frac{Y_1}{r}$ is strictly decreasing, so $\Vt_1$ is strictly convex.

\noindent (d) On $(\phi_L,\infty)$, $\Vt_1(\phi)-Y_1\phi/r=\frac{\gamma}{(1-\gamma)K}\phi^{1-1/\gamma}-(\phi/\phi_L)^{m_-+1}h(\phi_L)$ with $(\phi/\phi_L)^{m_-+1}\le 1$ (since $m_-+1<0$). The branch $(0,\phi_U)$ is analogous since $m_++1>0$. On the remaining interval $[\phi_U,\phi_L]$ (or $(0,\phi_L]$ if $\phi_U=0$, where the first branch of \eqref{eq:Vt2} applies on $(0,\min(\psi_L,\phi_L))$ and $[\psi_L,\phi_L]$ is compact), $\Vt_2(\phi)-Y_1\phi/r$ is bounded by $C(1+\phi+\phi^{1-1/\gamma})$.
\end{proof}

\begin{lem}[Duality upper bound]\label{lem:upper}
For every $w \gt \underline{w}_1$, every admissible control $(c,\pi,\tau)\in\mathcal{A}_1(w)$, and every $y \gt 0$,
\[
  \E\Big[\int_0^\tau e^{-\rho t}U_1(c_t)\,\mathrm{d}t+e^{-\rho\tau}V_2(W_\tau)\mathbf{1}_{\{\tau \lt \infty\}}\Big]
  \le\Vt_1(y)+yw .
\]
Consequently, $V_1(w)\le\Vh_1(w)$ for all $w \gt \underline{w}_1$.
\end{lem}

\begin{proof}
Let $\xi_t:=\exp(-(r+\frac{1}{2}\theta^2)t-\theta \beta_t)$ denote the state-price density and set $\Phi_t:=ye^{\rho t}\xi_t$, so that $\mathrm{d}\Phi_t=\Phi_t[(\rho-r)\mathrm{d}t-\theta \mathrm{d}\beta_t]$ with $\Phi_0=y$.

\paragraph{Step 1 (Budget constraint).} 
Define total wealth $\bar{W}_t:=W_t+Y_1/r \ge 0$ for $t\le\tau$, which evolves according to $\mathrm{d}\bar{W}_t=(r\bar{W}_t+\pi_t(\mu-r)-c_t)\mathrm{d}t+\sigma\pi_t\mathrm{d}\beta_t$. By It\^o's product rule, the deflated wealth-plus-consumption process $\xi_t\bar{W}_t+\int_0^t\xi_s c_s\,\mathrm{d}s$ is a nonnegative local martingale, and hence a supermartingale, on $[0,\tau]$. For any finite horizon $T \gt 0$, optional sampling yields
\begin{equation}\label{eq:budget}
  \E\Big[\int_0^{\tau\wedge T}e^{-\rho s}\Phi_s c_s\,\mathrm{d}s+e^{-\rho(\tau\wedge T)}\Phi_{\tau\wedge T}\bar{W}_{\tau\wedge T}\Big]\le y\bar{W}_0 .
\end{equation}

\paragraph{Step 2 (Fenchel duality).} 
By definition of convex conjugates, $U_1(c)\le\Ut(\Phi)+\Phi c$ for all $c\geq 0$, and on the event $\{\tau \lt \infty\}$, $V_2(W_\tau)\le\Vt_2(\Phi_\tau)+\Phi_\tau W_\tau$ by Lemma~\ref{lem:Vt2}(b). For any function $g$ on $(0,\infty)$, define $g^\sharp(\phi):=g(\phi)-Y_1\phi/r$. Then $\Vt_2(\Phi_\tau)+\Phi_\tau W_\tau=\Vt_2^\sharp(\Phi_\tau)+\Phi_\tau \bar{W}_\tau$.

\paragraph{Step 3 (Dual supersolution).} 
By Lemma~\ref{lem:cand}(b) and the identity $\Lt(Y_1\phi/r)=-Y_1\phi$, we have $\Lt\Vt_1^\sharp+\Ut\le 0$ almost everywhere on $(0,\infty)$. Moreover, $\Vt_1^\sharp\ge\Vt_2^\sharp$, and $\Vt_1^\sharp\in C^1(0,\infty)$ with a locally bounded second derivative. Applying It\^o's formula (with $\Vt_1''$ taken almost everywhere) to $e^{-\rho t}\Vt_1^\sharp(\Phi_t)$ gives
\[
  e^{-\rho(\tau\wedge T)}\Vt_1^\sharp(\Phi_{\tau\wedge T}) + \int_0^{\tau\wedge T} e^{-\rho s}\Ut(\Phi_s)\,\mathrm{d}s \le \Vt_1^\sharp(y) + \int_0^{\tau\wedge T} e^{-\rho s}\big(\Lt\Vt_1^\sharp+\Ut\big)(\Phi_s)\,\mathrm{d}s + M_{\tau\wedge T},
\]
where the stochastic integral $M_t := -\theta\int_0^t e^{-\rho s}\Phi_s \Vt_1^{\sharp\prime}(\Phi_s)\,\mathrm{d}\beta_s$ is a true martingale on $[0,T]$ because $\phi\Vt_1^{\sharp\prime}(\phi)$ exhibits polynomial growth and $\Phi$ is a geometric Brownian motion. Taking expectations and using $\Vt_1^\sharp\ge\Vt_2^\sharp$ on $\{\tau\le T\}$ yields
\begin{equation}\label{eq:dualineq}
  \E\Big[\int_0^{\tau\wedge T}e^{-\rho s}\Ut(\Phi_s)\,\mathrm{d}s+e^{-\rho\tau}\Vt_2^\sharp(\Phi_\tau)\mathbf{1}_{\{\tau\le T\}}\Big]
  \le\Vt_1^\sharp(y)+\E\big[e^{-\rho T}|\Vt_1^\sharp(\Phi_T)|\big].
\end{equation}
By Lemma~\ref{lem:cand}(d) and the moment identity $\E[e^{-\rho T}\Phi_T^p]=y^p e^{Q(p)T}$, where $Q(0)=-\rho \lt 0$, $Q(1)=-r \lt 0$, and $Q(1-1/\gamma)=-K \lt 0$, the last term converges to $0$ as $T\to\infty$.

\paragraph{Step 4 (Conclusion).} 
Combining \eqref{eq:budget}--\eqref{eq:dualineq} and dropping the nonnegative term $e^{-\rho T}\Phi_T \bar{W}_T\mathbf{1}_{\{\tau \gt T\}}\ge 0$ produces
\[
  \E\Big[\int_0^{\tau\wedge T}e^{-\rho s}U_1(c_s)\,\mathrm{d}s+e^{-\rho\tau}V_2(W_\tau)\mathbf{1}_{\{\tau\le T\}}\Big]
  \le\Vt_1^\sharp(y)+y\bar{W}_0+\varepsilon_T=\Vt_1(y)+yw+\varepsilon_T,
\]
where $\varepsilon_T:=\E[e^{-\rho T}|\Vt_1^\sharp(\Phi_T)|]\to 0$ as $T\to\infty$. Because $U_1$ and $V_2$ both share the sign of $1-\gamma$ (strictly positive for $\gamma \lt 1$ and strictly negative for $\gamma \gt 1$; see Lemma~\ref{lem:V2bounds}), letting $T\to\infty$ and applying the monotone convergence theorem (to the positive expressions when $\gamma \lt 1$ and to the negative expressions via $-U_1,-V_2\ge 0$ when $\gamma \gt 1$) establishes the desired upper bound. Taking the infimum over $y \gt 0$ and supremum over $(c,\pi,\tau)\in\mathcal{A}_1(w)$ yields $V_1(w)\le\Vh_1(w)$.
\end{proof}

\begin{lem}[Attainment of the dual bound]\label{lem:optimization}
For every $w \gt \underline{w}_1$, there exists an admissible policy $(c^*,\pi^*,\tau^*)\in\mathcal{A}_1(w)$ such that $J(w;c^*,\pi^*,\tau^*)=\Vh_1(w)$.
\end{lem}

\begin{proof}
Suppose first that $\phi_U \lt \phi_L$ and $w\in[w_L,w_U]\cap\mathbb{R}$. Set $\tau^*:=0$, and choose $\psi:=V_2'(w)\in[\phi_U,\phi_L]$ (if $w=\underline{w}_2$ and $\phi_L \gt \psi_\nu$, note that $V_2'(\underline{w}_2+)=\psi_\nu$ and choose any $\psi\in[\max(\psi_\nu,\phi_U),\phi_L]$). On this interval, $\Vt_1(\psi)=\Vt_2(\psi)$ and $\Vt_1'(\psi)=-w$. By convexity of $\Vt_1$, $\psi$ minimizes $z \mapsto \Vt_1(z)+zw$, so $\Vh_1(w)=\Vt_2(\psi)+\psi w=V_2(w)=J(w;c^*,\pi^*,0)$.

Otherwise, $w$ lies in the range of $X_L$ (namely $(\underline{w}_1,w_L)$), of $X_U$ (namely $(w_U,\infty)$), or of $X^{\mathrm{NR}}(\phi):=-\Vt^{\mathrm{NR}\prime}(\phi)=\frac{1}{K}\phi^{-1/\gamma}-\frac{Y_1}{r}$ (namely $(\underline{w}_1,\infty)$ when $\phi_L\le\phi_U$). Let $y \gt 0$ be the unique preimage satisfying $-\Vt_1'(y)=w$, which minimizes $z \mapsto \Vt_1(z)+zw$. Set $\Phi_t:=ye^{\rho t}\xi_t$, define $\tau^*:=\inf\{t\ge 0:\Phi_t\in[\phi_U,\phi_L]\}$ when $\phi_U \lt \phi_L$ (and $\tau^*:=\infty$ when $\phi_L\le\phi_U$), and for $t\le\tau^*$ define
\[
  W_t^*:=X(\Phi_t), \qquad c_t^*:=\Phi_t^{-1/\gamma}, \qquad \pi^*_t:=-\frac{\theta}{\sigma}\Phi_t X'(\Phi_t),
\]
where $X\in\{X_L,X_U,X^{\mathrm{NR}}\}$ denotes the active branch containing $y$.

Applying It\^o's formula to $W_t^*=X(\Phi_t)$ and substituting the ODE \eqref{eq:X_ODE_pre} confirms that
\[
  \mathrm{d}W^*_t=\big(rW_t^*+Y_1-c_t^*+\pi_t^*(\mu-r)\big)\mathrm{d}t+\sigma\pi^*_t\,\mathrm{d}\beta_t,
\]
so $W^*$ is indeed the wealth trajectory generated by $(c^*,\pi^*)$ starting from $W_0^*=X(y)=w$. Moreover, $W_t^* \gt \underline{w}_1$ for all $t\le\tau^*$, and on $\{\tau^* \lt \infty\}$ we have $W^*_{\tau^*}=\mathcal{W}_2(\Phi_{\tau^*})\ge \underline{w}_2$, ensuring admissibility $(c^*,\pi^*,\tau^*)\in\mathcal{A}_1(w)$. Along this candidate policy:
\begin{itemize}
\item the Fenchel inequalities in Step 2 of Lemma~\ref{lem:upper} hold with equality ($U_1(c_t^*)=\Ut(\Phi_t)+\Phi_t c_t^*$ and $V_2(W^*_{\tau^*})=\Vt_2(\Phi_{\tau^*})+\Phi_{\tau^*}W^*_{\tau^*}$ on $\{\tau^* \lt \infty\}$);
\item the dual inequality \eqref{eq:dualineq} holds with equality, with the boundary term $\E[e^{-\rho T}\Vt_1^\sharp(\Phi_T)\mathbf{1}_{\{\tau^* \gt T\}}]$ replacing $\E[e^{-\rho T}|\Vt_1^\sharp(\Phi_T)|]$;
\item the deflated budget process $\xi_t\bar{W}_t^*+\int_0^t\xi_s c_s^*\,\mathrm{d}s$ is a true martingale on $[0,T]$ because $\bar{W}_t^*=X^\sharp(\Phi_t)$ is polynomially bounded in $\Phi_t$, where $X^\sharp(\phi):=X(\phi)+Y_1/r$ satisfies $\phi X^\sharp(\phi)=\frac{1}{K}\phi^{1-1/\gamma}+B_L\phi^{m_-+1}$ on $(\phi_L,\infty)$ or $\frac{1}{K}\phi^{1-1/\gamma}+B_U\phi^{m_++1}$ on $(0,\phi_U)$; thus $\phi X^\sharp(\phi)\le C(1+\phi^{1-1/\gamma})$ on either branch, and the terminal error term $\E[e^{-\rho T}(|\Vt_1^\sharp|+\phi X^\sharp)(\Phi_T)\mathbf{1}_{\{\tau^* \gt T\}}]$ vanishes as $T\to\infty$.
\end{itemize}
Letting $T\to\infty$ therefore yields $J(w;c^*,\pi^*,\tau^*)=\Vt_1(y)+y\bar{W}_0-yY_1/r=\Vt_1(y)+yw=\Vh_1(w)$.
\end{proof}

\paragraph{Synthesis of Theorem~\ref{thm:main}.}
Combining Lemmas~\ref{lem:upper} and \ref{lem:optimization} establishes that $V_1=\Vh_1$, and that the strategy constructed in Lemma~\ref{lem:optimization} is optimal. For any $w \gt \underline{w}_1$, let $y=y(w)$ denote a minimizer of $z\mapsto\Vt_1(z)+zw$, so that $V_1(w)=\Vt_1(y)+yw$. For $w\ge\underline{w}_2$, Lemma~\ref{lem:Vt2}(b) implies
\begin{equation}\label{eq:V1V2gap}
  V_1(w)-V_2(w)\ge\Vt_1(y)-\Vt_2(y)=\widetilde{\Delta}(y).
\end{equation}

\begin{enumerate}
\item[(a)] \textbf{Never-retire regime ($\phi_L\le\phi_U$).} Here $\Vt_1=\Vt^{\mathrm{NR}}$, so $V_1=\Vh_1=V^{\mathrm{NR}}$. In Lemma~\ref{lem:optimization}, we have $\tau^*=\infty$ and $W^*_t=X^{\mathrm{NR}}(\Phi_t)$ with $X^{\mathrm{NR}}(\phi)=\frac{1}{K}\phi^{-1/\gamma}-\frac{Y_1}{r}$. Consequently, $c^*_t=\Phi_t^{-1/\gamma}=K(W^*_t+Y_1/r)$ and $\pi^*_t=-\frac{\theta}{\sigma}\Phi_t X^{\mathrm{NR}\prime}(\Phi_t)=\frac{\theta}{\gamma\sigma}(W^*_t+Y_1/r)$, proving that perpetual continuation with these Merton controls is optimal. By \eqref{eq:V1V2gap} and the fact that $\widetilde{\Delta}=h\ge 0$ (Lemma~\ref{lem:phiLU}(c)), it follows that $V^{\mathrm{NR}}\ge V_2$ on $[\underline{w}_2,\infty)$, with equality at $w$ requiring $h(y(w))=0$. Because $w\mapsto y(w)$ is injective and $h$ vanishes at most once, $V^{\mathrm{NR}}(w)=V_2(w)$ holds for at most one $w$; moreover, when $\phi_L \lt \phi_U$, $h$ is strictly positive everywhere (see the proof of Lemma~\ref{lem:phiLU}(c)), yielding $\mathcal{R}=\emptyset$.

\item[(b)(i)] \textbf{Retirement region.} Suppose $\phi_U \lt \phi_L$. If $w\in[w_L,w_U]\cap\mathbb{R}$, then $V_1(w)=V_2(w)$ by the first part of the proof of Lemma~\ref{lem:optimization}. Conversely, if $w\ge\underline{w}_2$ lies outside $[w_L,w_U]$, then $y(w)\notin[\phi_U,\phi_L]$, so \eqref{eq:V1V2gap} together with Lemma~\ref{lem:cand}(a) gives $V_1(w)-V_2(w)\ge\widetilde{\Delta}(y(w)) \gt 0$. Thus, $\mathcal{R}=[w_L,w_U]\cap\mathbb{R}$.

\item[(b)(ii)] \textbf{Value function decomposition.} On the continuation branches, $V_1(w)=\Vh_1(w)=\Vt_1(\phi)+\phi X(\phi)$ with $\phi=X^{-1}(w)$. Using the identity $\frac{m}{m+1}-1=-\frac{1}{m+1}$, this expression simplifies directly to $V_{1,L}(\phi)$ in \eqref{eq:db-low-value} on $(\underline{w}_1,w_L)$ and $V_{1,U}(\phi)$ in \eqref{eq:db-high-value} on $(w_U,\infty)$. The coefficients $B_L,B_U$ match \eqref{eq:BLU} by construction of $\Vt_1$, and $X_L' \lt 0$, $X_U' \lt 0$ follow from Lemma~\ref{lem:cand}(c).

\item[(b)(iii)] \textbf{Optimal controls.} Consider the optimal policy from Lemma~\ref{lem:optimization}. If $w\in\mathcal{R}$, then $\tau^*=0$. Otherwise, on $[0,\tau^*)$ we have $W^*_t=X(\Phi_t)$, where $X$ is either $X_L$ (a strictly decreasing bijection from $(\phi_L,\infty)$ onto $(\underline{w}_1,w_L)$) or $X_U$ (a strictly decreasing bijection from $(0,\phi_U)$ onto $(w_U,\infty)$). By part (b)(ii), $V_1'=X^{-1}$ on the respective continuation branch, so for all $t \lt \tau^*$,
\[
  \Phi_t=V_1'(W^*_t) \qquad\text{and}\qquad V_1''(W^*_t)=\frac{1}{X'(\Phi_t)}.
\]
Substituting these relations yields
\[
  c^*_t=V_1'(W^*_t)^{-1/\gamma} \qquad\text{and}\qquad \pi^*_t=-\frac{\theta}{\sigma}\Phi_t X'(\Phi_t)=-\frac{\theta}{\sigma}\frac{V_1'(W^*_t)}{V_1''(W^*_t)},
\]
showing that $(c^*,\pi^*)$ coincide with the feedback controls in \eqref{eq:pre_c_star_general} evaluated along $W^*$, and $W^*$ solves the closed-loop wealth SDE on $[0,\tau^*)$. Furthermore, for $t \lt \tau^*$, $\Phi_t\notin[\phi_U,\phi_L]$ implies $W^*_t\in(\underline{w}_1,w_L)\cup(w_U,\infty)=\mathcal{C}$. On $\{\tau^* \lt \infty\}$, $\Phi_{\tau^*}\in\{\phi_L,\phi_U\}$ implies $W^*_{\tau^*}=\mathcal{W}_2(\Phi_{\tau^*})\in\{w_L,w_U\}\subset\mathcal{R}$ (with $W^*_{\tau^*}=\underline{w}_2$ when $\Phi_{\tau^*}=\phi_L \gt \psi_\nu$). Consequently, $\tau^*=\inf\{t\ge 0:W^*_t\in\mathcal{R}\}$, establishing (b)(iii).

\item[(b)(iv)] \textbf{Smoothness and concavity.} By part (b)(ii), $V_1'=\phi$, so $V_1''=1/X' \lt 0$ on the continuation regions, while on $\mathcal{R}$, $V_1=V_2$ is strictly concave by Lemma~\ref{lem:V2concave}. Across interior free boundaries $w_L \gt \underline{w}_2$ and $w_U \lt \infty$, $V_1'$ is continuous because $X_L(\phi_L)=\mathcal{W}_2(\phi_L)$ and $X_U(\phi_U)=\mathcal{W}_2(\phi_U)$. When $\phi_L \gt \psi_\nu$, $\Vt_1$ is affine with constant slope $-\underline{w}_2$ across $[\max(\phi_U,\psi_\nu),\phi_L]$, which produces a concave kink in $V_1$ at $w_L=\underline{w}_2$ with left- and right-hand derivatives $V_1'(\underline{w}_2-)=\phi_L \gt \max(\phi_U,\psi_\nu)=V_1'(\underline{w}_2+)$.

\item[(b)(v)] \textbf{Local variational inequality.} For every $w\in\mathcal{R}\cap(\underline{w}_2,\infty)$, we have $V_2'(w)\in[\phi_U,\phi_L]\cap(0,\psi_\nu)\subset(u_1,u_2)$ by Lemma~\ref{lem:phiLU}(b). Applying Lemma~\ref{lem:S_meaning} and the definition of $S$ in \eqref{eq:S} therefore gives $\rho V_2(w)-\mathcal{H}_1V_2(w)=-S(V_2'(w)) \gt 0$.
\end{enumerate}

Finally, on the continuation region $\mathcal{C}$, $\mathcal{H}_1V_1=\rho V_1$ holds by the inverse ODE \eqref{eq:X_ODE_pre} (equivalently, via Lemma~\ref{lem:cand}(b) and the computation in Lemma~\ref{lem:S_meaning}). On $\mathcal{R}$, part (b)(v) holds with $V_1=V_2$, confirming that $V_1$ satisfies the variational inequality \eqref{Variational_inequality} classically away from the free boundaries.~~~\textbf{Q.E.D.}

\section{Proofs of Corollaries~\ref{cor:original} and \ref{cor:low_policy}} \label{app:proof_cor_original}

\noindent\textbf{Proof of Corollary~\ref{cor:original}.} By Theorem~\ref{thm:main}, $V_1(w)=\Vh_1(w)$ is finite for every $w>\underline{w}_1$, and the policy of Lemma~\ref{lem:optimization} is optimal for $V_1(w)$. Hence Proposition~\ref{prop:reduction}(a)--(b) gives the result.~~~\textbf{Q.E.D.}

\medskip
\noindent\textbf{Proof of Corollary~\ref{cor:low_policy}.} By Theorem~\ref{thm:main}(b)(ii), on $(\underline{w}_1,w_L)$ we have $V_1'(w)=\phi$ with $w=X_L(\phi)$ and $\phi\in(\phi_L,\infty)$, where $X_L$ is strictly decreasing; hence $\phi$ is strictly decreasing in $w$, and $\phi\to\infty$ as $w\downarrow\underline{w}_1$. By \eqref{eq:pre_c_star_general} and $V_1''=1/X_L'(\phi)$, $c_1^*=\phi^{-1/\gamma}$ and $\pi_1^*=-\frac\theta\sigma\phi X_L'(\phi)=\frac\theta\sigma\big(-m_-B_L\phi^{m_-}+\frac{1}{\gamma K}\phi^{-1/\gamma}\big)$. Since $B_L>0$ by \eqref{eq:BLU} and $m_-<0$, both terms of $\pi_1^*$ are positive and strictly decreasing in $\phi$, as is $c_1^*$. Hence $c_1^*$ and $\pi_1^*$ are strictly increasing in $w$, and both tend to $0$ as $\phi\to\infty$.~~~\textbf{Q.E.D.}

\section{Proofs of Lemma~\ref{lem:F_at_psiL} and Theorem~\ref{theorem kappa>1}} \label{app:proof_kappa_ge_one}

\noindent\textbf{Proof of Lemma~\ref{lem:F_at_psiL}.} By continuity, $F_{m_-}(\psi_L)$ is the first line of \eqref{eq:Fminus} evaluated at $\psi_L$. Let $F(\phi)$ denote the negative of the first line of \eqref{eq:Fminus}, so that $F_{m_-}(\psi_L)=-F(\psi_L)$. Substituting \eqref{eq:C1_post}, \eqref{eq:D1_case1_solved}, \eqref{eq:psi_nu_case1_solved}, and $\psi_L^{-1/\gamma}=L\kappa^{-\frac{1-\gamma}{\gamma}}$, we obtain
\begin{align*}
	F(\psi_L) & =-\frac{m_+-m_-}{(m_++1)(m_-+1)}D_1\psi_L^{m_+}-\frac{(\gamma m_-+1)(A-1)}{(m_-+1)(1-\gamma)K}\psi_L^{-\frac{1}{\gamma}} +\frac{m_-(Y_1-(L+Y_2))}{r(m_-+1)} \\
	          & =-\frac{m_+-m_-}{(m_++1)(m_-+1)} C_1\psi_L^{m_+}
	+\frac{(\gamma m_-+1)L}{\gamma^2m_+(m_++1)(m_-+1) K} \\
	          & \quad -\frac{(\gamma m_-+1)(A-1)\kappa^{-\frac{1-\gamma}{\gamma}}L}{(m_-+1)(1-\gamma)K}+\frac{m_-(Y_1-(L+Y_2))}{r(m_-+1)} \\
	          & =\frac{m_-(1-\nu)Y_2}{r(m_++1)(m_-+1)} \left(
	\frac{\Delta_L}{L}
	\right)^{-m_+/m_-} \\
	          & \quad +\dfrac{(\gamma m_-+1)L}{(m_-+1)K}\left[\frac{1}{\gamma^2m_+(m_++1)}-\frac{(A-1)\kappa^{-\frac{1-\gamma}{\gamma}}}{1-\gamma}\right] +\frac{m_-(Y_1-(L+Y_2))}{r(m_-+1)} \\
	          & =\dfrac{m_-}{r(m_-+1)}\left[\frac{(1-\nu)Y_2}{m_++1} \left(
		\frac{\Delta_L}{L}
		\right)^{-m_+/m_-}+Y_1-(L+Y_2)\right. \\
	          & \quad \left.+\frac{r(\gamma m_-+1)L\left\{1-\gamma-\gamma^2m_+(m_++1)(A-1)\kappa^{-\frac{1-\gamma}{\gamma}}\right\}}{(1-\gamma)\gamma^2m_+m_-(m_++1)K}\right] \\
	          & =\dfrac{m_-}{r(m_-+1)}\left[Y_1-Y_2\left\{1-\frac{1-\nu}{m_++1} \left(
		\frac{\Delta_L}{L}
		\right)^{-m_+/m_-}\right\}\right. \\
	          & \quad \left.-L\left\{1-\frac{r(\gamma m_-+1)\left\{1-\gamma-\gamma^2m_+(m_++1)(A-1)\kappa^{-\frac{1-\gamma}{\gamma}}\right\}}{(1-\gamma)\gamma^2m_+m_-(m_++1)K}\right\}\right] \\
	          & =\dfrac{m_-}{r(m_-+1)}\left(Y_1-\widetilde{Y}_1\right),
\end{align*}
where $\widetilde{Y}_1$ is defined in \eqref{eq:Y1_tilde_def}. None of these steps uses the sign of $A-1$, so the identity holds for every $\kappa>0$ and gives \eqref{eq:F_at_psiL}. Finally, $m_-<-1$ gives $\frac{m_-}{r(m_-+1)}>0$.~~~\textbf{Q.E.D.}

\medskip
\noindent\textbf{Proof of Theorem~\ref{theorem kappa>1}.} By Lemma~\ref{lem:S_sign}(a)--(b), $\mathcal{N}=(0,u_2)$. As in the proof of Lemma~\ref{lem:phiLU}, $I_-$ increases strictly from $-\infty$ on $(0,u_2)$ and is positive on $[u_2,\infty)$, so it has a unique zero $\phi^*\in(0,u_2)$, with $I_-<0$ on $(0,\phi^*)$ and $I_->0$ on $(\phi^*,\infty)$; and $I_+<0$ on $(0,u_2]$, so that $\phi_U=0$. By Proposition~\ref{prop:Fint}, $F_{m_-}>0$ on $(0,\phi^*)$, $F_{m_-}<0$ on $(\phi^*,\infty)$, and $\phi_L=\phi^*>0=\phi_U$. Theorem~\ref{thm:main}(b) with $w_U=\infty$ then gives $\mathcal{R}=[w^*,\infty)$, the representation of $V_1$, and the optimal controls; in particular $B_L>0$.

\noindent(i) By the sign pattern of $F_{m_-}$, $\phi^*<\psi_L$ if and only if $F_{m_-}(\psi_L)<0$, i.e., if and only if $Y_1>\widetilde{Y}_1$ by Lemma~\ref{lem:F_at_psiL}. In this case $\phi^*\in(0,\psi_L)$ is a zero of the first line of \eqref{eq:Fminus}, which is \eqref{eq:free_boundary_phi}. Its left-hand side is strictly decreasing on $(0,\infty)$: the coefficient of $\phi^{m_+}$ is negative since $m_-+1<0$ and $D_1>0$, and the coefficient of $\phi^{-1/\gamma}$ is nonnegative since $\gamma m_-+1<0$, $m_-+1<0$, and $(A-1)/(1-\gamma)\ge0$ for $\kappa\ge1$. Hence $\phi^*$ is the unique root in $(0,\psi_L)$. The formula for $w^*$ is $w^*=\mathcal{W}_2(\phi^*)=W_+(\phi^*)$, and $w^*>W_+(\psi_L)=w_{0,2}$ because $W_+$ is strictly decreasing. The parametric representation \eqref{pre-retirement value function solution}--\eqref{eq:X_solution_pre} is \eqref{eq:db-low-inverse}--\eqref{eq:db-low-value} with $B=B_L$, and \eqref{eq:BB} is \eqref{eq:Bmatch} with $m=m_-$ evaluated at $\phi^*$ using $\mathcal{W}_2(\phi^*)=W_+(\phi^*)$.

\noindent(ii)--(iii) Likewise, $\phi^*>\psi_\nu$ if and only if $F_{m_-}(\psi_\nu)>0$, where $F_{m_-}(\psi_\nu)$ is the third line of \eqref{eq:Fminus} at $\psi_\nu$ with $V_2(\underline{w}_2)$ from Lemma~\ref{lem:V2bounds}(i), which gives the stated formula. In this case $w^*=\mathcal{W}_2(\phi^*)=\underline{w}_2$, and the kink follows from Theorem~\ref{thm:main}(b)(iv). The remaining case $\psi_L\le\phi^*\le\psi_\nu$ is equivalent to $F_{m_-}(\psi_L)\ge0$ and $F_{m_-}(\psi_\nu)\le0$, i.e., to $Y_1\le\widetilde{Y}_1$ and $F_{m_-}(\psi_\nu)\le0$, and then $w^*=W_0(\phi^*)\in[\underline{w}_2,w_{0,2}]$.

The feedback controls follow from \eqref{eq:pre_c_star_general} and $V_1''=1/X_L'$; positivity of $\pi_1^*$ follows from $B_L>0$ and $m_-<0$. Finally, if $\kappa=1$ and $Y_1\ge L+Y_2$, then $\mathcal{N}=\emptyset$ by Lemma~\ref{lem:S_sign}(b), so $\phi_L=0<\infty=\phi_U$ by the proof of Lemma~\ref{lem:phiLU}, and Theorem~\ref{thm:main}(a) applies.~~~\textbf{Q.E.D.}

\section{Proof of Proposition~\ref{prop:db-high-wealth-continuation}} \label{app:proof_high_wealth_cont}

To establish that the semi-infinite retirement region $[w^*,\infty)$ breaks down when $0 \lt \kappa \lt 1$, it suffices to show that there exists a finite wealth level $\overline{w} \lt \infty$ such that
\begin{equation}\label{eq:db-high-wealth-continuation}
 V_1(w) \gt V_2(w), \qquad \text{for all } w \gt \overline{w}.
\end{equation}
Condition \eqref{eq:db-high-wealth-continuation} guarantees that continuing to work strictly dominates immediate retirement for all sufficiently wealthy agents, thereby ruling out any unbounded retirement region of the form $[w^*,\infty)$.

Because the option to retire voluntarily cannot reduce the agent's welfare, the pre-retirement value function always dominates the never-retiring benchmark value function: $V_1(w) \geq V^{\mathrm{NR}}(w)$ for all $w \gt \underline{w}_1$. For $w \gt w_{0,2}$ (equivalently, $0 \lt \psi \lt \psi_L$), inverting the post-retirement marginal-value relation \eqref{Wealth_post} as $w\to\infty$ (so $\psi\downarrow 0$) yields the asymptotic expansion
\[
\psi=\left(\frac{A}{K}\right)^\gamma w^{-\gamma}+o\!\left(w^{-\gamma}\right), \qquad w\to\infty.
\]
Substituting this asymptotic behavior into \eqref{eq:db-never-retire-value} and \eqref{eq:post_value_positive_region_derivation} (noting that $m_+ + 1 \gt 1 - 1/\gamma$ and $A^\gamma = \kappa^{1-\gamma}$), we obtain
\begin{align*}
V^{\mathrm{NR}}(w)-V_2(w)&=\frac{1}{(1-\gamma)K^{\gamma}} \left(w+\frac{Y_1}{r}\right)^{1-\gamma}
-\left[\frac{m_+}{m_++1}D_1\psi^{m_++1} + \frac{A}{(1-\gamma)K}\psi^{-\frac{1-\gamma}{\gamma}}\right]\\
&=\frac{1}{(1-\gamma)K^{\gamma}} w^{1-\gamma} - \frac{A}{(1-\gamma)K}\psi^{-\frac{1-\gamma}{\gamma}}+o\!\left(w^{1-\gamma}\right)\\
&=\frac{1}{(1-\gamma)K^{\gamma}} w^{1-\gamma} - \frac{A^{\gamma}}{(1-\gamma)K^\gamma} w^{1-\gamma}+o\!\left(w^{1-\gamma}\right)\\
&=\frac{1-\kappa^{1-\gamma}}{(1-\gamma)K^{\gamma}} w^{1-\gamma}+o\!\left(w^{1-\gamma}\right), \qquad w\to\infty.
\end{align*}
When $0 \lt \kappa \lt 1$, the factor $\frac{1-\kappa^{1-\gamma}}{1-\gamma}$ is strictly positive for all $\gamma \neq 1$ (since $1-\kappa^{1-\gamma} \gt 0$ when $\gamma \lt 1$, whereas $1-\kappa^{1-\gamma} \lt 0$ and $1-\gamma \lt 0$ when $\gamma \gt 1$). Consequently, $V_1(w) \geq V^{\mathrm{NR}}(w) \gt V_2(w)$ holds for all sufficiently large $w$, completing the proof.~~~\textbf{Q.E.D.}

\section{Proof of Proposition~\ref{gpsdb:no-retirement-income}} \label{app:proof_no_retirement}

We divide the proof into two steps: first, we construct an alternative never-retiring policy that dominates any retirement strategy, establishing $V_1(w)=V^{\mathrm{NR}}(w)\ge V_2(w)$; second, we show that $\mathcal{R}$ contains at most one point.

\paragraph{Step 1: Dominance of perpetual continuation ($V_1 = V^{\mathrm{NR}} \ge V_2$).}
By Corollary~\ref{cor:original}, $V_1=V$, so it suffices to compare $V^{\mathrm{NR}}$ with the payoff of every policy in $\mathcal{A}(w)$. Let $(c_1,\pi_1)$ and $(c_2,\pi_2)$ denote admissible pre- and post-retirement consumption and portfolio policies associated with a voluntary retirement time $\tau$. Construct an alternative never-retiring strategy $(\widetilde{c},\widetilde{\pi})$ by setting
\[
\widetilde{c}_t=
\begin{cases}
c_{1,t}, & t \lt \tau,\\
c_{2,t}+Y_1-Y_2, & t\geq\tau,
\end{cases}
\qquad
\widetilde{\pi}_t=
\begin{cases}
\pi_{1,t}, & t \lt \tau,\\
\pi_{2,t}, & t\geq\tau.
\end{cases}
\]
Under $(\widetilde{c},\widetilde{\pi})$, the agent remains in the labor force indefinitely, maintains the original post-retirement portfolio allocation after $\tau$, and consumes the post-retirement amount plus the income differential $Y_1-Y_2 \gt 0$.

Let $W$ and $\widetilde{W}$ denote the wealth processes generated by the original and alternative policies, respectively, starting from the same initial wealth $w \gt \underline{w}_1$ under the same Brownian motion $\beta$. Clearly, $\widetilde{W}_t = W_t$ for all $t \le \tau$. For $t \geq \tau$, using the identity $Y_1-\widetilde{c}_t = Y_1-(c_{2,t}+Y_1-Y_2) = Y_2-c_{2,t}$, the alternative wealth process evolves according to
\begin{align}
\mathrm{d}\widetilde{W}_t
&=\left[r\widetilde{W}_t+Y_1-\widetilde{c}_t+(\mu-r)\widetilde{\pi}_t\right]\mathrm{d}t
+\sigma\widetilde{\pi}_t\,\mathrm{d}\beta_t \nonumber\\
&=\left[r\widetilde{W}_t+Y_2-c_{2,t}+(\mu-r)\pi_{2,t}\right]\mathrm{d}t
+\sigma\pi_{2,t}\,\mathrm{d}\beta_t,
\label{eq:Y2_c2}
\end{align}
which coincides with the post-retirement wealth dynamics of $W_t$. Since $\widetilde{W}_\tau=W_\tau$, uniqueness of strong solutions implies $\widetilde{W}_t=W_t$ for all $t\geq 0$. Moreover, because $\underline{w}_1 \lt \underline{w}_2$, we have
\[
\widetilde{W}_t = W_t \geq \underline{w}_2 \gt \underline{w}_1, \qquad t\geq\tau,
\]
so $(\widetilde{c},\widetilde{\pi})$ satisfies the pre-retirement borrowing constraint at all times, with strictly positive consumption $\widetilde{c}_t \geq Y_1-Y_2 \gt 0$ for $t\ge\tau$. Furthermore, the constant consumption shift preserves the admissibility integrability conditions.

Under the assumptions $0 \lt \kappa \lt 1$ and $Y_1-Y_2\geq\kappa L$, for every $t\geq\tau$ we have
\begin{equation}\label{eq:c1_c2}
\widetilde{c}_t-\kappa(c_{2,t}+L)
=(1-\kappa)c_{2,t}+Y_1-Y_2-\kappa L\geq 0.
\end{equation}
Because the CRRA utility function $x \mapsto x^{1-\gamma}/(1-\gamma)$ is strictly increasing, \eqref{eq:c1_c2} implies
\begin{equation} \label{eq:utility_comparison}
U_1(\widetilde{c}_t)
=\frac{\widetilde{c}_t^{1-\gamma}}{1-\gamma}
\geq \frac{[\kappa(c_{2,t}+L)]^{1-\gamma}}{1-\gamma}
=U_2(c_{2,t}), \qquad t\geq\tau.
\end{equation}
Thus, $(\widetilde{c},\widetilde{\pi})$ yields identical utility flow before $\tau$ and weakly higher utility flow at all times after $\tau$. Taking expectations gives
\[
\E\left[\int_0^\infty e^{-\rho t}U_1(\widetilde{c}_t)\,\mathrm{d}t\right]
\geq
\E\left[\int_0^\tau e^{-\rho t}U_1(c_{1,t})\,\mathrm{d}t+\int_\tau^\infty e^{-\rho t}U_2(c_{2,t})\,\mathrm{d}t\right].
\]
The payoff of the never-retiring policy $(\widetilde{c},\widetilde{\pi})$ is at most $V^{\mathrm{NR}}(w)$: repeating the proof of Lemma~\ref{lem:upper} with $\tau=\infty$ and $\Vt_1$ replaced by $\Vt^{\mathrm{NR}}$, which satisfies $\Lt\Vt^{\mathrm{NR}}+Y_1\phi+\Ut=0$, gives the bound $\Vt^{\mathrm{NR}}(y)+yw$ for every $y>0$, and $\inf_{y>0}[\Vt^{\mathrm{NR}}(y)+yw]=V^{\mathrm{NR}}(w)$. Taking the supremum over all admissible policies with a retirement decision yields $V^{\mathrm{NR}}(w)\geq V(w)=V_1(w)$. Since never retiring ($\tau=\infty$) is itself an admissible choice, $V_1(w)\geq V^{\mathrm{NR}}(w)$ also holds; hence
\[
V_1(w)=V^{\mathrm{NR}}(w), \qquad w \gt \underline{w}_1.
\]
In particular, for $w\geq\underline{w}_2$, immediate retirement is feasible, so $V^{\mathrm{NR}}(w)=V_1(w)\geq V_2(w)$.

\paragraph{Step 2: The retirement region contains at most one point.}
Suppose that $\phi_U<\phi_L$. By Theorem~\ref{thm:main}(b)(ii), for $w\in(\underline{w}_1,w_L)$, a non-empty interval since $w_L\ge\underline{w}_2>\underline{w}_1$, we have $V_1'(w)=\phi$ with $X_L(\phi)=w$, whereas $V^{\mathrm{NR}\prime}(w)=\phi'$ with $X^{\mathrm{NR}}(\phi')=w$, where $X^{\mathrm{NR}}(\phi)=\frac1K\phi^{-1/\gamma}-\frac{Y_1}{r}$. Since $V_1=V^{\mathrm{NR}}$ by Step 1, $\phi=\phi'$, and hence $X_L(\phi)-X^{\mathrm{NR}}(\phi)=B_L\phi^{m_-}=0$, contradicting $B_L>0$. Therefore $\phi_L\le\phi_U$, and Theorem~\ref{thm:main}(a) shows that $\mathcal{R}$ contains at most one point; in particular, immediate retirement cannot be optimal on an interior wealth interval of positive length.~~~\textbf{Q.E.D.}

\section{Proof of Lemma~\ref{lem:db-upper-boundary-admissibility}} \label{app:proof_upper_admissibility}

Recall from \eqref{A_kappa} that $A=\kappa^{\frac{1-\gamma}{\gamma}}$, so $\kappa^{1-\gamma}=A^\gamma$. Consequently, the critical threshold $\psi_L$ can be rewritten as
\begin{equation}\label{eq:db-psi-L-representation}
 \psi_L=\kappa^{1-\gamma}L^{-\gamma}
 =\left(\frac{A}{L}\right)^{\gamma}.
\end{equation}
Under the assumptions $0 \lt \kappa \lt 1$ and $L+Y_2 \gt Y_1$, both $\phi_U^\circ$ in \eqref{eq:db-explicit-upper-root} and $\psi_L$ in \eqref{eq:db-psi-L-representation} are strictly positive. Because $\gamma \gt 0$ and $L \gt 0$, raising both sides of the inequality $\phi_U^\circ \lt \psi_L$ to the power $1/\gamma$ preserves the inequality direction and yields
\[
 \frac{r(\gamma m_++1)(1-A)} {m_+(1-\gamma)K(L+Y_2-Y_1) } \lt \frac{A}{L},
\]
which rearranges directly to condition \eqref{eq:db-upper-boundary-parameter-condition}.~~~\textbf{Q.E.D.}

\section{Proof of Corollary~\ref{cor:configurations}} \label{app:proof_configurations}

Let $0<\kappa<1$. If $F_{m_+}$ takes a positive value, then $\{I_+<0\}\neq\emptyset$ by Proposition~\ref{prop:Fint}, so $\mathcal{N}\neq\emptyset$ and $u_1>0$ by Lemma~\ref{lem:S_sign}(c), and the proof of Lemma~\ref{lem:phiLU} shows that $\phi_U\in(u_1,u_2)$ is the unique zero of $I_+$ in $(u_1,u_2)$ with $I_+>0$ on $(0,\phi_U)$; hence $\phi_U$ is the smallest zero of $F_{m_+}$. Since $I_-$ is strictly increasing on $(u_1,u_2)$ and positive on $[u_2,\infty)$, $\phi_U<\phi_L$ if and only if $I_-(\phi_U)<0$, i.e., if and only if $F_{m_-}(\phi_U)>0$; and then $\phi_L$ is the unique zero of $I_-$, hence of $F_{m_-}$, in $(\phi_U,\infty)$. If $F_{m_+}$ takes no positive value, then $\phi_U=\infty$ and Theorem~\ref{thm:main}(a) applies. In the retirement case, $w_U<\infty$ by Proposition~\ref{prop:db-high-wealth-continuation}.

The configurations follow from Theorem~\ref{thm:main} and the monotonicity of $\mathcal{W}_2$, which is strictly decreasing on $(0,\psi_\nu]$ and equal to $\underline{w}_2$ on $[\psi_\nu,\infty)$: if $\phi_L\le\phi_U$, $\mathcal{R}$ has at most one point, and $\mathcal{R}=\emptyset$ unless $\phi_L=\phi_U$; if $\phi_L\ge\psi_\nu>\phi_U$, then $w_L=\underline{w}_2<w_U$; if $\phi_L<\psi_\nu$, then $\underline{w}_2<w_L<w_U$; and if $\phi_L>\phi_U\ge\psi_\nu$, then $w_L=w_U=\underline{w}_2$.~~~\textbf{Q.E.D.}

\section{Proof of Corollary~\ref{cor:explicit_criterion}} \label{app:proof_explicit_criterion}

On $(0,\psi_L)$, $F_{m_\pm}$ are given by the first lines of \eqref{eq:Fminus}--\eqref{eq:Fplus}. Since $(A-1)/(1-\gamma)<0$ for $0<\kappa<1$ and $\gamma m_++1>0$, the first line of \eqref{eq:Fplus} is strictly increasing in $\phi$, from $-\infty$ at $0+$ to $-\frac{m_+(Y_1-L-Y_2)}{r(m_++1)}$ at $\infty$. It therefore has a positive zero if and only if $L+Y_2>Y_1$, and this zero is $\phi_U^\circ$ in \eqref{eq:db-explicit-upper-root}.

\noindent(ii)$\Rightarrow$(i). Under \eqref{eq:four_conditions}, $F_{m_+}(\phi_U^\circ)=0$ because $\phi_U^\circ<\psi_L$. By Lemma~\ref{lem:F_at_psiL}, $F_{m_-}(\psi_L)<0<F_{m_-}(\phi_U^\circ)$. The intermediate value theorem gives $\phi_2\in(\phi_U^\circ,\psi_L)$ with $F_{m_-}(\phi_2)=0$, so that (i) holds with $\phi_1:=\phi_U^\circ$.

\noindent(i)$\Rightarrow$(ii). The zero $\phi_1$ of $F_{m_+}$ in $(0,\psi_L)$ forces $L+Y_2>Y_1$ and $\phi_1=\phi_U^\circ<\psi_L$. Since $F_{m_+}$ is strictly increasing on $(0,\psi_L)$, $F_{m_+}<0$ on $(0,\phi_1)$ and $F_{m_+}>0$ on $(\phi_1,\psi_L)$. Hence $\phi_U=\inf\{F_{m_+}>0\}=\phi_1$ by Definition~\ref{def:phiLU}, and $I_+(\phi_1)=0$. In particular $\mathcal{N}\neq\emptyset$, and because $I_+>0$ on $(0,u_1]$ (proof of Lemma~\ref{lem:phiLU}), $\phi_1>u_1$. Now $\phi_2>\phi_1>u_1$ is a zero of $I_-$. On $(u_1,u_2)$, $I_-$ is strictly increasing, and $I_->0$ on $[u_2,\infty)$. Hence $\phi_2\in(u_1,u_2)$ is the unique zero of $I_-$ in $(u_1,\infty)$, with $I_-<0$ on $(u_1,\phi_2)$ and $I_->0$ on $(\phi_2,\infty)$. By Proposition~\ref{prop:Fint}, $F_{m_-}(\phi_1)>0$ and $F_{m_-}(\psi_L)<0$; the latter means $Y_1>\widetilde{Y}_1$ by Lemma~\ref{lem:F_at_psiL}. This proves \eqref{eq:four_conditions}. It also shows $\phi_L=\sup\{F_{m_-}>0\}=\phi_2$ by Definition~\ref{def:phiLU}.

The remaining statements follow from Theorem~\ref{thm:main}(b): \eqref{eq:db-lower-boundary-equation} is the first line of \eqref{eq:Fminus}; the wealth boundaries follow from $\mathcal{W}_2=W_+$ on $(0,\psi_L)$, with $w_{0,2}<w_L$ because $\phi_L<\psi_L$; and \eqref{eq:db-coefficients} is \eqref{eq:Bmatch} evaluated with $\mathcal{W}_2=W_+$.~~~\textbf{Q.E.D.}

}
\end{document}